\documentclass[12pt]{article}
\usepackage{amsfonts,amsmath,amssymb}
\usepackage{enumerate}
\usepackage{hyperref}
\usepackage{bbm}
\usepackage{nicefrac}

\newcommand{\be}{\begin{equation}}
\newcommand{\ee}{\end{equation}}

\newcommand{\bea}{\begin{eqnarray}}
\newcommand{\eea}{\end{eqnarray}}

\newcommand{\bes}{\begin{subequations}}
\newcommand{\ees}{\end{subequations}}

\newcommand{\Tr}{\mbox{Tr}}

\def\prepG{\mathcal{G}}
\def\prepF{\mathcal{F}}

\newcommand{\Real}{\textrm{Re}}
\newcommand{\Imag}{\textrm{Im}}
\newcommand{\MM}{{\cal M}}
\newcommand{\KK}{{\cal K}}

\usepackage{multirow}
\usepackage{rotating}

\newcommand{\eq}[1]{\begin{equation}#1\end{equation}}
\newcommand{\spl}[1]{\begin{split}#1\end{split}}
\newcommand{\al}[1]{\begin{align}#1\end{align}}
\newcommand{\subeq}[1]{\begin{subequations}#1\end{subequations}}
\def\d{{d}}
\def\Re           {{\rm Re\hskip0.1em}}
\def\Im           {{\rm Im\hskip0.1em}}

\newcommand\bbone{\ensuremath{\mathbbm{1}}}

\newcommand{\sym}{\mathbb{C}}    % symplectic metric
\newcommand{\KOm}{K_\Omega}      % Kaehler potential for Omega deformations
\newcommand{\warp}{\Delta}       % warp factor

\begin{document}

\makeatletter
\renewcommand{\theequation}{\thesection.\arabic{equation}}
\@addtoreset{equation}{section}
\makeatother

\begin{titlepage}

$\,$

\vfill

\vfill

\begin{center}
   \baselineskip=16pt
   \begin{Large}\textbf{
All homogeneous {\emph N}$\,$=$\,$2$\,$ M-theory \\[8pt]
 truncations with supersymmetric AdS$_4$ vacua}
   \end{Large}
   		
\vspace{25pt}
		
{Davide Cassani$^{1}$, Paul Koerber$^{2}$ and Oscar Varela$^{3}$}
		
\vspace{25pt}

	\begin{small}

	{\it ${}^1$ Department of Mathematics, King's College London,\\
	The Strand, London WC2R 2LS, United Kingdom}

	{\tt davide.cassani at kcl.ac.uk}

	\vspace{15pt}

	{\it ${}^2$ Instituut voor Theoretische Fysica, Katholieke Universiteit Leuven\\
	Celestijnenlaan 200D, B-3001 Leuven, Belgium}

	{\tt koerber at itf.fys.kuleuven.be}
		
	\vspace{15pt}
		
	{\it ${}^3$	Institute for Theoretical Physics and Spinoza Institute,\\
	Utrecht University, 3508 TD Utrecht, The Netherlands}
	
	{\tt o.varela at uu.nl}
	
	\end{small}

\vskip 50pt

\end{center}

\begin{center}
\textbf{Abstract}
\end{center}

\begin{quote}

We study consistent truncations of M-theory to gauged $N=2$ supergravity in four dimensions, based on a large class of SU(3)-structures in seven dimensions. We show that the gauging involves isometries of the vector multiplet scalar manifold as well as the Heisenberg algebra and a special isometry of the hyperscalar manifold.
As a result, non-abelian gauge groups and new non-trivial scalar potentials are generated.
Then we specialize to all homogeneous SU(3)-structures supporting supersymmetric AdS$_4$ vacua. These are the Stiefel manifold $V_{5,2}$, the Aloff--Wallach spaces $N(k,l)$, the seven-sphere (seen as SU(4)/SU(3) or Sp(2)/Sp(1)) and the $M^{110}$ and $Q^{111}$ coset spaces. For each of these cases, we describe in detail the $N=2$ model and discuss its peculiarities.

\end{quote}

\vfill

\end{titlepage}

\tableofcontents

%\newpage

%%%%%%%%%%%%%%%%%%%%%%%%%%%%%%%%%%%%%%%%%%%%%%%%%%%%%%%%%%%%%%%%%%%%%%%%%%
\section{Introduction}

A fruitful spin-off  of the AdS/CFT-motivated classification of supersymmetric string and M-theory backgrounds containing AdS factors has been the establishment of consistent truncations of ten- and eleven-dimensional supergravity down to lower-dimensional gauged supergravities with AdS vacua \cite{buchelliu,vargaun1,Gauntlett:2006ai,Gauntlett:2007sm,vargaun2,ExploitingN=2,IIBonSE,vargaun3, LiuSzepietowskiZhao,T11red,granabena,OColgainSamtleben, caskoerbertrisak, OColgain:2011ng}. Consistent truncations in this context provide well-defined interacting theories for a finite number of modes on AdS, whose dynamics is guaranteed to also solve the higher-dimensional theory. Consistently truncated theories are thus very useful to economically describe dual field-theoretic phenomena away from the superconformal point, like RG evolution (see e.g.\cite{Girardello:1998pd,Freedman:1999gp,stretched2,Bobev:2009ms,gauntlettholo2}), finite temperature and chemical potential behaviour (e.g. \cite{Denef:2009tp, SupercSupers, gauntlettholo1, Aprile:2011uq, Aprile:2010ge, Bobev:2011rv,Herzog:2009gd}), or deformations into non-relativistic regimes (e.g. \cite{maldtach,Gregory:2010gx,gauntlettdonos2,cassanilifshitz}). The interest in consistent truncations is of course not limited to a holographic context: well-known early results include the maximally supersymmetric, spherical truncations to four dimensions of \cite{deWitS7} (recently updated in \cite{Nicolai:2011cy}); to seven dimensions of \cite{nasvam1,nasvam2}; and partial results to five dimensions in e.g.\ \cite{SO6truncationIIB}.

This paper is a step forward in the programme of classifying lower-dimensional supergravity theories that uplift to eleven-dimensional (11D) supergravity.
We establish all possible 4D, $N=2$ gauged supergravities that arise from left-invariant consistent truncation of 11D supergravity on 7D coset manifolds $M_7 = G/H$, and that contain supersymmetric AdS$_4$ vacua. These vacua can preserve either the full $N=2$ supersymmetry of the action, or spontaneously break it to $N=1$. By left-invariant (LI) truncation we mean that the Kaluza--Klein tower of M-theory on $M_7=G/H$ is truncated to the sector  invariant under the left-action of the group $G$. This ensures consistency by a symmetry principle: $G$-neutral modes cannot source $G$-charged fields.

The requirement that the truncated theory is $N=2$ supergravity implies that the coset space admits a left-invariant SU(3)-structure.
 This condition, together with the further requirement on the existence of a supersymmetric AdS$_4$ solution, specifies the possible internal manifolds to the list in table \ref{cosetlist}. Indeed, we have shown that, if  homogeneity of the internal space is assumed, the most general supersymmetric M-theory background containing AdS$_4$ is necessarily of Freund--Rubin type. In other words, homogeneity of $M_7$ rules out warp factors, trivially, and also internal fluxes of the 11D four-form $G_4$. Homogeneous supersymmetric AdS$_4 \times M_7$ Freund--Rubin solutions to 11D supergravity were classified long ago \cite{casromoverview,KKreview}, and the internal manifolds $M_7$ that also admit a left-invariant SU(3)-structure are precisely the Stiefel manifold $V_{5,2}$, the Aloff--Wallach spaces $N(k,l)$, the seven-sphere (seen as SU(4)/SU(3) or Sp(2)/Sp(1)) and the $M^{110}$ and $Q^{111}$ coset spaces listed in table \ref{cosetlist}.

\begin{table}
\begin{center}
\begin{tabular}{clccccc}
\hline
	& $M_7$       &    	$G$				&				$H$				&		$B_6$					&				$N$	 AdS$_4$
\\
\hline
\\[-7pt]
	&  $S^7$	&    			SU(4)	&			SU(3)		&		$\mathbb{CP}^3$					&			$2$
\\
	&  $M^{110}$  &   	SU(3)$\,\times\,$SU(2)	\qquad &	SU(2)$\,\times\,$U(1)\;\,	&	 $\mathbb{CP}^2 \times \mathbb{CP}^1 $		 &	 $2$
\\
	& $Q^{111}$   &   SU(2)$^3$	&				U(1)$^2$	 &	 $\mathbb{CP}^1 \!\times \mathbb{CP}^1\! \times \mathbb{CP}^1 $	&	 	  $2$
\\%[5pt]
	&  $V_{5,2}$     & 	SO(5)	&				SO(3)	&	  $Gr(5,2) $			&	$2$
\\[2pt]
\hline
\\[-12pt]
	&  $S^7$	&  		Sp(2)	&			Sp(1)		&		  $\mathbb{CP}^3$					&				$1$ and $2$	
\\
	&  $N(1,1)$	   &  	SU(3)	&				U(1)	&	  $F(1,2) $		&	$1$ and $2$		
\\[2pt]
\hline
\\[-12pt]
%%%
%
	& $N(k,l)$	   & SU(3)	&				U(1)	&	   $F(1,2) $		&		$1$
\\%[-3pt]
%
%%%
%
	& $N(1,-1)$	&   	SU(3)	&				U(1)	&	 $F(1,2) $		&	$1$	
\\
\end{tabular}
\caption[coset list]{All coset manifolds $M_7=G/H$ admitting a $G$-invariant SU(3)-structure and
supporting a supersymmetric, homogeneous AdS$_4 \times M_7$ compactification of 11D supergravity. These are also all compact 7D coset manifolds  that admit a $G$-invariant SU(3)-structure and do not have U(1) factors. The fourth column denotes the 6D base $B_6$ of the 7D manifold (cf.\ section \ref{subsec:SU3str}); here, $Gr$ denotes the Grassmannian and $F$ the flag manifold. The last column gives the amount of supersymmetry of the AdS$_4$ solutions within our $N=2$ models.}\label{cosetlist}
\end{center}
\end{table}

The list of \cite{casromoverview,KKreview} actually comprises other homogeneous supersymmetric solutions of 11D supergravity to which a LI consistent truncation can be associated, but for these instances the latter is not based on an SU(3)-structure (hence does not lead to $N=2$ supergravity) and has already been covered by previous work.
Indeed, going through the list of \cite{casromoverview,KKreview}, we find that SO(5)/SO(3)$_{\rm max}$ and Spin(7)/G$_2$ admit no new LI truncation to $N=1$ supergravity beyond the universal weak G$_2$ truncation of \cite{vargaun2}.\footnote{Only recently have compact, homogeneous $G$-invariant G$_2$-structure seven-folds been classified in the mathematics literature \cite{Reidegeld09}. The results agree with \cite{casromoverview,KKreview}.} All $N=4$ and $N=3$ LI cases were covered in the universal tri-Sasakian truncation of \cite{caskoerbertrisak}. Moreover, Freund--Rubin backgrounds with $4 \leq N \leq 8$ can only arise when $M_7$ is (an orbifold of) the round  $S^7$ and, as in the $N=8$ truncation of \cite{deWitS7}, the associated Kaluza--Klein ans\"atze necessarily involve inhomogeneous deformations. We can thus conclude that the consistent truncations in this paper, along with those of \cite{vargaun2,caskoerbertrisak} are, up to subtruncation, all the possible LI consistent truncations of 11D supergravity on homogeneous 7D manifolds with supersymmetric AdS$_4$ vacua. Note that this conclusion crucially relies on left-invariance: we can make no claims about further possible consistent truncations which, like the $N=8$ truncation on $S^7$, deform inhomogeneously the internal space.

Our results actually reach further generality in the following way. Rather than analysing the possible $N=2$ theories associated to the cosets of table \ref{cosetlist}  on a case-by-case basis, we instead introduce a family of manifolds with SU(3)-structure that contains the cosets as particular cases, and perform the dimensional reduction on this larger general class. In order to do this, we borrow some usual techniques from the flux compactification literature. Namely, we will proceed similarly to the familiar, and closely related, compactifications of type II supergravity on SU(3)-structure six-folds, which themselves generalise the Calabi--Yau compactifications with background fluxes (see e.g.\ \cite{fluxrev1} for a review). Following \cite{louismicu2,granaN2part1,minpoor,Micu:2006ey}, we assume that the 7D manifold is equipped with a set of forms, typically not harmonic but instead postulated to obey certain algebraic and differential conditions. We then use this set of forms to expand the supergravity three-form and the SU(3)-structure forms, whose corresponding moduli we show to be governed by special K\"ahler geometry.
The resulting truncated theory is $N=2$ supergravity, coupled to vector and hypermultiplets, with a rich family of both electric and magnetic gaugings inherited from the intrinsic torsion of the SU(3)-structure and from the four-form flux.
We find that in general these gaugings involve both isometries of the vector multiplet and of the hypermultiplet scalar manifolds, leading to a non-abelian gauge algebra, and to new, non-trivial scalar potentials.
Examples of M-theory reductions to $N=2$ supergravity with charged vector multiplet scalars are known \cite{Aharony:2008rx,Looyestijn:2010pb}, and we extend these results to a larger class of gaugings, including a special isometry (also studied in \cite{Looyestijn:2010pb}) and the Heisenberg algebra of the hyperscalar manifold.

An important feature of our reduction scheme is that the algebraic restrictions on the expansion forms will be imposed here with no integration over the internal manifold. This stronger requirement ensures  that reduction on our general class of manifolds can in many cases be performed not only at the level of the action but also at the level of the equations of motion. The latter feature implies consistency of the truncation. The benefits of this approach are manifold. It not only allows us to simultaneously manage the reduction on the cosets of table \ref{cosetlist} in a systematic and unified way, but it also allows to include further consistent truncations on a larger class of internal spaces not any longer required to be homogeneous. Our general class of internal spaces contains, for example, any Sasaki--Einstein (SE$_7$) space. Accordingly, our 4D $N=2$ family of supergravities includes the universal truncation of 11D supergravity on SE$_7$ \cite{vargaun2}. It also includes  the largest $N=2$ subtruncation of the universal $N=4$ truncation on tri-Sasakian manifolds \cite{caskoerbertrisak,triS}. Moreover, our family of $N=2$ theories also contains the models arising from consistent truncation of type IIA on nearly-K\"ahler and homogeneous SU(3)-structure six-folds~\cite{ExploitingN=2,poorNK}, in the limit of vanishing Romans mass. As an aside, it is interesting to note that our family of $N=2$ gaugings does also allow for a further extension, no longer immediately compatible with a consistent 11D uplift, by an additional parameter that in type IIA admits a straightforward interpretation as a Romans mass.

Contact between our general family of SU(3)-structure manifolds and the cosets of table \ref{cosetlist} is achieved when the postulated set of forms is taken to be the set of LI forms on each coset. In fact, the assumptions on the forms defined on our general class of manifolds are just abstractions of those defined by the LI forms. Obtaining the particular theory arising for each coset from our general family of $N=2$ gauged theories is just a matter of parameter-fixing.

%%%%%%%%%

\begin{table}[t!]
\centering
%\small
\tabcolsep=0.15cm
\begin{tabular}{lcccccc}
\hline
$M_7$ &	$n_V$	& $n_H$ & VM s.m. & $\mathcal{F}$ &  HM s.m. & $\mathcal{G}$ \\[2pt]
\hline
\\[-10pt]
$S^7\!\!=\! \frac{{\rm SU}(4)}{{\rm SU}(3)}$	& $1$ &	$1$ &	$\frac{{\rm SU}(1,1)}{{\rm U}(1)}$ & $- \frac{(X^1)^3}{X^0}$ & $\frac{{\rm SU}(2,1)}{{\rm S(U}(2) \times {\rm U}(1))}$	
& --
\\[10pt]
$M^{110}$ & $2$ & $1$ &	$\left(\frac{{\rm SU}(1,1)}{{\rm U}(1)}\right)^2$ & $-\frac{(X^1)^2 X^2}{X^0}$ & $\frac{{\rm SU}(2,1)}{{\rm S(U(2) \times U(1))}}$ & --
\\[10pt]
$Q^{111}$ & $3$	& $1$ &	$\left(\frac{{\rm SU}(1,1)}{{\rm U}(1)}\right)^3$ & $-\frac{X^1 X^2 X^3}{X^0}$ & $\frac{{\rm SU}(2,1)}{{\rm S(U(2) \times U(1))}}$ & --
\\[10pt]
$V_{5,2}$ & $1$	& $2$ &	$\frac{{\rm SU}(1,1)}{{\rm U}(1)}$ & $-\frac{(X^1)^3}{X^0}$ & $\frac{{\rm G}_{2(2)}}{{\rm SO}(4)}$	 & $-\frac{(Z^1)^3}{3 \sqrt{3} Z^0}$
\\[5pt]
\hline
\\[-10pt]
$S^7 \!\!=\!\frac{{\rm Sp}(2)}{{\rm Sp}(1)}$ & $2$ & $2$ & $\left(\frac{{\rm SU(1,1)}}{{\rm U(1)}}\right)^2$ & $-\frac{(X^1)^2 X^2}{X^0}$ & $\frac{{\rm SO}(4,2)}{{\rm SO}(4) \times {\rm SO}(2)}$ &
$\frac{(Z^0)^2 -(Z^1)^2}{2i}$
\\[10pt]
$N(1,1)$ & $3$ & $2$ &$\left(\frac{{\rm SU}(1,1)}{{\rm U}(1)}\right)^3$	& $-\frac{X^1 X^2 X^3}{X^0}$ & $\frac{{\rm SO}(4,2)}{{\rm SO}(4) \times {\rm SO}(2)}$
& $\frac{(Z^0)^2 -(Z^1)^2}{2i}$			
\\[5pt]
\hline
\\[-10pt]
$N(k,l)$ & $3$ & $1$ &	$\left(\frac{{\rm SU}(1,1)}{{\rm U}(1)}\right)^3$ & $-\frac{X^1 X^2 X^3}{X^0}$ & $\frac{{\rm SU}(2,1)}{{\rm S(U(2) \times U(1))}}$ & --
\\[10pt]
$N(1,-1)$ & $5$ & $1$ &	\!\!$\frac{{\rm SU}(1,1)}{{\rm U}(1)} \times \frac{{\rm SO}(4,2)}{{\rm SO}(4) \times {\rm SO}(2)}$ &
$\frac{X^3[-X^1X^2+(X^4)^2+(X^5)^2]}{X^0}$\ & $\frac{{\rm SU}(2,1)}{{\rm S(U(2) \times U(1))}}$ & --
\\[5pt]
\hline
\end{tabular}
\\[10pt]	
\caption[coset summary]{Summary of some relevant features of the $N=2$ gauged supergravities obtained from LI consistent truncation of 11D supergravity on $M_7$. The table shows the number of vector multiplets ($n_V$) and hypermultiplets ($n_H$), and the respective scalar manifolds (VM s.m. and HM s.m.) with the associated special K\"ahler prepotentials ($\mathcal F$ and $\mathcal{G}$).}\label{cosetsummary}
\end{table}

The cosets of table~\ref{cosetlist} admit a LI moduli space of SU(3)-structures, where two significant ones characterized by specific torsion classes may arise, defining the preserved supersymmetry of the corresponding AdS vacuum. Either one or both structures may be present for each coset. These special LI SU(3)-structures are, on the one hand, the well-known SE$_7$ structure and, on the other hand, a certain, U(1)-invariant, type of weak G$_2$ structure, which can be regarded as a circle fibration over (the vanishing Romans mass limit of) the class of half-flat six-folds discussed in \cite{lt,tomasiellocosets,cosets}. Accordingly, although both types of manifolds give rise to $N=2$ supergravity actions, the former preserve $N=2$ in the vacuum while the latter spontaneously breaks \hbox{$N=2 \rightarrow N=1$}.\footnote{See \cite{Louis:2009xd,Louis:2010ui} for a recent general discussion of spontaneous supersymmetry breaking in gauged $N=2$ supergravity.}
Referring to table~\ref{cosetlist}, the first six cosets can be endowed with a SE$_7$ structure, and in fact exhaust the list of all possible homogeneous SE$_7$ spaces, see e.g.~\cite{Sparks:2010sn}. The last four cosets admit the second type of structure. The two cosets admitting both types of SU(3)-structures in fact admit a LI tri-Sasaki structure, which allows both for a SE$_7$ and (when squashed) for a distinct weak G$_2$ structure.

We have summarized in table~\ref{cosetsummary} the main features of the $N=2$ effective theories arising from LI consistent truncation on the cosets of table \ref{cosetlist}. As noted in~\cite{caskoerbertrisak}, the LI truncation on SU(4)/SU(3) coincides with the universal SE$_7$ truncation of~\cite{vargaun2}, which comprises one vector multiplet and one hypermultiplet. The truncations on $M^{110}$ and $Q^{111}$ merely add one and two Betti vector multiplets, respectively, to the universal SE$_7$ truncation, while that on the Stiefel manifold $V_{5,2}$ includes an additional hypermultiplet. The $N=2$ models on Sp(2)/Sp(1) and $N(1,1)$ that we present here are actually subtruncations of the full $N=4$ tri-Sasakian truncation of \cite{caskoerbertrisak} obtained by imposing a certain $\mathbb{Z}_2$ symmetry, but were not discussed in that paper. In particular, the $N=2$ theory based on Sp(2)/Sp(1) is, like its $N=4$ parent, universally valid for any tri-Sasakian space.
In the symplectic frame specified by the truncation, the gauging is both electric and magnetic across all models (a summary is given in tables \ref{GeomFluxTable}, \ref{GeomFluxCntdTable} below). The gauge group is always abelian and only the hyperscalars pick up charges, except for the Aloff--Wallach space $N(1,-1)$, where there is also an interesting, non-abelian gauging in the vector multiplet sector. It should be noted that some partial results about these 4D truncations have already appeared in the literature: in \cite{Micu:2006ey,Karthauser:2006wb} for the $V_{5,2}$ and the $N(k,l)$ models, and in \cite{Karthauser:2006wb,Klebanov:2010tj} for the $M^{110}$ and $Q^{111}$ models. Reference \cite{Micu:2006ey} also discussed some general features of the dimensional reduction of M-theory on SU(3)-structure seven-folds. However, none of these references specifies both the full overarching $N=2$ theory and the details of each model as we do in the present work.

The structure of the rest of the paper is as follows. In section~\ref{sec:SU3str} we describe the two canonical SU(3)-structures supported by the cosets of table \ref{cosetlist}, and introduce the general class of SU(3)-structure manifolds compatible with these, but not limited to them. We study its moduli space and show it to be governed by special geometry. The reduction on our general class of spaces is performed in section~\ref{sec:DimRed}, where the resulting general family of 4D theories is also presented. This is shown to be compatible with gauged $N=2$ supergravity in section~\ref{sec:N=2sugra}. In section~\ref{sec:TheModels} we retrieve the explicit models of table~\ref{cosetsummary} from our general reduction, and we further discuss their salient features. Section~\ref{sec:discussion} concludes with an outlook. The appendix contains the proof that homogeneous supersymmetric AdS$_4$ solutions to 11D supergravity are of Freund--Rubin type (appendix \ref{sec:susyconds}), an algorithm to compute the prepotential on the LI almost complex structure moduli space of our internal manifolds (appendix~\ref{howtoprepot}), as well as technical details on the dimensional reduction (appendix~\ref{sec:DetailsDimRed}), and matching with $N=2$ supergravity (appendix~\ref{CheckScalarPot}). We have also included a self-contained description of the LI geometry on the relevant cosets  (appendix~\ref{sec:cosets}). Finally, we give (appendix \ref{sec:susyspectra}) the spectrum of each coset model at every supersymmetric critical point, and its arrangement into OSp multiplets.

%%%%%%%%%%%%%%%%%%%%%%%%%%%%%%%%%%%%%%%%%%%%%%%%%%%%%%%%%%%%%%%%%%%%%%%%%%
\section{SU(3)-structures and their moduli space}   \label{sec:SU3str}

\subsection{SU(3)-structures in seven dimensions}\label{subsec:SU3str}

An SU(3)-structure on a 7D manifold $M_7$ is specified by a real one-form $\eta$, a real two-form $J$ and a complex decomposable three-form $\Omega$.\footnote{A complex three-form is decomposable if it can be written as the wedge product of three complex one-forms.} The one-form $\eta$ comes with a dual vector, $X$, such that $\iota_X\eta = 1$.
These must be globally defined, and need to satisfy the algebraic constraints
\begin{eqnarray} \label{SU3stralg}
\iota_X J= \iota_X  \Omega = 0\,,\qquad \Omega \wedge J = 0\,,\qquad \Omega \wedge \overline\Omega \neq 0 \, ,\qquad J \wedge J \wedge J \neq 0\,.
\end{eqnarray}
We will normalize $\Omega$ such that $\Omega \wedge \overline\Omega = -\tfrac{4i}{3} J \wedge J \wedge J$.

Since SU(3)$\,\subset\,$SO(7), an SU(3) structure determines a metric on $M_7$. This is constructed as follows. Locally, $X$ specifies a 1D subspace, while $\eta$ identifies a 6D, by definition orthogonal, subspace $B_6$ (by requiring its tangent space to be
spanned by all vectors $Y$ satisfying $\iota_Y \eta = 0$). From the first condition in \eqref {SU3stralg}, it follows that $J$ and
$\Omega$ live on $B_6$. They will then induce a metric $g_6$ on $B_6$ just like a standard SU(3)-structure in 6D (see e.g.~\cite{gengeomreview}).
Denoting the line element of $g_6$ by $ds^2(B_6)$, the 7D line element can in the end be written as
\begin{equation} \label{ds7split}
ds^2(M_7) \,=\, ds^ 2(B_6) + \eta^2 \, ,
\end{equation}
and the associated volume form is ${\rm vol_7} = \frac{1}{6} J\wedge J\wedge J \wedge \eta$.

SU(3)-structures are classified by their intrinsic torsion. This splits in torsion classes, which transform in irreducible representations of SU(3), and
parameterize the exterior derivative of the invariant forms as \cite{Dall'Agata:2003ir,BehrndtCveticLiu}
\eq{\spl{\label{SU3tcl}
d \eta &\,=\, R \, J + T_1 + 2{\rm Re}(\bar{V}_1  \lrcorner \Omega )  + \eta \wedge W_0\, , \\[3pt]
d J &\,=\, \tfrac{3}{2}\, \Imag (\bar{W}_1 \Omega) + W_4 \wedge J +  W_3
+ \eta \wedge \big[ \,\tfrac{2}{3} {\rm Re}E \, J + T_2 + 2 {\rm Re}(\bar{V}_2  \lrcorner \Omega) \,\big]  \,, \\[3pt]
d \Omega &\,=\, W_1 J \wedge J + W_2 \wedge J + \bar{W}_5 \wedge \Omega  + \eta \wedge \big( E \,\Omega -4V_2 \wedge J + S \big) \,.
}}
Here, the real scalar $R$ and the complex scalars $W_1$, $E$ are SU(3)-singlets; $T_1$, $T_2$ are real two-forms, while $W_2$ is a complex two-form, all living on $B_6$ and transforming in the $\mathbf{8}$ of SU(3); $W_3$ is a real three-form, of type $(2,1)+(1,2)$ with respect to the almost complex structure $I$, and transforming in the $\mathbf{6}+\bar{\mathbf{6}}$; $S$ is a complex $(2,1)$-form in the $\mathbf{6}$. Finally,
$V_1$, $V_2$, $W_5$ are $(1,0)$-forms in the $\mathbf{3}$, and $W_0$, $W_4$ are real one-forms in the $\mathbf{3}+\bar{\mathbf{3}}$. The five torsion classes $W_k,$ $k=1,\ldots 5$, correspond to the better known torsion classes of 6D SU(3)-structures \cite{ChiossiSalamon}.

We remark that these torsion classes are not completely independent, as they must satisfy certain consistency constraints coming from
the nilpotency of the exterior derivative (see e.g.~\eqref{tclconstr1}).
It will also turn out that for the manifolds considered in this paper all the SU(3) vector torsion classes, namely $V_1,V_2,W_0,W_4$ and $W_5$, vanish.\footnote{The appearance of such vector torsion classes on a homogeneous manifold would necessarily imply extra left-invariant vectors, and
further reduce the SU(3)-structure.}

A given manifold $M_7$ can admit a whole space of SU(3)-structures, which are distinguished by the different values of their torsion classes. When this is used as a compactification space of 11D supergravity, some specific configurations of the torsion classes are particularly relevant, in that they correspond to supersymmetry being preserved on an AdS$_4 \times M_7$ background. Indeed, the conditions for a supersymmetric AdS vacuum can be expressed as differential conditions on the structure forms, which in turn translate into restrictions on the torsion classes. As we prove in appendix~\ref{sec:susyconds}, when $M_7$ is homogeneous the only allowed maximally symmetric, supersymmetric solutions are of the Freund--Rubin type, which means the four-form flux of 11D supergravity is purely along AdS, and the metric on $M_7$ is Einstein. Within this class of solutions, for the manifolds of table \ref{cosetlist} two classes of supersymmetric conditions arise, leading to vacua preserving a different fraction of supersymmetry.

The first class preserves $N=2$ supersymmetry in 4D, and is characterized by non-vanishing singlet classes only, constrained in such a way that
\begin{eqnarray} \label{SE7stru}
d\eta = 2 J \,, \qquad  dJ= 0\,,\qquad d \Omega = 4i\, \eta \wedge \Omega .
\end{eqnarray}
This defines a Sasaki--Einstein structure on $M_7$. Note that for a solution, $X$ preserves $\eta$, $J$ and the almost complex structure $I$. Hence it is a Killing vector of the metric \eqref{ds7split}. However, by evaluating $\mathcal L_X \Omega = \iota_X d\Omega =  4i \Omega$, we see that $X$ rotates $\Omega$ by a phase; this corresponds to the R-symmetry of the $N=2$ solution.

The other class leads to solutions that are invariant under the U(1) generated by $X$, and which preserve $N=1$ supersymmetry only. The SU(3) torsion classes are constrained as
\eq{\spl{\label{halfflat7d}
d \eta =  \frac i4  W_1 J  + i W_2  \, ,\qquad
d J = \frac{3i}{2}\, W_1\Re \Omega\,,\qquad
d \Omega = W_1 J \wedge J + W_2 \wedge J \,,
}}
with $\Re W_1 = \Re W_2 = 0$. A solution corresponds to a particular $X$-invariant weak G$_2$ holonomy, and can be described as a U(1)-fibration (specified by $\eta$) over a 6D half-flat manifold whose SU(3)-structure is described by $J$ and $\Omega$. The derivation of the conditions \eqref{SE7stru}, \eqref{halfflat7d} is given in appendix \ref{sec:susyconds}.

In the last column of table \ref{cosetlist}, we specified which cosets allow which class of vacua. Interestingly, there are some cosets admitting vacua that belong to both classes. This happens because actually the manifold admits a further reduced structure (namely, a tri-Sasakian structure) preserving an enhanced amount of supersymmetry ($N=3$, see section~\ref{sec:TheModels} for details).

\subsection{A finite-dimensional space of SU(3)-structures} \label{subsec:expansionforms}

In order to dimensionally reduce 11D supergravity on SU(3)-structure manifolds, we need to specify our truncation ansatz. Following previous studies in the literature (see e.g.\ \cite{louismicu2, granaN2part1, Micu:2006ey}), we introduce a finite basis of differential forms in which the SU(3)-structure forms and the 11D supergravity three-form are to be expanded. Since the 7D metric follows from the SU(3)-structure, this will completely fix the truncation ansatz.

We assume that there exists a set of real differential forms on $M_7$, made of: a single one-form $\theta$, $n_V$ two-forms $\omega_i$, $2n_H$ three-forms $\alpha_A$, $\beta^A$, $n_V$ four-forms $\tilde{\omega}^i$, and a single six-form $\tilde \omega^0$, with the range of the indices being $i=1,\ldots, n_V$ and $A=0,1, \ldots, n_H-1$. The values of $n_V$ and $n_H$ are not fixed a priori in our general discussion, and will turn out to correspond to the number of vector multiplets and hypermultiplets in the 4D supergravity theory, respectively.
Using these forms, we parameterize the family of SU(3)-structures on $M_7$ as
\begin{eqnarray} \label{LISU3str}
\eta = e^V \theta , \;\qquad J= e^{-V}  v^i \omega_i \,, \;\qquad \Omega = e^{-\frac{3}{2}V}( Z^A \alpha_A - \prepG_A \beta^A)\,,
\end{eqnarray}
where $V, v^i$ are real and $Z^A$, $\prepG_A$ are complex constant parameters.\footnote{The factors of $e^V$ in the expansions of $J$ and $\Omega$ have been chosen for convenience, in order to ensure that $v^i$ and $Z^A$ become, respectively, the almost symplectic and almost complex structure moduli of the base $B_6$. These naturally arise within a type IIA string frame setup.} When in the next section we will discuss the dimensional reduction of 11D supergravity, these parameters will be promoted to scalar fields depending on the external space-time coordinates. Note that the expansion forms have no dependence on the parameters.\footnote{See \cite{minpoor} for a 6D analysis in which the basis forms are allowed to depend on the moduli.}

We now list a set of sufficient conditions on the forms ensuring that \eqref{LISU3str} indeed defines a family of SU(3)-structures, and that the dimensional reduction of 11D supergravity on $M_7$ goes through, leading to $N=2$ gauged supergravity in 4D.
Similar relations have been discussed previously in the literature (see e.g.\ \cite{louismicu2, granaN2part1, Micu:2006ey}), therefore our presentation will be quick. However, contrarily to e.g.\ \cite{louismicu2, Micu:2006ey}, our constraints will be given with no integration over $M_7$. This strong assumption ensures in many cases that the reduction goes through at the level of the equations of motion as well, which implies consistency of the truncation. For the moment one should take our relations as assumptions on the geometry of a generic compact manifold $M_7$. When in section~\ref{sec:TheModels} we will focus on coset manifolds $G/H$ admitting SU(3)-structure, we will see that the assumptions we are making are all satisfied if one identifies the expansion forms precisely with the forms being invariant under the left-action of the group $G$. However, we stress that the consistent truncation we will establish is valid for some classes of non-homogeneous SU(3)-structures as well.
For instance, a generic Sasaki--Einstein or tri-Sasakian structure also provides a set of forms with the correct properties. Likewise, the fibrations of Calabi--Yau three-folds over a circle studied in \cite{Aharony:2008rx,Looyestijn:2010pb} satisfy our constraints and thus lead to a consistent truncation, at least when just the universal volume modulus and no complex structure deformations are included in the truncation (which leads to $n_V =n_H =1$).

The one-form $\theta$ comes with a dual vector $k$, such that $\iota_k \theta= 1$; hence we have the identification $X = e^{-V}k$.
We assume that all the basis forms apart from $\theta$ live on the 6D space orthogonal to $k$,
\begin{eqnarray} \label{comp1}
\iota_k\omega_i = \iota_k \alpha_A = \iota_k \beta^A = \iota_k \tilde{\omega}^i = 0\,.
\end{eqnarray}
Furthermore, we require
\begin{eqnarray} \label{comp2}
\omega_i \wedge \alpha_A  = \omega_i \wedge \beta^A =  \tilde{\omega}^i \wedge \alpha_A  = \tilde{\omega}^i \wedge \beta^A  = \alpha_A \wedge \alpha_B = \beta^A \wedge \beta^B = 0\, ,
\end{eqnarray}
and
\eq{ \label{normforms}
\omega_i \wedge \tilde{\omega}^j = -\delta_i^j \,\tilde\omega^0 , \qquad \alpha_A \wedge \beta^B = -\delta_A^B \,\tilde\omega^0\,.}
These conditions on the basis forms imply that $\eta$, $J$ and $\Omega$ constructed as in (\ref{LISU3str}) obey the algebraic constraints (\ref{SU3stralg}). The 7D metric eq.~\eqref{ds7split} can be written as
\begin{eqnarray} \label{LImetint}
ds^2(M_7) \,=\, e^{-V} d {s}^2(B_6) + e^{2V} \theta^2\,,
\end{eqnarray}
where $ds^2(B_6)$ is determined by $J$ and $\Omega$. Note that this 6D metric is actually rescaled with respect to the one given in \eqref{ds7split}.
Eqs.~\eqref{comp2}, \eqref{normforms} also imply that there is a symplectic Sp$(2n_H, \mathbb{R})$ structure on the space of three-forms in our basis. The quantities in our dimensional reduction coming from an expansion in the three forms will inherit this symplectic structure. Collecting the three-forms in a vector $\Sigma$ and defining the Sp$(2n_H, \mathbb{R})$ metric $\sym$ as
\eq{
\Sigma = \binom{ \beta^A }{ \alpha_A  }\,,\qquad\qquad
\sym = \left(\!\! \begin{array}{cc} 0 & \delta_A{}^B \\ -\delta^A{}_B & 0 \end{array} \!\!\right),}
we can adopt a convenient symplectic notation, in which the $A,B$ indices are suppressed.
Introducing furthermore the symplectic vector
\eq{
Z = \binom{ Z^A }{ \prepG_A  }\,,
}
we can for instance write the expansion of $\Omega$ in \eqref{LISU3str} as $\Omega \,=\, e^{-\frac{3}{2}V} Z^T \sym \Sigma\,.$ This symplectic notation will be used throughout the paper.

In addition to the above requirements, we also need that in our basis the wedge product of a pair of two-forms can be expanded in the four-forms, with constant coefficients ${\cal K}_{ijk}\,$:
\eq{
\label{omomK}
\omega_i \wedge \omega_j \,=\, {\cal K}_{ijk}\, \tilde{\omega}^k \qquad\Rightarrow\qquad
\omega_i \wedge  \omega_j \wedge \omega_k \,=\, -{\cal K}_{ijk} \,\tilde\omega^0 \,.
}
Let us furthermore introduce the following useful contractions
\be \label{ModintNos}
\KK = \tfrac{1}{6}\mathcal K_{ijk} v^iv^jv^k, \qquad \KK_i  = \mathcal K_{ijk} v^jv^k , 
\ee
in terms of which the 7D volume form can be written as
\eq{{\rm vol}_7 = -e^{-2V} \KK\,  \tilde\omega^0\wedge \theta.}
Sometimes we need a canonical volume form independent of the choice of a particular
metric and then we will take $-\tilde \omega^0 \wedge \theta$.

Moreover, we need that the set of basis three-forms be closed under the action of the 7D Hodge dual associated with the metric (\ref{LImetint}). While this is automatic for the even forms (see eq.~\eqref{Hodgeomegai} below), for the three-forms $\Sigma$ we need to assume that the following relation holds:
\be
\label{hodgethree}
* \Sigma \,=\, - \mathbb{M} \,\Sigma \wedge \eta\,,
\ee
where the matrix of coefficients $\mathbb M$ is required not to depend on the coordinates on $M_7$, but can depend on the metric parameters. Recalling $*^2= id$, one can show that $\mathbb M$ satisfies $\mathbb M^2 = - \bbone$ and that it is symplectic
(i.e.\ $\mathbb M^T \sym \mathbb M = \sym$), hence it can be written as
\eq{\label{symplmatrM}
\mathbb{M} \,=\,
\left( \begin{array}{cc} (\Imag \MM)^{-1}\Real \MM   \,&\, -(\Imag \MM)^{-1} \\ [1mm]
\Imag \MM  + \Real \MM (\Imag \MM)^{-1} \Real \MM \,&\, -\Real \MM  (\Imag \MM)^{-1}\end{array} \right)  ,
}
where $\mathcal M_{AB}$ is a complex, symmetric matrix.

Finally, we require that the set of basis forms is closed under exterior derivative:
\eq{\spl{ \label{GenExtAlg}
d \theta \,&=\, m^i \omega_i\,,\\[2pt]
d\omega_i \,&=\, q_i{}^j \theta \wedge \omega_j + Q_i^T \sym \Sigma\,,\qquad
d \tilde{\omega}^i \,=\, -  \theta \wedge \tilde{\omega}^j q_j{}^i \,,\qquad d \tilde \omega^0 = 0\,,\\[2pt]
d \Sigma \,&=\, \theta \wedge \mathbb U\, \Sigma  \,-\, \tilde \omega^i Q_i \,,
}}
where we defined
\eq{\label{GeomFluxMatr}
Q_i = \binom{ m_i{}^A }{ e_{iA}  } \,,\qquad
\mathbb{U} = \left(\!\! \begin{array}{cc} v^A{}_B  & t^{AB} \\ s_{AB} & u_A{}^B \end{array} \!\!\right) \, ,
}
and $m^i$, $q_i{}^j$, $m_i{}^A$, $e_{iA}$,  $u_A{}^B$,  $s_{AB} $, $t^{AB} $, $v^A{}_B$ are real constants, with $s_{AB}  = s_{BA} $,  $t^{AB}  =  t^{BA} $, and $v^A{}_B = - u_B{}^A$. It follows that $\mathbb{U}^T \sym + \sym \mathbb{U} = 0$, i.e.\ $\mathbb U$ is in the $\mathfrak{sp}(2n_H, \mathbb{R}$) algebra.
Nilpotency of the exterior derivative imposes the constraints
\eq{\spl{\label{nild2z}
& m^i Q_i = 0 \; , \qquad  m^i q_i{}^j = 0 \, ,\qquad
\mathbb{U}Q_i + q_i{}^j Q_j \,=\, 0 \, , \\[2pt]
& Q_i^T \sym Q_j  - q_{[i}{}^k {\cal K}_{j]kl} m^l=0 \,,
}}
where the expression in the last line is obtained after noting that the exterior derivative of the expression on the right in \eqref{omomK}, together with the fact that on a compact manifold the volume-form cannot be exact, yields
\eq{\label{qKvanishes} q_{(i}{}^l {\cal K}_{jk)l}=0\,.
}
These constants are going to be the geometric fluxes in our dimensional reduction. While for the first part of the paper we will keep them generic, in the second part, when we will turn to our concrete examples, they will take specific (possibly zero) values.

The above relations completely characterize the SU(3)-structure. The SU(3) torsion classes can be evaluated in terms of the parameters $V,v^i,Z^A, \prepG_A$ and of the geometric fluxes. For instance, we find that the scalar torsion classes are given by
\eq{\label{TorsionParam}
R  = \frac{e^{2V} m^i \mathcal{K}_i}{6 \,\mathcal{K}} \, ,\qquad
W_1  = 
\frac{\sqrt 2\,e^{\frac{\KOm}{2}+\frac{V}{2}}}{3 \,{\mathcal{K}}^{1/2}} v^i Q_i^T \sym Z \, ,\qquad
E  = -i \, e^{\KOm -V} Z^T \sym \mathbb{U} \bar{Z} \, ,
}
where $\KOm$ is given in eq.~\eqref{csKP} below.
The differential constraints on the basis forms also ensure that the SU(3) vector torsion classes $V_1,V_2,W_0,W_4,W_5$ all vanish. More generally, the constraints ensure that no SU(3) triplets appear in the reduction
(see e.g.\ \cite{granaN2part1} for a discussion on the importance of this fact).

\subsection{Special K\"ahler geometry on the space of deformations} \label{subsec:SKdef}

In this section we will briefly discuss how the finite-dimensional space of SU(3)-structures defined above is described by special K\"ahler geometry. More thorough discussions can be found e.g.\ in \cite{granaN2part1,minpoor}. For the purposes of this section, we can just focus on the forms $J$ and $\Omega$ living on $B_6$, since the scale of the one-form $\eta$ turns out to decouple in the dimensional reduction and corresponds to a trivial extra deformation associated with the 4D dilaton.
It is known that at each point of a 6D space, the deformations of the almost complex structure induced by a decomposable three-form $\Omega$ span a special K\"ahler space \cite{hitchinfunc}, and an analogous result holds for the deformations of the (complexified) almost symplectic form $J$. When the assumptions of the previous section are satisfied, these properties extend globally on the 6D (and 7D) manifold, in analogy with the well-known case of the moduli space of Calabi--Yau three-folds \cite{candelas}.

Let us first consider the almost complex structure deformations.
Given an expansion of $\Omega$ in terms of a symplectic basis as in \eqref{LISU3str}, the $Z^A$, $A=0,1,\ldots,n_H$, are projective coordinates on the space of deformations, which has complex dimension $n_H$, while the $\prepG_A$ are holomorphic functions of the $Z^A$. The $\prepG_A$ are related to the prepotential function $\prepG(Z)$ as $\prepG_A = \partial{\prepG}/\partial{Z^A}$.
Contrarily to the case of $J$-deformations (described below), constructing the prepotential requires some non-trivial extra work. In appendix~\ref{howtoprepot} we solve this problem by showing how the special K\"ahler data can concretely be extracted given a symplectic basis of expansion forms. Namely, we provide a simple algorithm for obtaining the dependent variables $\prepG_B$ as holomorphic functions of the independent variables $Z^A$, as well as the prepotential $\prepG(Z)$.
Having this useful result at hand, the K\"ahler potential $\KOm$ is constructed as
\eq{ \label{csKP}
e^{-\KOm}\, \tilde\omega^0 \,=\,  -i\, e^{3V} \Omega \wedge \bar{\Omega}
\,=\,  i \left( \bar{Z}^A \prepG_A -Z^A \bar{\prepG}_A \right) \tilde\omega^0\,,
}
and the K\"ahler metric $g_{a\bar b}$ on the space of deformations is deduced by introducing (away from the $Z^0=0$ locus) special coordinates $z^a = Z^a/Z^0$, $a=1,\ldots,n_H$, and computing $g_{a\bar b} = \frac{\partial}{\partial z^a}\frac{\partial}{\partial\bar z^{\bar b}} \KOm$.

The period matrix $\mathcal{M}_{AB}$, which relates the upper and lower components of the symplectic section as
\eq{\label{defM}
 \prepG_A = \mathcal{M}_{AB} Z^B \, , \qquad
 D_{c} \prepG_A = \overline{\mathcal{M}}_{AB} D_{c} Z^B \, ,
}
(where $D_c$ is the K\"ahler covariant derivative) is computed in terms of the prepotential via the standard special K\"ahler geometry formula
\eq{ \label{csMAB}
\MM_{AB} = \overline{\prepG}_{AB} + 2i \frac{(\Imag \prepG)_{AC} Z^C (\Imag \prepG)_{BD} Z^D}{(\Imag \prepG)_{CD} Z^C Z^D }\,,
}
where $\prepG_{AB} = \partial^2{\prepG}/\partial{Z^A}\partial{Z^B}$. One can prove that the matrix $\mathcal M$ introduced in the expression \eqref{hodgethree} for the Hodge dual of the basis three-forms, is precisely identified with this period matrix
(see e.g.\ \cite{minpoor}).

We now turn to the moduli space of the (complexified) almost symplectic form.
Given the assumptions of the previous section, we show in the following that locally this is also a special K\"ahler manifold. The proof is analogous to the one for the K\"ahler structure moduli space of Calabi--Yau three-folds \cite{StromingerYukawa,candelas}, with the important difference that we do not need to integrate over the compact space due to the strong assumptions on the basis forms we made above.

A metric on the moduli space of the almost symplectic structure $J$ (which, as we will see, arises naturally in the dimensional reduction) is defined by
\eq{\label{defgij}
g_{ij} \,=\, \frac{\omega_i \wedge * \omega_j}{ 4\,e^{2V}{\rm vol}_7} \,.
}
This can be computed using the identity
\eq{\label{Hodgeomegai}
 * \omega_i \,=\,  - \eta \wedge J \wedge \omega_i + \frac{3}{2}\, \frac{\omega_i \wedge J \wedge J}{J\wedge J\wedge J}\, \eta \wedge J\wedge J \,,
}
which follows from the fact that we have an SU(3)-structure and that the basis two-forms are of type (1,1) with respect to $I$ (since \eqref{comp2} implies $\Omega \wedge \omega_i = 0$).
Plugging in the expansion of $J$ and using the properties of the basis forms, we obtain
\be\label{metricvectorscal}
g_{ij} = -  \frac{1}{4\mathcal K}\mathcal K_{ijk} v^k  + \frac{1}{16\mathcal K^2} \mathcal K_{ikl}v^kv^l \, \mathcal K_{jmn}v^mv^n\,.
\ee
We see that $g_{ij}$ depends on the $v^i$ metric parameters only, and therefore is indeed a metric on the space of deformations of $J$, expanded as in \eqref{LISU3str}.
In the dimensional reduction, the $J$-moduli $v^i$ are naturally complexified with the real parameters $b^i$ arising from expanding the 11D supergravity three-form $A_3$ as $A_3 = b^i \omega_i \wedge \theta + \ldots$ (see eq.~\eqref{threeformexp}), and one is led to introduce the complex combination
\eq{ \label{compJmoduli}
t^i = b^i + i\, v^i\,.
}
One immediately checks that \eqref{metricvectorscal} is a special K\"ahler metric with respect to these coordinates, namely
$g_{ij} = \frac{\partial}{\partial t^i}\frac{\partial}{\partial\bar t^j} K$,
where the K\"ahler potential $K$ is
\be\label{KahlerPotJ}
e^{-K} \,=\, \frac{4}{3}\,\mathcal K_{ijk} v^iv^jv^k \,=\, 8\,\mathcal K\,.
\ee
This is the standard K\"ahler potential following from the cubic prepotential
\be\label{cubicprepot}
\prepF=-\frac{1}{6} \KK_{ijk} \frac{X^i X^j X^k}{X^0}\,,
\ee
through the special K\"ahler geometry formula
$e^{-K} =  i(\,\bar X^I\prepF_I - X^I\bar{\prepF}_I\,),$
once the projective coordinates $X^I$ are identified as
\be
X^I \equiv (X^0,X^i) = (X^0, X^0 t^i)\,,
\ee
and with ${\prepF}_I={\partial \prepF(X)}/{\partial X^I}$.\footnote{It is interesting to note that the deformations of $J$ and $\Omega$ can be described on the same footing by using the formalism of generalized geometry. Indeed, after introducing the zero-form $\omega_0 =1$ and the index $I=(0,i)$, the symplectic expansion of $\Omega$ in \eqref{LISU3str} finds a counterpart in the polyform $e^{t^i\omega_i} = X^I \omega_I - \prepF_I \tilde \omega^I$. Moreover, a symplectic structure on the space of even forms is defined by $\langle \omega_I, \tilde\omega^J\rangle = -\delta_I^J \tilde\omega^0$, where $\langle \,,\,\rangle$ is the Mukai pairing, which in six dimensions is antisymmetric. It follows that the K\"ahler potential is given by $e^{-K}\tilde\omega^0 = -i \langle e^{t^i\omega_i}, e^{\bar t^i \omega_i}\rangle$, in analogy with \eqref{csKP}. See e.g.\ \cite{granaN2part1} for details.
}

In the dimensional reduction, the period matrix $\mathcal N_{IJ}$ relating $X^I$ and $\prepF_J$ will also appear, playing the role of the gauge kinetic matrix. Starting from the prepotential \eqref{cubicprepot} and implementing the formula (\ref{csMAB}), we obtain the explicit expressions
\eq{\spl{\label{eq:ReNImN}
\Re \,\mathcal N_{00} &=   -\frac{1}{3} \mathcal K_{ijk}b^ib^jb^k\,,\qquad\;\;
\Re \,\mathcal N_{0i} = \frac{1}{2}\mathcal K_{ijk}b^jb^k\,,\qquad
\Re \,\mathcal N_{ij} = -\mathcal K_{ijk} b^k ,\\[2pt]
\Im \,\mathcal N_{00} &=  - \mathcal K(1
+ 4g_{ij}b^ib^j)\,,\qquad
\Im \,\mathcal N_{0i} = 4 \mathcal Kg_{ij} b^j\,,\qquad
\Im \,\mathcal N_{ij} = -4 \mathcal K g_{ij} \,.
}}

%%%%%%%%%%%%%%%%%%%%%%%%%%%%%%%%%%%%%%%%%%%%%%%%%%%%%%%%%%%%%%%%%%%%%%%%%%
\section{The dimensional reduction}\label{sec:DimRed}

In this section, we present the reduction of the bosonic sector of 11D supergravity on any 7D manifold admitting a basis of differential forms of the type described in section~\ref{subsec:expansionforms}. In the following we provide in turn our truncation ansatz and the resulting 4D action, while the technical details of the reduction are given in appendix~\ref{sec:DetailsDimRed}. In section~\ref{sec:N=2sugra}, we will prove that the reduced action agrees with the formalism of 4D $N=2$ gauged supergravity coupled to $n_V$ vector multiplets and $n_H$ hypermultiplets.
Then in section~\ref{sec:TheModels} we will apply our general results to the coset manifolds of table~\ref{cosetlist}.

\subsection{The truncation ansatz}

The bosonic content of 11D supergravity \cite{CJS11D} includes the metric and a three-form potential $A_3$, with field strength $G_4 = dA_3$. The action is
\eq{\label{11daction}
S_{11} \,=\, \frac{1}{2\kappa_{11}^2} \int \Big( R *1 -\frac{1}{2} G_4 \wedge * G_4 -\frac{1}{6} A_3 \wedge G_4 \wedge G_4  \Big)\,.
}

For the 11D metric we take the following truncation ansatz
\eq{\label{KKmetric}
ds^2 \,=\, e^{2V} {\cal K}^{-1} ds^2_4 +  e^{-V} ds^2(B_6) + e^{2V} \big(\theta + A^0 \big)^2 .
}
where $ds^2_4$ is a metric on the 4D space-time, while the 7D part is like the one introduced in the previous section, with the parameters $v^i, z^a, V$ now promoted to scalar fields on the 4D space-time. We are also gauging the reparametrizations of the coordinate associated with the vector dual to $\theta$ by a 4D one-form $A^0$. Finally, we introduced a Weyl factor in front of the external metric in order to end up in 4D Einstein frame.

The ansatz for the three-form potential $A_3$ is defined by the most general expansion in the basis forms:
\eq{
\label{threeformexp}
A_3 = C_3 + B \wedge (\theta + A^0) - A^i \wedge \omega_i + \xi^A \alpha_A - \tilde \xi_A \beta^A + b^i \omega_i \wedge (\theta + A^0),
}
where $C_3$ is a three-form, $B$ is a two-form, the $A^i$ are one-forms, and $b^i$, $\xi^A$, $\tilde{\xi}_A$ are real scalars, all defined on the 4D space-time. Note that $\xi^A, \tilde \xi_A$ can be assembled in a symplectic vector
\be
\xi = \binom{\xi^A}{\tilde \xi_A}\,.
\ee
For the field strength $G_4$ we take
\be\label{ansatzG4}
G_4 \,=\, dA_3 + G_4^{\rm flux}\,,
\ee
with
\eq{\label{G4flux}
G_4^{\rm flux} \,=\, p^A \alpha_A \wedge \theta - q_A \beta^A \wedge \theta + e_i\, \tilde\omega^i\,,
}
where $e_i$, $p^A$, $q_A$ are constants parameterizing the flux of the four-form on the internal manifold. Note that these are only
non-trivial if they correspond to non-exact parts of $G_4$; otherwise, they can be reabsorbed through redefinitions of the fields $\xi, \tilde{\xi}, b$.
It will be convenient to arrange the four-form fluxes $p^A$, $q_A$ together with the geometric fluxes $Q_i$ introduced in \eqref{GenExtAlg}, \eqref{GeomFluxMatr} in a larger matrix $Q_I$, where we define a new index $I=(0,i)= 0,1, \ldots, n_V$. Namely, we introduce
\eq{
Q_0 = {\binom{ p^A }{ q_A  }}\,,\qquad \text{and}\qquad
Q_I = (Q_0, Q_i) = \binom{p^A\; \; m_i{}^A}{q_A\;\; e_{iA}}\,.
}

The Bianchi identity $dG_4 = 0$ imposes the constraints
\begin{eqnarray}\label{bianchcond}
q_i{}^j e_j =0 \; , \qquad  Q_0^T \sym Q_i = 0\, .
\end{eqnarray}

The three-form $C_3$ is non-dynamical in four dimensions; after having worked out the 4D action we will dualize it to a constant $e_0$. We can then introduce
\eq{e_I = (e_0,e_i)}
and, for convenience,
\eq{m^I = (0,m^i)\,.}

The dimensional reduction is then performed by plugging the above ansatz in the 11D action \eqref{11daction}. The details on the reduction of the 11D Einstein--Hilbert term are given in section~\ref {subsec:CalcRicciScalar} of the appendix, where we provide a formula for the reduction of the higher-dimensional Ricci scalar (eq.~\eqref{Ricciscalar}) that also applies to more general setups than the one of interest in this paper. Note that for the reduction to work, we need to assume that the vector $k$ dual to $\theta$ is Killing for some canonical metric.
An important piece of the computation is the evaluation of the internal Ricci scalar $R_7$ in terms of the 4D scalars and of the geometric fluxes, which yields a very non-trivial contribution to the 4D scalar potential. In order to do so, we first found a formula for $R_7$ purely in terms of SU(3)-structure data; this is given in~\eqref{R7onlysingletclasses}. Then we plugged in this formula the expansions defined in section~\ref{subsec:expansionforms}, which lead in the end to the result displayed in eq.~\eqref{Vgeo}.
The details on the reduction of the kinetic and Chern--Simons terms for the 11D three-form are given in section~\ref{DetailsRedForm} of the appendix. In particular, the expansion of the 11D field strength $G_4$ and the result of the dualization of $C_3$ are given in
eqs.~\eqref{G4KK}-\eqref{dualH4}.
Adding up all the contributions, we find that 11D supergravity reduces to the 4D action below.

%%%%%%%%%%%%%%%%%%%%%%%%%%%%%%%%%%%%%%%%%%%%%%%%%%%%%%%%%%%%%
\subsection{The 4D action}\label{sec:4Daction}

The full 4D action is obtained by adding the contributions from the 11D Einstein--Hilbert term, eq.~\eqref{Sgeo}, and from the form sector, eq.~\eqref{SG4}. It reads
\be
S \,=\, \frac{1}{\kappa_4^2} \int  \left[\, \tfrac{1}{2}R_4*\!1  + \mathcal L_{\rm kin} + \mathcal L_{\rm top} -V *\!1 \,\right]\,,
\ee
where the 4D gravitational coupling constant is $\kappa^{-2}_4 = \kappa_{11}^{-2} \int (-\tilde\omega^0 \wedge \theta) $, and the Hodge star uses the 4D metric $ds^2_4$.
The kinetic terms are
\eq{\spl{\label{Lkin}
\mathcal L_{\rm kin} &= \tfrac{1}{4}\, e^{-4\phi} dB \wedge *dB  + \tfrac{1}{4}\, {\rm Im}\,\mathcal N_{IJ} F^I \wedge *F^J + \tfrac{1}{4}\, {\rm Re}\,\mathcal N_{IJ} F^I \wedge F^J \\[2mm]
& + g_{ij}Dt^i\wedge * D\bar t^{j} + g_{a \bar b} Dz^a \wedge *D{\bar z}^{\bar b} +  d\phi\wedge* d\phi  -\tfrac{1}{4} e^{2\phi} (D\xi)^T \!\wedge *  (\sym \mathbb{M} D\xi),\qquad
}}
while the topological terms, coming from the 11D Chern--Simons term, are
\eq{\spl{ \label{lagtopG4}
{\cal L}_{\textrm{top}}  = &
-\tfrac{1}{4} dB \wedge \left[ \left(2e_I + Q^T_I \sym \xi \right)A^I  -\xi^T \sym D\xi \right]\\[3pt]
& +\tfrac{1}{4}\, q_{i}{}^l \KK_{jkl} A^i \wedge A^j \wedge (dA^k -  m^k B)
-\tfrac{1}{4} e_I m^I B \wedge B\,.
}}
Here, $A^I = (A^0, A^i)$ are the gauge vectors, with generalized field strength
\eq{\label{N=2fieldstrengths}
F^I \,=\, DA^I - m^I B\,,
}
where
\eq{\label{gaugecovder}
DA^0=dA^0\,,\qquad DA^i = dA^i - A^0 \wedge A^j q_j{}^i \,.}
The gauge kinetic matrix $\mathcal N_{IJ}$ was given in \eqref{eq:ReNImN}.
Further, we recall that $g_{ij}$ is the special K\"ahler metric~\eqref{metricvectorscal} on the moduli space of the complexified almost symplectic structure parameterized by the $t^i$, while $g_{a \bar b}$ is the special K\"ahler metric on the space of almost complex structures parameterized by the $z^a$. Furthermore the symplectic matrix $\mathbb{M}$ appearing in the last term of \eqref{Lkin} was defined in \eqref{symplmatrM}.

The covariant derivatives of the scalars arising from the internal metric involve just the gauge field $A^0$, and read
\eq{\spl{\label{covderv}
D v^i \,&=\, d v^i -  v^j q_j{}^i  A^0 \, , \\[2pt]
D Z^A \,&=\, d Z^A + \left(-u_B{}^A Z^B + t^{AB}\prepG_B\right) A^0 \, ,\\[2pt]
D \prepG_A \,&=\, d \prepG_A + \left(s_{AB} Z^B + u_A{}^B \prepG_B  \right) A^0 \,.
}}
Note that $Dz^a$ immediately follows from $DZ^A$, recalling that we identify $z^a = Z^a/Z^0$.\footnote{\label{ftnt:prepotinv}
The transformation of $(Z^A,\prepG_A)$ that $A^0$ is gauging as expressed above, is
manifestly symplectic. Note that the transformation of the $Z^A$ already
implies a certain transformation of the $\prepG_A$, since the $\prepG_A$ are just
functions of the $Z^A$. A priori this does not take the same form as
above, the consistency condition being
\eq{\label{consistencyU}
(s_{AB} Z^B + u_A{}^B \prepG_B) = \delta \prepG_A = \frac{\partial \prepG_A}{\partial Z^C} \delta Z^C =  \,\prepG_{AC} (- u_B{}^C Z^B + t^{CB} \prepG_B)  \, .
}
However, in our case this will work out exactly, since, as we
show in the appendix around \eqref{deltaOmega}, the transformation is a proper
complex deformation.
}
The covariant derivatives of the scalars coming from the 11D three-form are
\eq{\spl{\label{covDers}
D b^i \,& = \, db^i - A^0\, b^jq_j{}^i +  A^j q_j{}^i \, , \\[2pt]
D\xi \,&=\, d\xi + A^I Q_I + A^0\, \mathbb{U} \xi\,.
}}
Recalling that we have $t^i = b^i + i v^i$, it follows that
\eq{ \label{covDersFinal}
Dt^i \, \equiv \,  Db^i + i\, Dv^i \,=\, dt^i - (t^j A^0 -A^j)\,q_j{}^i\,.
}
The real scalar $\phi$, sometimes called the 4D dilaton, is defined as
\eq{\label{4Ddilaton}
e^{2\phi} = e^{3V} {\cal K}^{-1},
}
and replaces in the 4D action the degree of freedom associated with the rescalings of $\eta$.

Finally, the scalar potential is $V = V_{\rm geo} + V_{G_4}$, where $V_{\rm geo}$ is the contribution from the internal Ricci scalar, given in \eqref{Vgeo}, and $V_{G_4}$ is the contribution from the kinetic term of the M-theory three-form, given in \eqref{formSPotRewr}. The expression we eventually find is:
\eq{\spl{\label{fullVfromDimRed}
V  \,= &\,\;  8\,e^K g_{ij} v^k q_k{}^i v^l q_l{}^j -2\,e^{K+2\phi} \left( X^I Q_I^T + \xi^T\mathbb{U}^T \right) \sym \mathbb{M} \left( Q_J \bar X^J + \mathbb{U}\xi\right) \\[3pt]
+ &\; 4\,e^{K + \KOm + 2\phi}  (g^{ij}- 4 v^iv^j)\left(Q_i{}^T \sym Z\right) \left(Q_j{}^T \sym \bar{Z}\right) \\[3pt]
+ &\; 8i\,e^{K + \KOm}  \, Z^T \mathbb{U}^T \sym \mathbb{U} \bar{Z}
+8\; e^{K + 2\KOm} (Z^T \sym \mathbb{U} \bar{Z})^2 + 4\,e^{K + \KOm + 2\phi} (m^i \mathcal{K}_i)(Z^T \sym \mathbb{U} \bar{Z}) \\[3pt]
- & \; \tfrac{1}{4}e^{4\phi} \left[ m^I \Im \mathcal N_{IJ} m^J +   \big({\cal E}_I -\Re{\cal N}_{IK} m^K \big)  (\Imag {\cal N})^{-1 IJ} \big({\cal E}_J - \Re{\cal N}_{JL} m^L \big) \right],
}}
where we have defined
\eq{\label{defE_I}
{\cal E}_I \,=\, e_I  + Q_I^T \sym \xi - \tfrac 12\delta_I^0\,  \xi^T \sym \mathbb U \xi\,.
}

Let us remark a few interesting features of the action above, also comparing with previous work in the literature. Whenever the charges $q_i{}^j$ are non-vanishing, we see that the vector covariant derivatives \eqref{gaugecovder} have a non-abelian structure. Note that the $q_i{}^j$ charges arise in \eqref{GenExtAlg} from a non-trivial dependence of the forms $\omega_i$, $\tilde \omega^i$ on the coordinate on $M_7$ parameterizing the direction of the vector $X$. It follows that this non-abelian structure is a specific feature of M-theory on 7D manifolds with SU(3)-structure. This was first found in \cite{Aharony:2008rx}, and further studied in \cite{Looyestijn:2010pb}, for M-theory compactifications on Calabi--Yau three-folds fibered over a circle. Ref.\ \cite{Looyestijn:2010pb} also studied the couplings of the scalars associated with the hypermultiplet sector (which was not analysed in \cite{Aharony:2008rx}), including the charges appearing in our matrix $\mathbb U$. Here, we are extending these results to more general internal manifolds, and our 4D action reproduces the ones in \cite{Aharony:2008rx,Looyestijn:2010pb} once we take $Q_I= e_I = m^I = 0$ and the two-form $B$ is dualized to a scalar. The full gauge algebra of our model will be discussed in section~\ref{sec:thegaugings}, while in section \ref{sec:TheModels} we will see that one of our coset consistent truncations (the one on $N(1,-1)$) exhibits the non-abelian structure discussed here.

We can also make contact with similar reductions of type IIA supergravity on 6D manifolds with SU(3)-structure, leading to $N=2$ supergravity \cite{louismicu2,gaugingHeisenberg,housepalti,granaN2part1,ExploitingN=2,Cassani:2009na}. In order to do this, we consider a reduction along the vector $k$ dual to $\theta$. This is allowed if $k$ preserves our truncation ansatz, or at least a consistent subtruncation of it. Actually, since we want off-shell supersymmetry to be preserved, not only the metric, but also our SU(3)-structure forms $\eta$, $J$ and $\Omega$ need to be preserved, i.e.\ we have to impose $\mathcal L_k \eta = \mathcal L_k J = \mathcal L_k \Omega = 0$. Let us consider e.g.\ $J\,$: vanishing of the Lie derivative leads to $v^{\hat \imath} q_{\hat \imath}{}^{\hat \jmath} = 0$, where we are appending a hat to the indices surviving the subtruncation; since we want the $v^{\hat \imath}$ to be independent, we conclude that the charges $q_{\hat \imath}{}^{\hat \jmath}$ need to vanish. Similar reasoning leads to require that the $\mathbb U$ charges associated with the fields in the subtruncation have to vanish. It follows that in the type IIA reduction ansatz the $v$, $b$, $z$ scalars become uncharged, and their covariant derivative becomes trivial.
As for the remaining charges, one has that $p^A$ and $q_A$ describe the type IIA NSNS three-form flux, while the $e_I$ correspond to the RR four-form flux, and $m^i$ is the RR two-form flux. This reproduces the action arising from dimensional reduction of type IIA on 6D SU(3)-structure manifolds, in the limit of vanishing Romans mass, $m^0=0$. In this context, switching on the Romans mass is trivially achieved by promoting $m^I = (0,m^i)$ to $(m^0,m^i)$. We will speculate about turning on $m^0$ in more general frameworks in section~\ref{sec:discussion}.
We conclude that with respect to type IIA on 6D SU(3)-structures, the present M-theory setup allows for extra couplings, which lead to non-abelian gauge groups and to a novel scalar potential including more terms than the ones previously appeared in the literature.

As a final remark, note that the whole action is invariant under Sp$(2n_H,\mathbb{R})$. Invariance under the electric-magnetic group Sp$(2n_V +2,\mathbb{R})$ is present just for some terms, like the last line of \eqref{fullVfromDimRed} (once the symplectic vector $(m^I, \mathcal E_I)^T$ is defined).

%%%%%%%%%%%%%%%%%%%%%%%%%%%%%%%%%%%%%%%%%%%%%%%%%%%%%%%%%%%%%%%%%%%%%%%%%%
\section{The $N=2$ gauging}\label{sec:N=2sugra}

\subsection{The action with magnetic vectors}

We are now going to show that the 4D action above is consistent with the bosonic sector of $N=2$ gauged supergravity.
A delicate point in this proof is how to treat the two-form $B$, since two-forms do not appear in standard matter-coupled $N=2$ supergravity with purely electric gauging (see \cite{reviewN2sugra} for a review).
This issue is not present when the parameters $m^I$ all vanish (i.e.\ the one-form $\theta$ is closed), since in this case $B$ enters in the action only through its field-strength $dB$, and therefore can easily be dualized to a scalar. After this transformation, the action is straightforwardly shown to be an electrically gauged $N=2$ supergravity coupled to $n_V$ vector multiplets and $n_H$ hypermultiplets. However, in all the concrete examples we will study in the next section, the $m^I$ are non-vanishing, implying that the potential $B$ appears explicitly in the action; this makes it more involved to dualize it, a preliminary electric-magnetic symplectic transformation on the vector fields being required. In order to avoid these complications, which would also affect the gauge kinetic matrix and the prepotential $\prepF$ of our model, we chose to remain in the same electric-magnetic frame naturally defined by the dimensional reduction, and to follow the approach of \cite{samtlebenmag} (see also \cite{ElecMagnN=2}). In this approach, the $m^I$ are seen as charges under magnetic vectors $\tilde A_I$, and there is no need to completely dualize the $B$-field away, since both the two-forms and their dual scalars play a role in the action. This is allowed because while the electric vectors and the scalars are dynamical fields with second-order equations of motion, both the magnetic vectors and the two-forms are treated as non-propagating, auxiliary fields.  The role of the $\tilde A_I$ is to make the magnetic gauging manifest, while the two-forms are required in order to ensure closure of the gauge algebra.\footnote{The appearance of magnetic charges and of two-forms in flux compactifications was first discussed in a type~II supergravity context, in \cite{Michelson} and in \cite{louismicu1}, respectively. An alternative way to match gauged $N=2$ supergravity would be to invoke the formalism of \cite{N=2tensor1,N=2tensor2}, where the coupling with tensor multiplets was constructed. In this approach, the two-forms are dynamical, the magnetic charges enter in their mass and topological terms, and their dual scalars do not appear in the action. These are precisely the features of the action we obtained from the dimensional reduction. The reason why we preferred the approach described above is that it makes the magnetic gauging manifest.}

In order to match the formalism of \cite{samtlebenmag}, we need to cast the action  obtained from dimensional reduction in an equivalent form, in which the two-form $B$ becomes non-dynamical, and its dual scalar, $a$, is introduced.  This is achieved by the new action
\eq{\label{new4Daction}
\widetilde S \,=\, \frac{1}{\kappa_4^2} \int  \left[\, \tfrac{1}{2}R_4*\!1  + \widetilde{\mathcal L}_{\rm kin} + \widetilde{\mathcal L}_{\rm top} -V *\!1 \,\right]\,,
}
in which $\mathcal L_{\rm kin}$ and $\mathcal L_{\rm top}$ of section~\ref{sec:4Daction} are replaced by
\eq{\spl{\label{LkinBis}
\widetilde{\mathcal L}_{\rm kin} &= \tfrac{1}{4}\, {\rm Im}\,\mathcal N_{IJ} F^I \wedge *F^J + \tfrac{1}{4}\, {\rm Re}\,\mathcal N_{IJ} F^I \wedge F^J + g_{ij}Dt^i\wedge * D\bar t^{j}\\[3pt]
& + g_{a \bar b} Dz^a \wedge *D{\bar z}^{\bar b} +  d\phi\wedge* d\phi  -\tfrac{1}{4} e^{2\phi} (D\xi)^T \!\wedge *  (\sym \mathbb{M} D\xi),\qquad\\[3pt]
& + \tfrac{1}{4} e^{4\phi} \left(Da + \tfrac{1}{2}\xi^T \sym D\xi \right)\wedge * \left(Da + \tfrac{1}{2}\xi^T \sym D\xi \right)
}}
and by
\eq{\label{lagtopG4bis}
\widetilde{{\cal L}}_{\textrm{top}}  =  \tfrac{1}{4}\, q_{i}{}^l \KK_{jkl} A^i \wedge A^j \wedge (dA^k - m^k B) -\tfrac{1}{4} e_I m^I B \wedge B -\tfrac{1}{2} B \wedge m^I d\tilde A_I\,,
}
where we defined
\eq{\label{exprDa}
Da  =  da - A^I \big( e_I + \tfrac{1}{2} Q_I^T \sym \xi \big) - \tilde{A}_I m^I \,,
}
$\tilde A_I\,$ being the magnetic vector fields. Note that the only changes in
$\widetilde{\mathcal L}_{\rm kin}$ and $\widetilde{\mathcal L}_{\rm top}$ with respect to ${\mathcal L}_{\rm kin}$ and ${\mathcal L}_{\rm top}$ are that the $dB$ terms have been substituted by the kinetic term for $a$ and a topological coupling between $B$ and the magnetic vectors, corresponding to the last terms of \eqref{LkinBis} and of \eqref{lagtopG4bis}, respectively.
The electric field strengths $F^I$ and the covariant derivatives of the other scalars displayed in the previous section remain the same, and the scalar potential \eqref{fullVfromDimRed} is also unmodified.
The relation between the new and the old action is seen by integrating out the vector field $m^I \tilde A_I$.
Indeed, its equation of motion reads
\eq{Da + \tfrac{1}{2}\xi^T \sym D\xi  \,=\, - e^{-4\phi} *dB\,, }
which states the duality relation between $B$ and $a$. Recalling \eqref{exprDa}, we can solve for $m^I\tilde A_I$. Substituting the solution in \eqref{LkinBis}, \eqref{lagtopG4bis}, the original action of section~\ref{sec:4Daction} is retrieved.
Note that, as promised, in the new action the two-form $B$ is non-propagating, in that its kinetic term has been replaced by the kinetic term for the scalar $a$. The equation of motion of $B$ is thus first-order, and gives the duality relation between the electric vectors $A^I$ and the magnetic vectors $\tilde A_I$, contracted with $m^I$.

We are now in the position of discussing consistency with $N=2$ supergravity in the presence of both electric and magnetic gaugings.
When the charges all vanish, our action is in straightforward agreement with the bosonic sector of ungauged $N=2$ supergravity, with the gravity multiplet --- comprising the 4D metric and the graviphoton $A^0$ --- coupled to $n_V$ vector multiplets $\{ A^i, t^i \}$ and $n_H$ hypermultiplets defined by the scalar fields $q^u = \{\phi, a, z^a, \xi^A, \tilde \xi_A\}$, with $u=1,\ldots,4n_H$.
From \eqref{LkinBis}, we see that the kinetic terms of the hyperscalars  define a $\sigma$-model with target space metric
\eq{ \label{cmapmetric}
h_{uv} dq^u dq^v \,=\,  d\phi^2 + g_{a \bar b} dz^a d{\bar z}^{\bar b}
+ \tfrac{1}{4} e^{4\phi} \left(da + \tfrac{1}{2}\xi^T \sym d\xi \right)^2 -\tfrac{1}{4} e^{2\phi} d\xi^T \sym \mathbb{M} d\xi\,,
}
which, as familiar from  Calabi--Yau compactifications, is the quaternionic-K\"ahler metric obtained from the special K\"ahler manifold spanned by the $z^a$ via the c-map construction of \cite{FS}.\footnote{From comparing our gauge kinetic terms with the ones in \cite{reviewN2sugra}, we see that the normalization of our vectors $A^I$ differs by a factor of $\sqrt 2$ from the one in \cite{reviewN2sugra}, namely $A^{\rm here} = \sqrt 2 A^{\rm there}$. From eq.~\eqref{generalscalarcovder} below we see that this affects the Killing vectors gauging the isometries of the scalar manifold: $k^{\rm here} = k^{\rm there}/\sqrt 2$. It follows that the general formula for the $N=2$ scalar potential, which is quadratic in these Killing vectors, acquires an extra factor of 2 with respect to the standard one, see eq.~\eqref{generalpotential}. \label{rescalingN=2vecs}}

Let us now turn to the non-vanishing charges.

\subsection{The gauging}\label{sec:thegaugings}

Even though a generic special K\"ahler manifold will not necessarily have isometries, we find that the ones associated with our effective action (\ref{new4Daction}),  which were discussed at length in section  \ref{sec:SU3str}, do have them. In fact, some of them are gauged. The corresponding Killing vectors and their prepotentials are related to the geometric and form fluxes, and determine  the coupling of the scalars to the vectors, as we will now describe, and the scalar potential, as we will show in appendix \ref{CheckScalarPot}, in a way compatible with gauged $N=2$ supergravity.

The general scalar covariant derivatives in $N=2$ gauged supergravity read
\eq{\spl{\label{generalscalarcovder}
Dt^j \,&=\, dt^j +  k^j_I A^I\,,\\
Dq^u \,&=\, dq^u + k^u_I A^I -\tilde k^{Iu}\tilde A_I\,.
}}
Here, we are allowing the isometries in the hyperscalar manifold to be gauged both electrically and magnetically by Killing vectors $k^u_I$ and $\tilde k^{Iu}$, respectively, while for our purposes it will be enough to assume that the isometries in the vector multiplet scalar manifold are gauged just by electric vectors $k^j_I$.

From \eqref{covDersFinal} we read off that the gauging of the vector multiplet scalars is described
by the holomorphic Killing vectors
\eq{\label{holoKillingvec} k_0^{\,i} \,=\, -\, t^j q_j{}^i \, , \qquad\qquad k_j^{\,i} \,=\, q_j{}^i  \, .
}
One can check that these vectors are Killing by using the constraint \eqref{qKvanishes}. Observe that these isometries are peculiar to special K\"ahler manifolds with a cubic prepotential, and therefore to dimensional reductions based on SU(3)-structures. In particular, $k_j^{\,i}$ generate shifts of the axions  $b^i$. See below equation (3.22) for further remarks about the origin of their gauging. The associated Killing prepotentials, satisfying \cite{reviewN2sugra}
\eq{
i \mathcal P_I \,=\, k^j_I\, \partial_j K \,=\, -\bar k^{\bar \jmath}_I\, \partial_{\bar\jmath} K\,,} are
\eq{
\mathcal{P}_0 \,=\, -\frac{b^j q_j{}^i\mathcal{K}_i}{4 \,\mathcal{K}} \, ,\qquad\qquad
\mathcal{P}_i \,=\,  \frac{q_i{}^j  \mathcal{K}_j}{4\, \mathcal{K}}\, ,
}
where to compute $\mathcal{P}_0$ we used $v^i q_j{}^i \mathcal K_i = 0$, which follows from~\eqref{qKvanishes}.

We now turn to the gauging of isometries in the hyperscalar manifold. From \eqref{covderv} we read off that the geometric $u, t$ and $s$-charges in the $\mathbb{U}$ matrix lead to gauging under the graviphoton $A^0$ of a Killing vector with action
\be\label{Uisometry}
\delta Z \,=\,  \mathbb{U} Z \,,\qquad\qquad \delta \xi \,=\,  \mathbb{U} \xi\,.
\ee
The vector can be written as
\eq{\label{vectorUisometry}
k_{\mathbb{U}} \,=\, (\mathbb{U}Z)^A \frac{\partial}{\partial Z^A} + (\mathbb{U}\bar Z)^A \frac{\partial}{\partial \bar Z^A} + (\mathbb{U} \xi)^A \frac{\partial}{\partial\xi^A} + (\mathbb{U} \xi)_A \frac{\partial}{\partial\tilde\xi_A}\,,
}
where for the special K\"ahler base of the quaternionic manifold we are using the projective coordinates $Z^A$ rather than the special coordinates $z^a = Z^a /Z^0$. As explained in footnote \ref{ftnt:prepotinv}, this transformation leaves the holomorphic prepotential invariant, and thus also the K\"ahler potential $\KOm$  and the metric derived from it, thus indeed leading to a Killing vector of the special K\"ahler base. By also working out the transformation of the matrix $\mathbb M$ under \eqref{Uisometry}, one can check that our vector is Killing for the full quaternionic metric \eqref{cmapmetric} (see \cite{SymmetrySpecialGeo,Looyestijn:2010pb} for more details).

By looking again at the scalar covariant derivatives, we infer that the remaining flux charges $m^I$, $e_I$, $Q_I$ are associated with a gauging of the Heisenberg algebra of isometries which exists for any quaternionic metric of the form \eqref{cmapmetric}. This algebra is generated by the Killing vectors
\eq{\spl{\label{Heisalg}
h^A = \frac{\partial}{\partial\tilde\xi_A} + \frac 12 \xi^A \frac{\partial}{\partial a} \,, \qquad h_A = \frac{\partial}{\partial\xi^A} - \frac 12\tilde\xi_A \frac{\partial}{\partial a} \,,\qquad h = \frac{\partial}{\partial a}\,,
}}
satisfying the Heisenberg commutation relations $[h_A,h^B] \,=\, \delta_A^B\, h\,$.
Also taking into account \eqref{vectorUisometry}, we find that the complete Killing vectors $k^u_I$ and $\tilde k^{Iu}$, respectively gauging the quaternionic isometries electrically and magnetically, read
\eq{\spl{\label{quatKillingV}
k_I^u\frac{\partial}{\partial q^u}  \,&=\,\delta_I^0 \, k_{\mathbb{U}} +   Q_I{}^A h_A  +  Q_{IA}h^A -e_I\, h \,,\\[2pt]
\tilde k^{Iu}\frac{\partial}{\partial q^u}\, &=\, - \, m^I h\,.
}}

We now compute the Killing prepotentials $\mathcal P^x_I$, $\tilde{\mathcal P}^{x\,I}$, $x=1,2,3$, associated with the quaternionic isometries being gauged. These are important data of $N=2$ supergravity, in that they appear both in the formula for the $N=2$ scalar potential and in the supersymmetry variations of the fermions.
In order to perform the computation, we need the connection $\omega^x$ on the SU(2)-bundle of the quaternionic-K\"ahler manifold with metric \eqref{cmapmetric}. We find that this is given by \cite{FS}
\eq{\spl{\label{SU2Connection}
\omega^1+i \omega^2 \;&=\; \sqrt 2\, e^{\frac{\KOm}{2} + \phi} \, Z^T \sym d \xi\,,  \\[3pt]
\omega^3 \;&=\; \frac{e^{2\phi}}{2} \left(d a +   \tfrac 12\xi^T \sym d \xi \right)  -\,2\, e^{\KOm} \Imag\left(Z^A\Imag \prepG_{AB} d\bar Z^B \right).
}}
One can verify that our Killing vectors preserve this connection, and hence the SU(2)-curvature $\Omega^x = \d \omega^x + \frac{1}{2} \epsilon^{xyz} \omega^y \wedge \omega^z\,$:
\eq{
\label{killingSU2conn}
\mathcal{L}_{k_I} \omega^x = 0 \qquad\Rightarrow\qquad
\mathcal{L}_{k_I} \Omega^x = 0 \,
}
(same for $\tilde k^{I}$). Due to this property, the general formula for the Killing prepotentials takes the particularly simple form:
\eq{
\mathcal P^x_I \,=\, k^u_I\omega^x_u\,,\qquad \qquad \tilde{\mathcal P}^{x\,I} \,=\, \tilde k^{u\,I}\omega^x_u\,.
}
This yields the following non-vanishing prepotentials:
\eq{\spl{
\mathcal P^1_I + i \mathcal P^2_I \,&=\, \sqrt 2 \, e^{\frac{\KOm}{2} + \phi}\, Z^T \sym \left(Q_I + \delta_I^0\, \mathbb{U} \xi\right) ,\\[4pt]
\mathcal P^3_I \,&=\, -\tfrac{1}{2} \,e^{2\phi}\left( e_I + Q_I^T\sym \xi  - \tfrac 12 \delta_I^0\, \xi^T \sym \mathbb{U} \xi \right) - \, \delta_I^0 e^{\KOm}Z^T \sym \mathbb{U} \bar Z  \,,\\[4pt]
\tilde{\mathcal  P}^{3I} \,&=\, - \tfrac{1}{2}\,  e^{2\phi}\, m^I \,.
}}

We have thus specified the isometries being gauged in the scalar covariant derivatives appearing in our action \eqref{new4Daction}. These unify gaugings previously discussed in the literature in different contexts: the gaugings in the vector multiplet sector generated by the Killing vectors \eqref{holoKillingvec} were discovered in \cite{Aharony:2008rx}, and further discussed in \cite{Looyestijn:2010pb}, for M-theory compactifications on Calabi--Yau three-folds fibered over a circle; the gauging of the isometry \eqref{Uisometry} was also described in \cite{Looyestijn:2010pb}. While the gauging of the Heisenberg algebra \eqref{Heisalg} was not discussed in \cite{Aharony:2008rx,Looyestijn:2010pb}, it is familiar from flux compactifications of type~II supergravity on 6D manifolds with SU(3) or SU(3)$\times$SU(3) structure, see e.g.\ \cite{gaugingHeisenberg,granaN2part1,granaN2part2,trigiante,cassanibilal}.
Above we have shown that for M-theory on generic SU(3)-structure manifolds, all these gaugings can possibly be switched on, provided the constraints \eqref{nild2z}, \eqref{qKvanishes} and \eqref{bianchcond} are satisfied. This leads to non-abelian gauge groups and new  scalar potentials.

The full Killing vectors gauging both the vector- and hypermultiplet sector are obtained by adding up the Killing vectors in
eq.~\eqref{holoKillingvec} and in eq.~\eqref{quatKillingV}:
\eq{
k_I \,=\, k_I^j \frac{\partial}{\partial t^j} + \bar k_I^{\bar\jmath} \frac{\partial}{\partial \bar t^{\bar\jmath}} + k_I^u\frac{\partial}{\partial q^u}\,,\qquad\qquad \tilde k^I = \tilde k^{u\,I} \frac{\partial}{\partial q^u}\,.
}
from these vectors we can infer the gauge algebra of our model. We already saw in section \ref{sec:4Daction} that when the charges $q_i{}^j$ are non-vanishing, the electric gauge vectors have non-abelian covariant derivatives \eqref{gaugecovder}. This non-abelian structure can be further specified by evaluating the commutators of our gauged isometries. Recalling the constraints \eqref{nild2z}, \eqref{qKvanishes} and \eqref{bianchcond}, we find that the non-vanishing commutators are:
\eq{\spl{\label{gaugealgebra}
[k_0,k_i] \,& = \, q_i{}^j k_j\,,\\[2pt]
[k_i,k_j] \,& =\, -q_{[i}{}^k \mathcal{K}_{j]kl} \,\tilde{k}^l \,.
}}

As for the two-form $B$, it enters in the generalized field strengths $F^I$ and in the topological term \eqref{lagtopG4bis} precisely as prescribed by \cite{samtlebenmag}. For vanishing $m^I$, a topological term like the one in \eqref{lagtopG4bis} was first found in $N=2$ supergravity in \cite{deWitLauwersVanProeyen}, and in our context it was also discussed in \cite{Aharony:2008rx,Looyestijn:2010pb}. Its role is to ensure gauge invariance by compensating the transformation of the gauge kinetic matrix $\Re \mathcal N$.

It remains to discuss the scalar potential.
In $N=2$ supergravity, this is completely determined by the gauging.
In appendix \ref{CheckScalarPot}, we prove that the expression for the scalar potential following from the gauging described above precisely matches the one in \eqref{fullVfromDimRed} obtained from the dimensional reduction.
We can therefore conclude that our general dimensional reduction is fully compatible with gauged $N=2$ supergravity.

%%%%%%%%%%%%%%%%%%%%%%%%%%%%%%%%%%%%%%%%%%%%%%%%%%%%%%%%%%%%%%%%%%%%%%%%%%
\section{Explicit models from coset reduction}\label{sec:TheModels}

\begin{figure}[t!]
	\centering
\includegraphics[width=14cm]{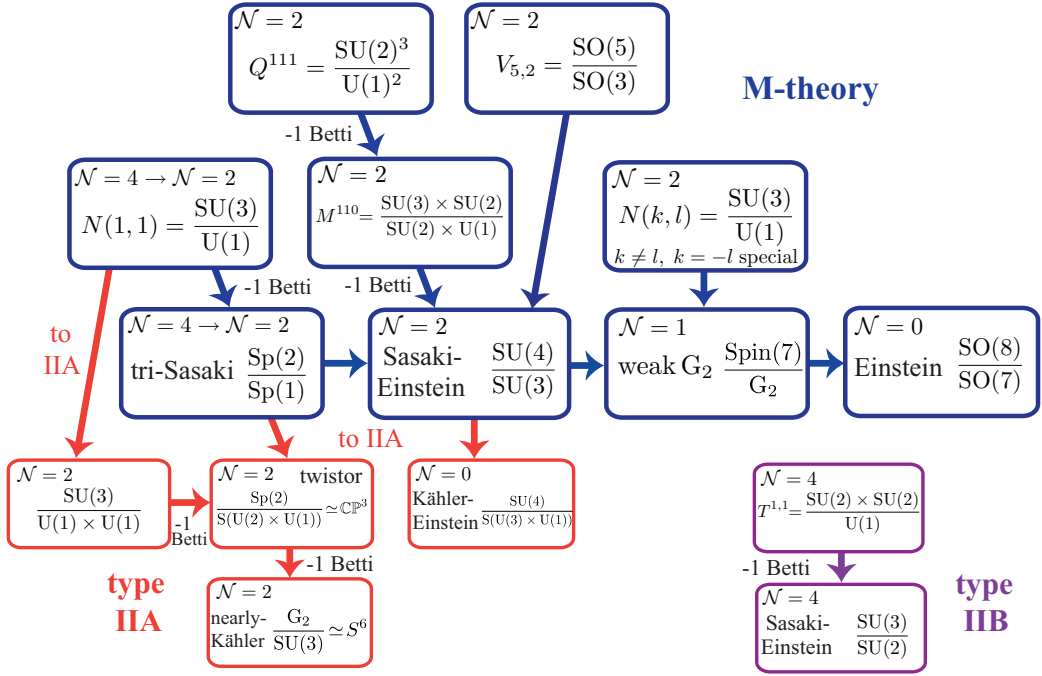}
\caption{\small Web of interrelated consistent truncations of 11D (blue),
type IIA (red) and type IIB (purple) supergravity. The diagram is based on figure 1 in \cite{caskoerbertrisak}, but updated with the coset models
discussed in this paper. In each box we display the geometric structure (if any) and the coset manifold
on which the truncation is based, as well as its amount of supersymmetry.
Each arrow denotes a consistent subtruncation. Note that the Nearly-K\"ahler truncation cannot be reached directly from the $N(1,1)$ and Sp(2)/Sp(1) models, since the respective choices of $m^i$ geometric fluxes are incompatible.}
\label{Diagram}
\end{figure}

Up to now we have kept the discussion general. We have shown that $D=11$ supergravity on any 7D manifold $M_7$, not necessarily homogeneous, equipped with the particular SU(3)-structure described in section \ref{sec:SU3str}, reduces to the general 4D $N=2$ supergravity described in sections \ref{sec:DimRed} and \ref{sec:N=2sugra}.
It is somewhat tedious, but otherwise straightforward, to verify that the set (or a subset thereof) of the LI forms on all the cosets of table \ref{cosetlist} fulfill all the requirements described in section \ref{sec:SU3str}. We have provided some details in appendix  \ref{sec:cosets}.
For an overview of all the models see figure \ref{Diagram}.
To determine the particular effective theory corresponding to each coset one only needs to specify the corresponding scalar manifolds, and the scalar charges imposed by the geometric and background fluxes. We analyse both issues in subsections \ref{subsec:SMCosets} and \ref{subsec:GaugingCosets}, where we will also explicitly write out the scalar potential for some of the models. Further comments on the individual models are made in subsection \ref{sec:features} (where, in particular, we find a new, non-supersymmetric solution). Finally, in subsection \ref{sec:susyVac} we describe the supersymmetric critical structure of our coset effective theories.

\subsection{The scalar manifolds}  \label{subsec:SMCosets}

Let us first discuss the vector multiplet and hypermultiplet scalar manifolds corresponding to the individual coset models, as given in table  \ref{cosetsummary}. As follows from our results of section \ref{sec:SU3str} and appendix \ref{sec:cosets}, the LI moduli spaces of K\"ahler and complex structure deformations are, respectively, special K\"ahler and special quaternionic. For our cosets, the respective complex and quaternionic dimensions, $n_V$ and $n_H$, are given by the number of LI two-forms and half the number of LI three-forms, respectively. These numbers are obtained on a case-by-case basis in appendix \ref{sec:cosets}, and we have summarised them in table \ref{cosetsummary}. Let us now further show that the corresponding spaces are the homogeneous spaces quoted in that table.

The vector multiplet scalars are governed by the cubic prepotential (\ref{cubicprepot}). Therefore, once the constant, symmetric tensor ${\cal K}_{ijk}$ has been specified for each model, the resulting special K\"ahler space in table \ref{cosetsummary} follows from the classification of \cite{deWit:1991nm}. To see this more explicitly, focus first on the cosets with $n_V = 1, 2, \textrm{or} \ 3$, that is, all except $N(1,-1)$. From the explicit parametrization of appendix \ref{sec:cosets}, where $v^i = e^{2u_i}$, it follows that the complexified K\"ahler moduli $t^i$ in (\ref{compJmoduli}) do take values on $n_V$ copies of the upper half plane, SU(1,1)/U(1). With the specific tensors ${\cal K}_{ijk}$  quoted in appendix \ref{sec:cosets}, it is then easy to evaluate the K\"ahler metric (\ref{metricvectorscal}). For $n_V=3$, the corresponding line element is
\begin{eqnarray} \label{LineElSKnV3}
ds^2 =  \sum_{i=1}^3 \big[ (du_i)^2 + \tfrac{1}{4} e^{-2u_i} (db^i)^2 \big] .
\end{eqnarray}
The cases $n_V=2$ and $n_V=1$ are recovered from  (\ref{LineElSKnV3}) by setting $t^3 = t^1$, and $t^3=t^1=t^2$, respectively. The only remaining case we need to consider is $N(1,-1)$, for which $n_V=5$. For this case, from the tensor ${\cal K}_{ijk}$ in (\ref{KijkN1m1}), and the generic expression  (\ref{cubicprepot}) for the prepotential, we obtain that quoted in table \ref{cosetsummary}. The corresponding K\"ahler potential is
\begin{eqnarray} \label{eq:KalPotN1m1}
K=-\log[i\left( \bar{t}^3-t^3 \right) \left((t^1-\bar{t}^1)(t^2-\bar{t}^2) -(t^4-\bar{t}^4)^2-(t^5-\bar{t}^5)^2\right)] ,
\end{eqnarray}
which is well-known (see e.g.\ \cite{Aharony:2008rx}) to give rise to the special K\"ahler manifold given in the table. The corresponding K\"ahler metric, which we omit, can be straightforwardly recovered from (\ref{eq:KalPotN1m1}). A real parametrization for this metric is provided in terms of  (\ref{N1m1moduli}). The rest of the vector multiplet quantities for all our models, namely, the gauge kinetic and topological terms, can be recovered for each coset model by inserting these metrics and the tensors ${\cal K}_{ijk}$ quoted in the appendices into the general expressions (\ref{eq:ReNImN}).

Let us now turn to the hypermultiplet sector. We found in section \ref{sec:N=2sugra} that, after dualisation, the hyperscalars are described by the metric (\ref{cmapmetric}) on a special quaternionic-K\"ahler manifold. Recall that these are quaternionic-K\"ahler manifolds in the image of the c-map, and are completely determined by their special K\"ahler submanifold. The latter is empty for $n_H=1$, as corresponds to reductions with rigid complex structure, and leads to the universal hypermultiplet moduli space, $\textrm{SU(2,1)/S(U(2)} \times \textrm{U(1))}$. For the cosets with $n_H=2$, the complex structure does admit LI fluctuations, and we find two possibilities. Both have special K\"ahler submanifold SU(1,1)/U(1), but equipped with either a cubic or a quadratic prepotential, which leads respectively to G$_{2(2)}$/SO(4) or SO(4,2)/(SO(4) $\times$ SO(2)) \cite{FS}. From the explicit calculation of the prepotentials in appendices \ref{howtoprepot} and \ref{sec:cosets}, we conclude that the $V_{5,2}$ model leads to the former case, while the Sp(2)/Sp(1) and $N(1,1)$ models lead to the second. We have quoted the prepotentials themselves in table \ref{cosetsummary}, written out in the symplectic frames introduced in appendices \ref{V52}, for $V_{5,2}$,  and  \ref{Sp2modSp1}, for Sp(2)/Sp(1) and $N(1,1)$.

We can now explicitly find the metrics on all three special quaternionic manifolds, that define the hypermultiplet non-linear sigma model after dualisation of the two-form $B$, by evaluating the general metric (\ref{cmapmetric}). For SU(2,1)/S(U(2)$\times $U(1)), $z^a=0$ and $\mathbb{M} = \mathbb{C}$,  so (\ref{cmapmetric}) reduces to
\begin{eqnarray} \label{cosetmetricSU21}
 h_{uv} dq^u dq^v  =
 (d\phi)^2
+ \tfrac{1}{4} e^{4\phi}  \big[ d a + \tfrac{1}{2} ( \xi d\tilde{\xi}- \tilde{\xi} d\xi ) \big]^2
+ \tfrac{1}{4} e^{2\phi}  (d\xi)^2
+ \tfrac{1}{4} e^{2\phi}  (d\tilde{\xi})^2 .
\end{eqnarray}
For the $n_H=2$ cases, we parametrize the (only) complex coordinate $z$ on the special K\"ahler submanifold SU(1,1)/U(1) in terms of real scalars $\chi$, $\varphi$ as in (\ref{eq:zV52}), for $V_{5,2}$, and (\ref {eq:zSp2Sp1}), for Sp(2)/Sp(1) and $N(1,1)$. Then we compute the matrix $\mathbb{M}$ entering the metric (\ref{cmapmetric}) in the specific symplectic frames introduced in appendices \ref{V52}, for $V_{5,2}$, and \ref{Sp2modSp1}, for Sp(2)/Sp(1) and $N(1,1)$. In order to do so, we first compute the matrix ${\cal M}$ in (\ref{csMAB}) with the help of the cubic, (\ref{CSprepotV52}), and quadratic, (\ref{eq:prepot}), prepotentials, and using the holomorphic sections $Z^A$, $G_A$ that can be read off from the expressions for $\Omega$ in (\ref{strV52}) and (\ref{strSp2}). After a final rearrangement, the end result is, for the metric on  G$_{2(2)}$/SO(4),
\begin{eqnarray} \label{cmapmetricV52}
&& h_{uv} dq^u dq^v = \nonumber \\
&& \ d\phi^2 + 3(d\varphi)^2 + \tfrac{3}{4} e^{4\varphi} (d\chi)^2  + \tfrac{1}{4} e^{4\phi} \left[ da + \tfrac{1}{2} ( \xi^0 d\tilde{\xi}_0 + \xi^1 d\tilde{\xi}_1 - \tilde{\xi}_0 d\xi^0 - \tilde{\xi}_1 d\xi^1 ) \right]^2 \nonumber \\
&& \ +\tfrac{1}{4} e^{2\phi-6\varphi} (d\xi^0)^2
+\tfrac{1}{4} e^{2\phi-2\varphi} \left[d\xi^1 -\sqrt{3} \chi d\xi^0 \right]^2 \nonumber \\
&& \ +\tfrac{1}{4} e^{2\phi+2\varphi} \left[ d\tilde{\xi}_1 - \sqrt{3} \chi^2 d\xi^0 + 2\chi d\xi^1  \right]^2 \nonumber \\
&& \ +\tfrac{1}{4} e^{2\phi+6\varphi} \left[ d\tilde{\xi}_0 + \sqrt{3} \chi d\tilde{\xi}_1 -\chi^3 d\xi^0 + \sqrt{3} \chi^2 d\xi^1  \right]^2
\end{eqnarray}
and for the metric on $SO(4,2)/(SO(4) \times SO(2))$,
\begin{eqnarray} \label{cmapmetricSp2}
&& h_{uv} dq^u dq^v = \nonumber \\
&& \ d\phi^2
+ \tfrac{1}{4} d\varphi^2
+ \tfrac{1}{4} e^{-2\varphi} (d\chi)^2  + \tfrac{1}{4} e^{4\phi} \left[ da + \tfrac{1}{2}( \xi^0 d\tilde{\xi}_0 + \xi^1 d\tilde{\xi}_1 - \tilde{\xi}_0 d\xi^0 - \tilde{\xi}_1 d\xi^1 ) \right]^2 \nonumber \\
&& \ +\tfrac{1}{8} e^{2\phi + \varphi} (d\xi^{0} +d\xi^{1} )^2
+\tfrac{1}{8} e^{2\phi + \varphi} (d\tilde{\xi}_{0} -d\tilde{\xi}_{1} )^2  \nonumber \\
&& \ +\tfrac{1}{8} e^{2\phi-\varphi} \left[ d\xi^{0} - d\xi^{1}  + \chi (d\tilde{\xi}_{0} -d\tilde{\xi}_{1} )  \right]^2 \nonumber \\
&& \ +\tfrac{1}{8} e^{2\phi-\varphi} \left[ d\tilde{\xi}_{ 0} + d\tilde{\xi}_{1}  - \chi (d\xi^{0} + d\xi^{1}  )  \right]^2 .
\end{eqnarray}
We have verified that these three metrics are Einstein, as corresponds to qua\-ter\-nio\-nic-K\"ahler spaces, with normalisation such that the Ricci tensor equals $-2(n_H+2)$ times the metric. We have also doublechecked that these are the (only, once the normalisation has been fixed) $G$-invariant metrics on the corresponding non-compact homogeneous spaces $G/H$ by explicitly constructing the metric from the Maurer-Cartan form in the Iwasawa parametrization.

\subsection{The gaugings}  \label{subsec:GaugingCosets}

As shown by the individual analysis of appendix \ref{sec:cosets}, the LI forms on all cosets close under exterior differentiation into an algebra of the form (\ref{GenExtAlg}) dependent on a set of parameters, the so-called geometric fluxes. Together with the Freund-Rubin parameter $e_0$, these are then responsible for the gauging of the effective theory. Note that since for all our
coset models the Betti number $b_3=0$, all the fluxes $e_i, p^A$ and $q_A$ are trivial. To obtain the scalar charges and the potential for each particular coset model, we just need to evaluate the general expressions in sections \ref{sec:DimRed} and \ref{sec:N=2sugra} using the specific geometric fluxes for each coset. In order to avoid overloading the paper with lenghty individual formulae, we only tabulate a consistent set of geometric fluxes for each one. These allow for the reconstruction of all relevant gauging-related quantities of the specific models using the general expressions of the previous two sections.

\begin{table}
\begin{center}
\begin{tabular}{lccc}
\hline
$M_7$ 	&	$m^i$	&	$q_i{}^j$	&	$m_i^A$ 	
\\
\hline
\\%[5pt]
$S^7 = \frac{{\rm SU}(4)}{{\rm SU}(3)}$	&	{\footnotesize $m^1=2$}	& 	0	&	0 	 	
\\[10pt]
$M^{110}$		
& {\footnotesize $m^1=m^2=2$}	
& 	0	&	0 	%&	0 	

\\[10pt]
$Q^{111}$	
&	{\footnotesize $m^1=m^2=m^3=2$}	
& 	0	&	0 		

\\[10pt]
$V_{5,2}$		
&	{\footnotesize $m^1=2$}	
& 	0	
&	0 	
\\
\hline
\\[-7pt]
%
%
%
%\\[10pt]
%
%
$S^7 = \frac{{\rm Sp}(2)}{{\rm Sp}(1)}$	
& {\footnotesize $m^1=m^2=2$}	
& 	0
& 	{\footnotesize $\begin{array}{l}
m_1{}^0 = m_2{}^0 = 0 \\[2pt]
m_1{}^1 = -m_2{}^1 = -4 \\[2pt]
\end{array}$	}
\\[10pt]
$N(1,1)$	
&	{\footnotesize $m^1=m^2=m^3=2$}
& 	0	
& 	{\footnotesize $\begin{array}{l}
m_1{}^0 = m_2{}^0  = m_3{}^0 = 0 \\[2pt]
2m_1{}^1 = 2m_2{}^1 = -m_3{}^1 = 4 \\[2pt]
\end{array}$	}
%
%&	0
%%	
%
%
%
\\
\hline
\\[-7pt]
%
%
%
%
%\\[10pt]
%
%
$N(k,l)$	
& 	{\footnotesize $\begin{array}{l}
m^1=\tfrac{l}{2}  , \;	m^2=\tfrac{k}{2},  \\[3pt]
m^3=-\tfrac{k+l}{2}   \\
\end{array}$	}
& 	0
& 	{\footnotesize $\begin{array}{l}
m_1{}^0=m_2{}^0
=m_3{}^0=1 \\
\end{array}$	}
\\[10pt]
$N(1,-1)$	&  	{\footnotesize $\begin{array}{l} m^1=-\tfrac{1}{2}, \, m^2=\tfrac{1}{2},\\ m^3=m^4=m^5=0 \end{array}$}
& 	{\footnotesize $\begin{array}{l}
q_5{}^4 = -q_4{}^5 = 3, \\
\text{all other } q_i{}^j=0
\end{array}$	}
& 	{\footnotesize $\begin{array}{l}
m_1{}^0=m_2{}^0
=m_3{}^0=1, \\
m_4{}^0=m_5{}^0 = 0
\end{array}$	}
\\
\hline
\end{tabular}
\\[10pt]	
\caption[Geom fluxes]{Geometric fluxes $m^i, q_i{}^j$ and $m_i^A$ for each of the coset models. For our choice of expansion forms, the parameters $e_{iA}$ vanish in all models.}\label{GeomFluxTable}
\end{center}
\end{table}

%%%%%%%%%%%%%%%%%%
%%%%%%%%%%%%%%%%%%

\begin{table}
\begin{center}
\begin{tabular}{lccc}
\hline
$M_7$ 	&	$u_A{}^B$	&				$s_{AB}$	&	$t_{AB}$
\\
\hline
\\%[5pt]
$S^7 = \frac{{\rm SU}(4)}{{\rm SU}(3)}$	&	0	& 	{\footnotesize $s_{00}=-4$}	& 	{\footnotesize $t^{00}=4$}
\\[10pt]
$M^{110}$		
&  	0	& 		{\footnotesize $s_{00}=-4$}	& 	{\footnotesize $t^{00}=4$}
\\[10pt]
$Q^{111}$	
& 	0	& 		{\footnotesize $s_{00}=-4$}	& 	{\footnotesize $t^{00}=4$}
\\[10pt]
$V_{5,2}$		
& 	{\footnotesize $\begin{array}{ll}
u_0{}^0 = 0,  				&	u_0{}^1 = -\tfrac{4}{\sqrt{3}},   \\
u_1{}^0 = \tfrac{4}{\sqrt{3}},  	& 	u_1{}^1 = 0   \\[2pt]
\end{array}$	}
& 	{\footnotesize $\begin{array}{ll}
s_{00} = 0,  				&	s_{01} = 0,  \\
s_{10} = 0,					& 	s_{11} = -\tfrac{8}{3}    \\[2pt]
\end{array}$	}
& 	{\footnotesize $\begin{array}{ll}
t^{00} = 0 , 				&	t^{01} = 0 , \\
t^{10} = 0,					& 	t^{11} = \tfrac{8}{3}    \\[2pt]
\end{array}$	}
\\[5pt]
\hline
\\[-10pt]
%
%
%
%\\[10pt]
%
%
$S^7 = \frac{{\rm Sp}(2)}{{\rm Sp}(1)}$	
& 	0
& 	{\footnotesize $\begin{array}{ll}
s_{00} = -4,				&	s_{01} = 0,  \\
s_{10} = 0,					& 	s_{11} = 0   \\[2pt]
\end{array}$	}
& 	{\footnotesize $\begin{array}{ll}
t^{00} = 4,				&	t^{01} = 0,  \\
t^{10} = 0,					& 	t^{11} = 0      \\[2pt]
\end{array}$	}
\\[10pt]
$N(1,1)$	
& 	0
& 	{\footnotesize $\begin{array}{ll}
s_{00} = -4,  				&	s_{01} = 0,  \\
s_{10} = 0,					& 	s_{11} = 0    \\[2pt]
\end{array}$	}
& 	{\footnotesize $\begin{array}{ll}
t^{00} = 4,  				&	t^{01} = 0,  \\
t^{10} = 0,					& 	t^{11} = 0     \\[2pt]
\end{array}$	}
\\[5pt]
\hline
\\[-10pt]
%
%
%
%
%\\[10pt]
%
%
$N(k,l)$	
& 	0
& 	0
&	0	
\\[10pt]
$N(1,-1)$	
& 	0
& 	0
&	0
\\
\hline
\end{tabular}
\\[10pt]	
\caption[Geom fluxes]{Geometric fluxes $u_A{}^B, s_{AB}$ and $t_{AB}$ for each of the coset models.}\label{GeomFluxCntdTable}

\end{center}

\end{table}

The set of specific geometric fluxes, $m^i$, $q_i{}^j$, $m_i{}^A$, $e_{iA}$,  $u_A{}^B$,  $s_{AB} $, $t^{AB}$, that we have brought to tables \ref{GeomFluxTable} and \ref{GeomFluxCntdTable} are those that can be read off by comparing the general algebra (\ref{GenExtAlg}) with the individual expressions given in appendix  \ref{sec:cosets}. Although these parameters do not have an intrinsic meaning (they rather depend on the specific duality frame of the set of forms used to compute them) they are enough to characterize the charges and the scalar potential for the individual models.

We conclude this section by explicitly evaluating the scalar potentials for some of the models in table \ref{cosetsummary}, using the scalar manifold parametrization of appendix  \ref{sec:cosets} and the geometric fluxes of tables \ref{GeomFluxTable} and \ref{GeomFluxCntdTable}. Some further massaging is required to bring the potentials in the form written below. The first one is the scalar potential for the $Q^{111}$ model:
{\setlength\arraycolsep{2pt}
\begin{eqnarray} \label{potQ111}
V &=&
-8 e^{2\phi}  \big( e^{-2u_1}  + e^{-2u_2}  + e^{-2u_3}  \big)
 + e^{4\phi} \big( e^{-2u_1+2u_2+2u_3}  + e^{2u_1-2u_2+2u_3}  + e^{2u_1+2u_2-2u_3}  \big)
 \nonumber \\
 &&
+ e^{4\phi-2u_1-2u_2-2u_3} \Big[ e^{4 u_1} \big( b^2 + b^3 )^2
+ e^{4u_2} \big( b^1 + b^3)^2
+ e^{4u_3} \big( b^1 + b^2 )^2  \Big]
 \nonumber \\
&&
+ \frac{1}{4} e^ {4\phi-2u_1-2u_2-2u_3}  \Big[e_0 +2 b^1 b^2 +2 b^1 b^3 +2 b^2 b^3  + 2 (\xi^0)^2  +2 (\tilde\xi_0)^2 \Big]^2
 \nonumber \\
&&
+ 4 e^{2\phi-2u_1-2u_2-2u_3} \big( ( \xi^0 )^2 + (\tilde{\xi}_0 )^2 \big) ,
\end{eqnarray}
}where the scalars $u_i$, $i=1,2,3$, have already been introduced above equation (\ref{LineElSKnV3}). Here we have set the form fluxes $p^A, q_A, e_i$ to zero without loss of generality: we have checked that this can always be done by a field and $e_0$ redefinition. The potentials for the $M^{110}$ and the SU(4)/SU(3) models can be obtained from (\ref{potQ111}) by setting $t^3=t^1$ and $t^3=t^1=t^2$ (recall that $t^i = b^i + i e^{2u_i}$). This just reflects the consistent subtruncation pattern of removing Betti vector multiplets. The potential for the SU(4)/SU(3) model can be seen to coincide with that for the universal SE$_7$ truncation \cite{vargaun2}.

The second one is the scalar potential for the $V_{5,2}$ model, where we have also set to zero all the form fluxes, except $e_0$, without loss of generality:
{\setlength\arraycolsep{2pt}
\begin{eqnarray} \label{potV52}
V &=&
-24e^{4\phi+2u_1} -\tfrac{16}{3} e^{-6u_1}   +\frac{1}{3} \Big( 9 e^{2\phi+u_1} -2 e^{-\varphi-3u_1}  \big(1+\chi^2 +e^{2\varphi}  \big)  \Big)^2
 + 12 e^{4\phi-2u_1} (b^1)^2
 \nonumber \\
&&
+ \frac{1}{4} e^{4\phi-6u_1}  \Big[e_0 +6 (b^1)^2 +\tfrac{4}{3}  (\tilde\xi_1)^2 +\tfrac{4}{\sqrt{3}} \tilde{\xi}_1 \xi^0 -\tfrac{4}{\sqrt{3}} \tilde{\xi}_0 \xi^1  + \tfrac{4}{3} (\xi^1)^2 \Big]^2 .
 \nonumber \\
&&
+\frac{4}{9}  e^{2\phi-6u_1} \Big[   3e^{3\varphi} (\xi^1)^2
   +  e^{\varphi} \Big( \sqrt{3} \xi^0 + 2\tilde\xi_1 +3\chi \xi^1\Big)^2
\\
&&
\quad \qquad \qquad+   e^{-\varphi} \Big( \sqrt{3} \tilde\xi_0 -2\xi^1 + 2\chi(2 \tilde\xi_1 +\sqrt{3} \xi^0) +3 \chi^2 \xi^1 \Big)^2
 \nonumber  \\
&& \quad \qquad \qquad +  e^{-3\varphi} \Big( -\sqrt{3} \tilde\xi_1 + \chi( 3\tilde\xi_0 -2\sqrt{3} \xi^1) +\chi^2 ( 2\sqrt{3} \tilde\xi_1 + 3\xi^0) +\sqrt{3} \chi^3 \xi^1  \Big)^2    \Big] . \nonumber
\end{eqnarray}
}Here we have used, again, the scalar $u_1$ introduced above equation (\ref{LineElSKnV3}), and $\varphi, \chi$ in (\ref{eq:zV52}). The $V_{5,2}$ effective theory can be further truncated consistently to the universal SE$_7$ model, by getting rid of the additional hypermultiplet. Accordingly, the universal SE$_7$ potential can be recoved from (\ref{potV52}) by switching off the complex structure deformations, $\varphi=0$, $\chi=0$, and two of the Darboux scalars $\xi^A$, $\tilde{\xi}_A$. Not all symplectic frames for the latter allow for the truncation to go through consistently, however: for example, it is inconsistent to set $\xi^1=\tilde\xi_1 =0$ in the frame used in (\ref{potV52}). To truncate consistently to the universal model, one should first symplectically rotate
 \begin{eqnarray} \label{eq:sympV52}
&  \xi^0 =\frac{1}{2} (\xi^{\prime 0} - \sqrt{3} \xi^{\prime 1}), \qquad
 & \xi^1 =\frac{1}{2} (-\sqrt{3}\tilde{\xi}_0^{\prime} -\tilde\xi_1^{\prime}), \nonumber \\
&  \tilde\xi_0 =\frac{1}{2} (\tilde{\xi}_0^{\prime} -\sqrt{3} \xi_1^{\prime}),  \qquad
 & \tilde\xi_1 =\frac{1}{2} (\sqrt{3} \xi^{\prime 0} + \xi^{\prime 1}).
 \end{eqnarray}
and then do set $\xi^{1\prime}=\tilde\xi_1^\prime =0$. Although no prepotential for the complex structure deformations turns out to exist in the primed basis, the latter is  preferred by the mass matrix: the primed scalars are the states of definite mass (see table \ref{table:SEspectra}).

Finally, we provide an explicit expression for the $N(1,1)$ potential:
{\setlength\arraycolsep{0pt}
\begin{eqnarray} \label{potN11}
V &=&
-12 e^{2\phi} \big(e^{-2u_1} +e^{-2u_2} \big)
 -4 e^{-2 u_1 - 2 u_2 - 2 u_3}
 +e^{-2\varphi-2 u_1 - 2 u_2 - 2 u_3} \big(1 + \chi^2+e^{2\varphi} \big)^2
 \nonumber \\
&&
  +\frac{1}{2} e^{2\phi-\varphi} \big( 1+\chi^2 +e^{2\varphi}    \big) \big( e^{2 u_1 - 2 u_2 - 2 u_3} + e^{-2 u_1 + 2 u_2 - 2 u_3} +  4 e^{-2 u_1 - 2 u_2 + 2 u_3} - 6 e^{-2 u_3} \big)
 \nonumber \\
&& + e^{4\phi} \big( e^{-2u_1+2u_2+2u_3}  + e^{2u_1-2u_2+2u_3}  + e^{2u_1+2u_2-2u_3}  \big) \nonumber \\
&& + e^{4\phi-2u_1-2u_2-2u_3} \Big[ e^{4 u_1} \big( b^2 + b^3 - \tilde{\xi}_1)^2
+ e^{4u_2} \big( b^1 + b^3 -\tilde{\xi}_1)^2
+ e^{4u_3} \big( b^1 + b^2 +2\tilde{\xi}_1)^2  \Big]
 \nonumber \\
&&
+ \frac{1}{4} e^{4\phi-2u_1-2u_3-2u_3}  \Big[e_0 \! + \! 2b^1b^2 \! + \! 2b^1b^3 \! + \! 2b^2b^3 \! - \! 2(b^1+b^2-2b^3) \tilde\xi_1  \! + \! 2(\xi^0)^2 \!+ \! 2(\tilde{\xi}_0)^2 \Big]^2
 \nonumber \\
&&
+\frac{1}{2}  e^{2\phi-2u_1-2u_2-2u_3} \Big[   4e^{\varphi} (\xi^0)^2 + e^{\varphi} (b^1+b^2-2b^3 - 2\tilde{\xi}_0)^2    %
 \nonumber \\
&&
\;\qquad
 + e^{-\varphi} (b^1+b^2-2b^3 +2\tilde{\xi}_0 -2 \chi \xi^0)^2
 + e^{-\varphi} \big( 2\xi^0 - \chi ( b^1+b^2-2b^3 -2\tilde{\xi}_0  )  \big)^2 \Big] \nonumber  \\
\end{eqnarray}
}The squashed $S^7$ model is obtained from here by setting $u_2 =u_1$, $b^2=b^1$ (and then redefining $u_3 \rightarrow u_2$, $b^3 \rightarrow b^2$). The universal SE$_7$ model is further obtained by setting $u_3=u_2=u_1$, $b^3=b^2=b^1$, and $\tilde{\xi}_1 =\xi^1=0$.

\subsection{Salient features}
\label{sec:features}

We would now like to make some further comments on each individual model, completing the description given at a glance in tables \ref{cosetsummary}, \ref{GeomFluxTable} and \ref{GeomFluxCntdTable}, and the case-by-case coset construction of appendix \ref{sec:cosets}.  In particular, we comment on subtruncation patterns and non-supersymmetric AdS vacua, and leave to next subsection the description of the supersymmetric points.

\subsubsection*{$M^{110}$, $Q^{111}$ and $V_{5,2}$}

The reductions on $M^{110}$ and $Q^{111}$ correspond to the addition of one and two Betti vector multiplets to the SU(4)/SU(3) model. As pointed out in \cite{caskoerbertrisak}, the forms defining a Sasaki--Einstein structure have the
same algebraic and differential properties as the LI forms on SU(4)/SU(3). Accordingly, the reduction on SU(4)/SU(3)
can be extended to a universal reduction on any Sasaki--Einstein manifold, which coincides with that of \cite{vargaun2}.
Similarly, one can replace the $\mathbb{CP}^2$ in the base of $M^{110}$ by any 4D compact K\"ahler-Einstein
manifold and obtain a consistent reduction of the same form. The $\mathbb{CP}^2$ can be replaced by $\mathbb{CP}^1 \times \mathbb{CP}^1$, leading to $Q^{111}$, or by a del Pezzo surface.

Since all these models contain the universal SE$_7$ truncation, all AdS solutions of the latter are also solutions. These include the $N=2$ supersymmetric Sasaki--Einstein solution (and its skew-whiffed counterpart),
the non-supersymmetric Pope--Warner solution \cite{popewarner} (which is stable within these truncations) and the Englert solution \cite{englert} (which is unstable, as is already the case for the universal SE$_7$ truncation). This also holds for the model on $V_{5,2}$.

In addition, we find a new non-supersymmetric AdS family of solutions of the $M^{110}$ and $Q^{111}$ models, which are not solutions of the universal SE$_7$ model. It can be checked that the following,
\eq{\spl{ \label{eq:newAdSSol}
&e^{2U_1} = \left(\tfrac{9}{5}\right)^{1/3} a \, , \quad
e^{2U_2} = \left(\tfrac{9}{5}\right)^{1/3} a \, , \quad
e^{2U_3} = \left(\tfrac{9}{5}\right)^{-2/3} a \, , \quad
e^{2V} = \tfrac{2}{7} 15^{2/3} a \, , \\
&b^1= b^2 = b^3=0 \, , \quad
(\xi^0)^2 + (\tilde{\xi}_0)^2 = \tfrac{172}{49} a^3 \, , \quad
e_0 = - \tfrac{540}{49} a^3 \, , \quad
\, ,
}}
is an AdS critical point of the $Q^{111}$ potential (\ref{potQ111}), which acquires the value $V = - 12 \sqrt{\tfrac{2}{7}}a^{9/2}$. Here, $a>0$ is just an overall scale, and the scalars $U_1, U_2, U_3$ are those introduced in appendix \ref{Q111}. They are related to the scalars $u_1, u_2, u_3$ used in  (\ref{potQ111}) via the change (\ref{ChangeSusyQ111}). Permutations of the scalars $U_1, U_2, U_3$ in (\ref{eq:newAdSSol}) are also solutions. The corresponding solution of the $M^{110}$ model follows (up to permutation) from the truncation $U_1 = U_2$, $b^1=b^2$, consistently valid across the full moduli space. We have verified that all the masses are above the BF bound, which indicates that the solution is stable within the $Q^{111}$ and $M^{110}$ truncations.

\subsubsection*{Sp(2)/Sp(1) and $N(1,1)$}

These models support a tri-Sasakian structure. Moreover, it was shown in \cite{caskoerbertrisak} that the reduction on the coset space Sp(2)/Sp(1) to an $N=4$ theory can be extended
to a reduction on a generic tri-Sasakian manifold. In appendix \ref{Sp2modSp1} we construct the maximal
$N=2$ subtruncation of the $N=4$ model, which contains $n_V=2$ vector multiplets and $n_H=2$ hypermultiplets.
Since this subtruncation can be obtained by imposing invariance under a certain
$\mathbb{Z}_2$ symmetry on the expansion forms and since this symmetry can be expressed in terms of the tri-Sasakian
structure, it follows that this subtruncation is automatically consistent and moreover that it can also be
extended to a generic tri-Sasakian manifold.

One interesting point is that for every choice of metric there are two possible SU(3)-structures
in the model, labeled by $\epsilon = \pm 1$ in the appendix.\footnote{Strictly speaking, if one goes from the SU(3)-structure with $\epsilon=1$ to the one with $\epsilon=-1$, the orientation is also changed, so that in order to obtain exactly
the same solution one has to compensate this by setting $J \rightarrow -J$ and $\Omega \rightarrow \bar{\Omega}$.}
The $N=2$ model therefore displays two separate branches. These have the same bosonic sector (with different range of the scalars, however),
but are expected to have a different fermionic sector. Indeed, both of these branches contain all of
the solutions, but since they are described with a different SU(3)-structure, a different amount of
supersymmetry will be visible.
For instance, all of the AdS vacuum solutions of the $N=4$ model of \cite{caskoerbertrisak} will be present:
the round solution (with manifest $N=2$ for $\epsilon=1$
and $N=1$ for $\epsilon=-1$), the non-supersymmetric associated Englert solution and the Pope-Warner solution (the latter being already unstable within this truncation).
Furthermore, there are the following solutions, which are rotationally symmetric in the $\eta^I$-space: the squashed solution (with $N=0$ for $\epsilon=1$ and
$N=1$ for $\epsilon=-1$), and the associated Englert solution.

We note two main subtruncations. First, one obtains the Sasaki-Einstein subtruncation of \cite{vargaun2}, with $n_V=n_H=1$
by putting $U_1 = V_1 = V_2$ and $z=i$. As vacuum AdS solutions it contains the round solution (with $N=2$), the associated
Englert solution and the Pope-Warner solution. Second there is the truncation with $n_V=2, n_H=1$ that one obtains by keeping only the
expansion forms that are invariant under the action of $X$. This amounts to putting $z=-i$.
The reduced theory is equivalent to the type IIA theory of \cite{ExploitingN=2} on Sp(2)/S(U(2)$\times$U(1)) with Romans mass $m=0$ \cite{ExploitingN=2}.
As vacuum solutions, this theory has the round solution (with $N=1$), the squashed solution and the Englert solution associated
to the latter. Note that the geometric fluxes $m^i$ that this $n_V=2$, $n_H=1$ truncation necessarily possesses, prevent a further truncation to the $n_V=1$, $n_H=1$ model corresponding to (massless) type IIA on G$_2$/SU(3) \cite{ExploitingN=2} or, equivalently, IIA on any nearly-K\"ahler six-fold \cite{poorNK}. Finally, a different further subtruncation is the $N=1$ SO(3)$_R$-invariant truncation described in section 7.2 of \cite{caskoerbertrisak}.

Finally we note that the $N=2$ truncation on $N(1,1)$ can be found by adding a Betti vector multiplet to the truncation on Sp(2)/Sp(1).

\subsubsection*{$N(k,l)$ and $N(1,-1)$}

For generic $k,l$ the truncation on $N(k,l)$, which has $n_V=3, n_H=1$, has two $N=1$ vacuum solutions of the type of eq.~\eqref{susyN=1}. As indicated
in diagram \ref{Diagram}, these models do not have a Sasaki--Einstein subtruncation, but rather a weak G$_2$-subtruncation.

A special point is $k=-l=1$. In this case the two $N=1$ vacuum solutions collapse to one solution. This manifold
has two extra LI two-forms, so that the consistent reduction enhances to one with $n_V=5, n_H=1$. This allows
for an extra non-supersymmetric LI Einstein solution with off-diagonal metric entries, which was first found in \cite{Einstein7}.

The model on $N(1,-1)$ is also interesting because it is the only one with non-trivial $q$-fluxes. According to the general analysis this implies
a gauging of the scalars in the vector multiplet sector as well as the presence of a non-abelian gauge group.

\subsection{The supersymmetric vacuum structure} \label{sec:susyVac}

All models admit supersymmetric AdS critical points which, by a suitable choice of geometric fluxes, can be always brought to the origin of the parametrizations of appendix \ref{sec:cosets} or, alternatively, arises at the points given in the appendices. The SU(4)/SU(3), $M^{110}$, $Q^{111}$ and $V_{5,2}$ models have only one $N=2$ point, that renders those spaces Sasaki-Einstein. The models based on the squashed $S^7$ and $N(1,1)$ posess two supersymmetric critical points, with $N=2$ and $N=1$. The generic $N(k,l)$ models display two $N=1$ critical points (denoted by I and II in \cite{KKreview}). These coalesce for $N(1,-1)$ on a single $N=1$ point. At the supersymetric critical points we find that the spectrum of each model organizes itself in OSp$(4|N)$ multiplets, as it should. We have summarized the type of OSp multiplets that arise for each model at each supersymmetric point in tables \ref{table:SummaryN=2Points} and \ref{table:SummaryN=1Points}. Appendix \ref{sec:susyspectra} contains the full spectra at these points. All these supersymmetric points partner with their corresponding skew-whiffed, non-supersymmetric counterparts.

%%%%%%%%%%%%%%%%%%%%%%%%%%%%%%%%%%%%%%%%%%%%%%%%%%%%%%%%%%%%%
%%%%%%%%%%%%%%%%%%%%%%%%%%%%%%%%%%%%%%%%%%%%%%%%%%%%%%%%%%%%%
%%%%%%%%%%%%%%%%%%%%%%%%%%%%%%%%%%%%%%%%%%%%%%%%%%%%%%%%%%%%%
%%%%%%%%%%%%%%%%%%%%%%%%%%%%%%%%%%%%%%%%%%%%%%%%%%%%%%%%%%%%%
\begin{table}%[!h]
\begin{center}
\tabcolsep=0.15cm
{\scriptsize
\scalebox{1}{
\begin{tabular}{p{3cm}|cccccc}
\hline

& 	
$S^7=\frac{{\rm SU}(4)}{{\rm SU}(3)}$ \rule{0em}{1.3em}
&   	
$M^{110}$
&	
$Q^{111}$
&
$V_{5,2}$
&
$S^7=\frac{{\rm Sp}(2)}{{\rm Sp}(1)}$
&			
$N(1,1)$
\\
\\[-14pt]

& 	

&   	

&	

&

&
(round)
&
(round)\rule{0em}{1.3em}

\\%[-14pt]
\hline
\\[-10pt]
Gravity
&
1, $(2,0)$
&
1, $(2,0)$
&
1, $(2,0)$
&
1, $(2,0)$
&
1, $(2,0)$
&
1, $(2,0)$
\\[5pt]
Long vector
&
1, $(4,0)$
&
1, $(4,0)$
&
1, $(4,0)$
&
1, $(4,0)$
&
2, $(4,0)$, $(3,0)$
&
2, $(4,0)$, $(3,0)$
\\[5pt]
Massless vector
&
0
&
1, $(1,0)$
&
2, $(1,0)$, $(1,0)$
&
0
&
0
&
1, $(1,0)$
\\[5pt]
Chiral
&
0
&
0
&
0
&
2, $(\tfrac{2}{3}, -\tfrac{2}{3})$, $(\tfrac{2}{3}, \tfrac{2}{3})$
&
0
&
0
\\[4pt]
\hline
\end{tabular}
}
\caption[N=2susyPoints]{Summary of OSp$(4|2)$ multiplets at the $N=2$ point of each model. The format of each entry is $n, \; (\Delta_1, R_1), \ldots,  (\Delta_1, R_n)$,
where $n$ is the number of multiplets of each type, and $(\Delta_i, R_i)$ are the conformal dimension and U(1) R-charge of the lowest component of each multiplet.}\label{table:SummaryN=2Points}
}\normalsize
\end{center}
\end{table}
%%%%%%%%%%%%%%%%%%%%%%%%%%%%%%%%%%%%%%%%%%%%%%%%%%%%%%%%%%%%%
%%%%%%%%%%%%%%%%%%%%%%%%%%%%%%%%%%%%%%%%%%%%%%%%%%%%%%%%%%%%%
%%%%%%%%%%%%%%%%%%%%%%%%%%%%%%%%%%%%%%%%%%%%%%%%%%%%%%%%%%%%%
%%%%%%%%%%%%%%%%%%%%%%%%%%%%%%%%%%%%%%%%%%%%%%%%%%%%%%%%%%%%%

%%%%%%%%%%%%%%%%%%%%%%%%%%%%%%%%%%%%%%%%%%%%%%%%%%%%%%%%%%%%%
%%%%%%%%%%%%%%%%%%%%%%%%%%%%%%%%%%%%%%%%%%%%%%%%%%%%%%%%%%%%%
%%%%%%%%%%%%%%%%%%%%%%%%%%%%%%%%%%%%%%%%%%%%%%%%%%%%%%%%%%%%%
%%%%%%%%%%%%%%%%%%%%%%%%%%%%%%%%%%%%%%%%%%%%%%%%%%%%%%%%%%%%%
\begin{table}%[!h]
\begin{center}
{\scriptsize
\scalebox{1.1}{
\begin{tabular}{p{3cm}|cccc}
\hline

& 	
$S^7=\frac{{\rm Sp}(2)}{{\rm Sp}(1)}$
&   	
$N(1,1)$
&	
$N(k,l)_{I, II}$
&
$N(1,-1)$
\\
\\[-12pt]

& 	
(squashed)
&   	
(squashed)
&	

&

\\
\hline
\\[-10pt]
Gravity
&
1, $(2)$
&
1, $(2)$
&
1, $(2)$
&
1, $(2)$
\\[5pt]
Massive gravitino
&
1, $(4)$
&
1, $(4)$
&
1, $(4)$
&
1, $(4)$
\\[5pt]
Massive vector
&
0
&
0
&
0
&
2, $(1+\tfrac{\sqrt{421}}{8})$
\\[5pt]
Massless vector
&
1, $(\tfrac{3}{2})$
&
2, $(\tfrac{3}{2}, \tfrac{3}{2})$
&
2, $(\tfrac{3}{2}, \tfrac{3}{2})$
&
2, $(\tfrac{3}{2}, \tfrac{3}{2})$
\\[5pt]
Chiral
&
5, $(5, \tfrac{10}{3},  \tfrac{10}{3},  \tfrac{10}{3},  \tfrac{5}{3})$
&
6, $(5, \tfrac{11}{3}, \tfrac{10}{3},  \tfrac{10}{3},  \tfrac{10}{3},  \tfrac{5}{3})$
&
4
&
4
\\[4pt]
\hline
\end{tabular}
}
\caption[N=1susyPoints]{Summary of OSp$(4|1)$ multiplets at the $N=1$ point of each model. The format of each entry is $n, \; (\Delta_1 , \ldots ,  \Delta_n)$, where $n$ is the number of multiplets of each type, and $\Delta_i$ is the conformal dimension of the lowest component of each multiplet. The $\Delta_i$ have been omitted for the chiral multiplets of  $N(k,l)$ and $N(1,-1)$.}\label{table:SummaryN=1Points}
}
\normalsize
\end{center}
\end{table}
%%%%%%%%%%%%%%%%%%%%%%%%%%%%%%%%%%%%%%%%%%%%%%%%%%%%%%%%%%%%%
%%%%%%%%%%%%%%%%%%%%%%%%%%%%%%%%%%%%%%%%%%%%%%%%%%%%%%%%%%%%%
%%%%%%%%%%%%%%%%%%%%%%%%%%%%%%%%%%%%%%%%%%%%%%%%%%%%%%%%%%%%%
%%%%%%%%%%%%%%%%%%%%%%%%%%%%%%%%%%%%%%%%%%%%%%%%%%%%%%%%%%%%%

%%%%%%%%%%%%%%%%%%%%%%%%%%%%%%%%%%%%%%%%%%%%%%%%%%%%%%%%%%%%%%%%%%%%%%%%%%
\section{Outlook}\label{sec:discussion}

In this paper, we established a general dimensional reduction of 11D supergravity on 7D manifolds with SU(3) structure, leading to $N=2$ gauged supergravity in four dimensions. We discussed how our reduction encompasses and generalizes several results previously derived in the literature on flux compactifications of 11D and type~II supergravity.
In the second part of the paper, we specialized to the coset manifolds $V_{5,2}$ (the Stiefel manifold), $N(k,l)$ (the Aloff--Wallach spaces), SU(4)/SU(3), Sp(2)/Sp(1) (which are two versions of the seven-sphere), $M^{110}$ and $Q^{111}$, which represent all the homogeneous SU(3)-structures admitting supersymmetric AdS$_4$ vacua of 11D supergravity. For each of them, we established a consistent truncation of 11D supergravity to a specific gauged $N=2$ supergravity model, which we analysed in detail. The main features of these models having already been summarized in the Introduction, here we speculate about how our general reduction could possibly be further extended.

We start remarking that the gaugings specified by our dimensional reduction are not the most general ones allowed by $N=2$ supersymmetry. For instance, our vector of magnetic charges is $m^I = (0,m^i)$, implying that the magnetic dual of the graviphoton never participates in the gauging. From a purely 4D perspective, there is no obstruction against introducing a non-vanishing parameter $m^0$ so that $m^I$ gets completed to $m^I = (m^0,m^i)$. Our $N=2$ action of section \ref{sec:N=2sugra} is already suitable for accomodating the extra charge: the only modification required is that the expression of the scalar potential gets generalized as discussed in appendix \ref{CheckScalarPot}.
When the $q_i{}^j$ and $\mathbb U$ charges vanish, the 11D supergravity reduction has a  type IIA counterpart (see discussion at the end of section~\ref{sec:DimRed}), and the addition of $m^0$ is interpreted as switching on the Romans mass. However, when a type IIA picture is not available, the higher-dimensional interpretation of $m^0$ is not clear. One way to incorporate it in an M-theory reduction might be to see it as a non-geometric flux: in the spirit of \cite{granaN2part2}, one could generalize the first differential relation in \eqref{GenExtAlg} to $\mathcal D\theta = m^I \omega_I$, where $\mathcal D$ is a generalized differential operator, of the form $\mathcal D = d + Q\cdot$, with $Q$ being a non-geometric flux which acts by index contraction.

From a 4D point of view, there are further gaugings that could naturally be switched on. Indeed, our flux matrices transform symplectically under Sp$(2n_H,\mathbb R)$, but are symplectically incomplete as far as Sp$(2n_V + 2,\mathbb R)$ is concerned.
One could speculate about a possible higher-dimensional interpretation of these extra gaugings as non-geometric fluxes
(see e.g.\ \cite{granaN2part2} for a similar discussion in a type II framework).

\section*{Acknowledgments}

We would like to thank Bernard de Wit, Mike Duff, Jerome Gauntlett, Tom\'as Ort\'\i n, Erik Plauschinn, Maria J.\ Rodriguez, Dan Waldram, Nick Warner and, especially, Stefan Vandoren, for helpful discussions. O.V. further wishes to thank Eoin \'O Colg\'ain for collaboration during the early stages of this work. D.C. is supported by an STFC grant ST/J002798/1.  P.K.\ is a Postdoctoral Fellow of the FWO -- Vlaanderen. The work of
P.K.\ is further supported in part by the FWO -- Vlaanderen project
G.0651.11 and in part by the Federal Office for Scientific,
Technical and Cultural Affairs through the `Interuniversity
Attraction Poles Programme Belgian Science Policy' P6/11-P. O.V. is supported in part by the Netherlands Organisation for Scientific Research (NWO) under the VICI grant 680-47-603, and by the Spanish Government research grant FIS2008-01980. O.V.\ finally wishes to thank the Simons Center, Stony Brook, for hospitality as this work was being finished during his participation at the summer workshop.

\appendix

%%%%%%%%%%%%%%%%%%%%%%%%%%%%%%%%%%%%%%%%%%%%%%%%%%%%%%%%%%%%%%%%%%%%%%%%%%
\section{Supersymmetry conditions for AdS$_4$ vacua}
\label{sec:susyconds}

In this appendix we analyse the supersymmetry conditions for vacuum AdS$_4$ solutions to 11D supergravity, and in particular
use them to show that homogeneous supersymmetric solutions are always of Freund--Rubin type. This theorem and its proof
is similar to an analogous theorem for type IIA supergravity that likewise forbids static SU(2) or genuine SU(3)$\times$SU(3)-structure
supersymmetric AdS$_4$ solutions on homogeneous 6D compactification manifolds \cite{effectivecosets,gengeomreview}.
Furthermore, we use the relation between the respective torsion classes of a G$_2$-structure and SU(3)-substructures to construct the supersymmetry conditions for SU(3)-structure vacua. For the manifolds studied in this paper, they lead to two important subcases.

\subsection{Homogeneous supersymmetric solutions are of Freund--Rubin type}

The supersymmetry conditions for a bosonic configuration in 11D supergravity are obtained by requiring that
the variation of the gravitino, given by
\eq{
\delta \Psi_M \,=\, \nabla_M \epsilon + \frac{1}{288} \left(\Gamma_M{}^{NPQR} - 8 \, \delta^N_M \Gamma^{PQR} \right) G_{NPQR} \epsilon \, ,
}
vanishes for some 11D Majorana--Weyl spinor $\epsilon$.

Let us first set up the reduction ansatz for vacuum solutions of the type AdS$_4 \times M_7$, which is simpler than
the off-shell reduction ansatz used elsewhere in the paper, but introduces a warp factor for completeness.
For the metric we take
\eq{
\label{gravitinovar}
\d s^2 = e^{2\warp} g_{(4)\mu\nu} \d x^{\mu} \d x^{\nu} + g_{mn} \d y^m \d y^n \, ,
}
where $\warp(y)$ is the warp factor, $g_{(4)\mu\nu}$ is the AdS$_4$ metric
and $g_{mn}$ the metric on the internal space $M_7$. The ansatz for the four-form flux is
\eq{
\label{ansatzRR}
G_4 = e^{4\warp} \text{vol}_4  f + F \, ,
}
where $\text{vol}_4$ is the volume-form built from $g_{(4)\mu\nu}$.
To study the supersymmetry generator we introduce the following compactification ansatz for the gamma-matrices
\eq{
 \Gamma_{\mu} = e^{A} \gamma_{\mu} \otimes \bbone \, , \qquad
 \Gamma_{m} = -i \gamma_{0123} \otimes \gamma_m \, ,
}
where $\gamma_{\mu}$ and $\gamma_m$ are the 4D and 7D gamma-matrices respectively (constructed using vielbeins associated to
the metrics $g_{mn}$ and $g_{(4)\mu\nu}$ respectively). We also take $\gamma_m^\dagger= \gamma_m$ and $\gamma_{1\ldots 7}=i$, where the indices in the latter and in $\gamma_{0123}$ above are
flat.

We will construct the supersymmetry conditions for having minimal $N=1$ supersymmetry in 4D (which includes $N>1$ as subcases).
The general $N=1$ ansatz for the supersymmetry generator is\footnote{For an analysis of supersymmetry ans\"atze in this context see e.g.\ \cite{BehrndtCveticLiu}.}
\eq{
\epsilon \,=\, e^{\warp/2} \psi_+ \otimes \zeta + e^{\warp/2} \psi_- \otimes \zeta^c \, .
}
Here, $\psi_{\pm}$ are 4D Weyl spinors, related by $\psi_- = D_4 \psi_+^*$, with $D_4$ such that $\gamma_\mu^* = D_4^{-1} \gamma_{\mu} D_4$. So together they define four supercharges ($N=1$).
$\zeta$ is a 7D {\em Dirac} spinor, with complex conjugate $\zeta^c = D_7 \zeta^*$, where $D_7$ is such that $\gamma_m^* = - D_7^{-1} \gamma_m D_7$.

Now we are ready to plug the compactification ansatz in \eqref{gravitinovar}. We will not aim here at a complete analysis, but mostly
work towards proving the theorem stated above. First, the 4D spinors
should satisfy the AdS Killing spinor equation
\eq{
\nabla_{\mu} \psi_- = \frac{1}{2} w\, \gamma_{\mu} \psi_+ \, ,
}
where the complex constant $w$ is associated to the AdS radius $\ell$:
$|w|=1/\ell$. Second, instead of studying the spinor equations directly it will be convenient to follow the $G$-structure approach and introduce differential forms constructed in terms of spinor bilinears as follows
\eq{
\zeta^{c\,\dagger} \gamma_{(l)} \zeta = \frac{1}{l!} \, \zeta^{c\,\dagger} \gamma_{m_1\ldots m_l} \zeta \, \d y^{m_1} \wedge \cdots \wedge \d y^{m_l} \, ,
\qquad l = 0,3,4,7 \, .
}
Using the properties under complex conjugation of the spinors and gamma-matrices one can show that the above bilinears
are zero for other values of $l$. Moreover, we can also introduce
\eq{
\zeta^\dagger \gamma_{(l)} \zeta \, ,
}
which is real for $l=0,1,4,5$ and imaginary for $l=2,3,6,7$. From the vanishing of the gravitino variation one finds
the following properties for the exterior derivatives of these spinor bilinears:
\subeq{\al{
& \d \left( e^{3\warp} \zeta^{c\,\dagger} \zeta \right) - 2 \, e^{2\warp} w\, \zeta^\dagger \hat{\gamma}_{(1)} \zeta = 0 \, , \label{susycond2} \\[2pt]
& \d \left( e^{3\warp} \zeta^{c\,\dagger} \hat{\gamma}_{(3)} \zeta \right) - 2 \, e^{2\warp} w\, \zeta^\dagger \hat{\gamma}_{(4)} \zeta + e^{3\warp} F \,\zeta^{c\,\dagger} \zeta = 0 \, , \label{susycond4} \\[2pt]
& \d \left( e^{4\warp} \zeta^\dagger \hat{\gamma}_{(2)} \zeta \right) + 3 i \, e^{3\warp} \Im \left( w\,  \zeta^\dagger \hat{\gamma}_{(3)} \zeta^c \right) - i\, e^{4\warp} * F (\zeta^\dagger \zeta) = 0 \, \label{susycond5} \, ,
}}
These equations can also be found from plugging our reduction ansatz in the spinor bilinear equations of \cite{gauntlettMkilling}. For the proof of our theorem we need some more relations, which follow from the original gravitino equation, namely
\eq{
\partial_m \left( e^\warp \zeta^\dagger \hat{\gamma}^m \zeta \right) - \frac{5}{6} e^\warp \zeta^\dagger F \zeta = 0 \label{extrasusycond2} \, ,
}
and
\subeq{\al{
& \frac{1}{2} \, e^{-\warp} \Im (w \,\zeta^\dagger \zeta^c) + \frac{f}{6} \zeta^\dagger \zeta = 0 \, , \label{algsusycond1} \\
& \frac{1}{2} \, \zeta^\dagger \d \warp \zeta + \frac{1}{2} \, e^{-\warp} \Re (w\, \zeta^\dagger \zeta^c) + \frac{1}{12} \zeta^\dagger F \zeta = 0  \, . \label{algsusycond2}
}}
In these relations, contraction of forms with gamma-matrices (the Dirac slash) is implicitly understood.

The proof now goes as follows. For a homogeneous solution all scalars have to be constant. It follows then from eq.~\eqref{susycond2} that
\eq{
\zeta^\dagger \gamma_{(1)} \zeta = 0 \, .
\label{condg1}
}
This condition is equivalent to the statement that $\zeta$ and $\zeta^c$ are proportional\footnote{One pedestrian way to show this goes as follows. Go to an explicit representation
of the 7D gamma-matrices in terms of 6D gamma-matrices. Write $\zeta$ as a sum of a 6D positive- and a negative-chirality spinor: $\zeta = \zeta_+^1 + \zeta_-^2$.
Now expand $\zeta_-^2 = c \, \zeta_-^1 + c^i \gamma_i \zeta_+^1$. Plugging this into \eqref{condg1} one finds $c^i=0$ and $|c|^2=1$, from which in the end \eqref{propcond} follows.}
\eq{
\zeta^c = e^{i\beta} \zeta \, ,
\label{propcond}
}
or alternatively the statement that $e^{i\beta/2} \zeta$ becomes a Majorana spinor.
Plugging \eqref{condg1} into eqs.~\eqref{extrasusycond2} and \eqref{algsusycond2} implies then $\Re (w \zeta^\dagger \zeta^c) = 0$,
or using eq.~\eqref{propcond},
\eq{
\Re (w \, e^{i\beta}) = 0 \, .
\label{phaseval}
}

Eq.~\eqref{propcond} also implies $\zeta^\dagger \gamma_{(2)} \zeta = 0$, which we can use in eq.~\eqref{susycond5} together with \eqref{phaseval} and again \eqref{propcond}
to find finally $F = 0$. From eq.~\eqref{ansatzRR} we see that this means that the flux $G_4$ is completely along the 4D volume, which is the hallmark of a Freund--Rubin solution. We can find the Freund--Rubin parameter $f$ from \eqref{algsusycond1},
\eq{
f = -3 \, e^{-\warp} \Im (w\, e^{i \beta}) \, .
\label{frpara}
}
Furthermore, the warp factor will be constant even for a non-homogeneous solution. This concludes the proof.

\subsection{Relation between G$_2$-structure and SU(3)-structure}

In the regime of the previous subsection, where eq.~\eqref{propcond} holds, we find that there has to be
a G$_2$-structure on $M_7$, associated to the Majorana spinor $e^{i\beta/2} \zeta$. Indeed, we can associate the only
surviving non-trivial spinor bilinears to the structure form $\phi$ of the G$_2$-structure and its Hodge dual
as follows:
\eq{\spl{
 \phi &\, =\, - i \zeta^\dagger \gamma_{(3)} \zeta \, , \\
*_7 \phi &\, =\, \zeta^\dagger \gamma_{(4)} \zeta \, .
}}
We find then from eq.~\eqref{susycond4}
the conditions for $N=1$ supersymmetry imply the condition for weak G$_2$ holonomy:
\eq{
\d (e^{3\warp} \phi) \,=\, 2 \, e^{2\warp} \tilde{w}  *_7 \phi \, ,
\label{weakG2}
}
with $\tilde{w} = - i e^{i\beta} w$. It is convenient to express this in terms of the G$_2$ torsion classes, which correspond
to the different representations of the exterior derivative of the G$_2$-structure form and its Hodge dual:
\eq{\label{G2tcl}\spl{
 \d \phi &= \tau_0 *_7\!\phi + 3 \, \tau_1 \wedge \phi + *_7 \tau_3 \, , \\
 \d *_7\! \phi &= 4 \, \tau_1 \wedge *_7 \phi + \tau_2 \wedge \phi \, ,
}}
where $\tau_0, \tau_1, \tau_2$ and $\tau_3$ transform under G$_2$ as $\bf{1}, \bf{7}, \bf{14}$ and $\bf{27}$ respectively.
We find then immediately that we must have
$\tau_0 = 2 \, \tilde{w}$ and $\tau_1 = \tau_2 = \tau_3=0$. In the Freund--Rubin setup, one can show that eqs.~\eqref{frpara} and \eqref{weakG2} (with $\d \warp=0$) are
sufficient for a supersymmetric solution.

Now, assume we have an SU(3) structure on $M_7$. Then the supersymmetry condition \eqref{weakG2} can be translated into conditions on the SU(3)-structure. In order to do this, note that having an SU(3)-structure means there exists a globally defined Dirac spinor on $M_7$, which yields a U(1) worth of Majorana spinors.  Correspondingly, we can construct a one-parameter family of $\text{G}_2$-structures containing our SU(3)-structure as
\eq{\label{G2fromSU3}\spl{
 & \phi  = \eta \wedge J + \Re \!\left(e^{i \alpha }\Omega \right) \, , \\
 & \Rightarrow  *_7 \phi  = \frac{1}{2} J \wedge J - \eta \wedge \Im \left(e^{i \alpha} \Omega \right)\, ,
}}
where the parameter $\alpha$ is related to the choice of the Majorana spinor defining the G$_2$-structure starting from the Dirac spinor defining the SU(3)-structure. The next step is to construct a relation between G$_2$ and SU(3)-structure torsion classes. To simplify the expressions we put the
vector SU(3) torsion classes $V_1,V_2,W_0,W_4$ and $W_5$, which never appear in this paper, to zero. We find then
\eq{\label{G2tclSU3}\spl{
\tau_0  \, = &\,\, \frac{1}{7}\left[6 \, R + 12 \, \Re (e^{i\alpha} W_1) + 4 \, \Im E\right] \, , \quad
\tau_1  =  \left[\frac{1}{3} \Re E + \frac{1}{2} \Im (e^{i\alpha} W_1) \right]\eta \, , \\
\tau_2  \,= &\,\, T_2 + \Im (e^{i\alpha} W_2) \, , \\
\tau_3  \,= & \,\,\frac{1}{14}\left[4 R + \Re (e^{i\alpha} W_1) -2 \Im E\right] \left[4\eta \wedge J - 3 \Re (e^{i\alpha}\Omega)\right]\\ & \,\, - \eta \wedge \left[T_1 + \Re (e^{i\alpha} W_2)\right] -*_6 [\Re (e^{i\alpha} S) -W_3]  \, .
}}
We can now analyse the supersymmetry condition \eqref{weakG2} in terms of SU(3)-structure.

\subsection{SU(3)-structure supersymmetry conditions}

We are now ready to express the $N=1$ supersymmetry condition in terms of SU(3)-structure classes.\footnote{For a more thorough study of the AdS$_4$ supersymmetry conditions on an SU(3)-structure manifold, also outside the Freund--Rubin case, see \cite{lukassafin,BehrndtCveticLiu}. For a classification of $N=2$ solutions, see \cite{Gabella:2012rc}.} Using \eqref{G2tcl} and \eqref{G2tclSU3} in \eqref{weakG2} we find then (putting the warp factor to zero)
\be\begin{array}{cc}
 \tilde{w} = \frac{1}{14} \left[ 6 \, R + 4 \, \Im E + 12 \, \Re(e^{i\alpha}W_1) \right]  ,& \,\,\,4  R -2 \Im E + \Re (e^{i\alpha} W_1) = 0 \, , \\ [1mm]
  T_1 + \Re (e^{i \alpha} W_2) = 0 \, , &  T_2 + \Im (e^{i \alpha} W_2) = 0 \, ,\\ [1mm]
 \Re (e^{i \alpha} S) - W_3 = 0 \, ,& \frac{2}{3} \, \Re E + \Im (e^{i\alpha} W_1) = 0 \, .
\end{array}
\ee
In our case of homogeneous structures, the last equation actually splits in $\Re E = \Im (e^{i\alpha} W_1)=0$. Indeed, from \eqref{SU3tcl} we find the relation
$\d J^3 = 2 \, \Re E \, \eta \wedge J^3 $, which, since
on a compact manifold the volume-form cannot be exact, implies that $\Re E$ cannot be everywhere non-vanishing; if the structure is homogeneous, then we conclude that $\Re E = 0$. As for the solutions to the rest of the equations, we can distinguish two independent subcases, which play an important role
in the paper. First let us look at solutions with $N=2$. This means they should be valid for all $\alpha$, which
puts $W_1 = W_2 = S =0$. This leads immediately to the class of Sasaki--Einstein solutions:
\eq{
R = \tilde{w} \, , \qquad E = 2 \, \tilde{w}\, i \, , \qquad T_1 = T_2 = W_2 = S = W_3 = 0 \, ,
}
which are usually normalized by taking $\tilde{w}=2$, in such a way that the SU(3)-structure forms satisfy \eqref{halfflat7d} and the Einstein constant of the Sasaki--Einstein metric is 6. For these solutions, the U(1) generated by the vector dual
to $\eta$ plays the role of the $R$-symmetry, and the metric $ds^2(B_6)$ is K\"ahler--Einstein. The reduction to type IIA supergravity along this vector breaks all supersymmetry.

Second, we consider $N=1$ solutions for which the SU(3)-structure is
invariant under this U(1). In other words they must obey
$
\mathcal{L}_{X} J \,=\, \mathcal{L}_{X} \Omega \,=\, \mathcal{L}_{X} \eta \,=\, 0 \,$ .
This leads to
\eq{\label{susyN=1}\spl{
& \tilde{w} = - 3\, R \, , \qquad \Re (e^{i\alpha} W_1) + 4 \, R = 0 \, , \qquad T_1 + \Re (e^{i \alpha} W_2) = 0 \, , \\
& E = \Im (e^{i\alpha} W_1) = \Im (e^{i \alpha} W_2) = W_3 = T_2 = S = 0 \, .
}}
Choosing for instance $e^{i\alpha} = -i$, the exterior derivatives of the SU(3)-invariant forms read then as in \eqref{halfflat7d}. The solution takes the form of a U(1) fibration over 6D half-flat SU(3)-structure manifold. Since the structure is invariant under the vector $X$ dual to $\eta$, reduction to type IIA along it does not break supersymmetry. This leads to the solutions of \cite{lt,tomasiellocosets,cosets} in the limit
of zero Romans mass.

Note that a priori there could also be more general $N=1$ solutions which are not invariant under the U(1), but we did not find any among the coset manifolds we considered.

%%%%%%%%%%%%%%%%%%%%%%%%%%%%%%%%%%%%%%%%%%%%%%%%%%%%%%%%%%%%%%%%%%%%%%%%%%
\section{Calculation of the prepotential $\prepG$}\label{howtoprepot}

At each point of a 6D SU(3)-structure manifold, the space of deformations of the almost complex structure induced by the decomposable complex three-form $\Omega$ is special K\"ahler \cite{hitchinfunc}.
Sometimes this property extends globally, at least for a subset of all possible deformations. For instance, on coset manifolds this is the case as far as left-invariant deformations are concerned.
In this appendix, we elaborate on the results of \cite{hitchinfunc} (see also \cite{granaN2part1}) and show how the special K\"ahler geometry data can concretely be extracted starting from $\Omega$.
In particular, starting from an expansion of $\Omega$ in terms of a symplectic basis as in \eqref{LISU3str},\footnote{For simplicity we put in this appendix
$V=0$  in the expression \eqref{LISU3str} for $\Omega$. Since turning on $V$ only changes $\Omega$ by an overall factor, it has
no effect on the expression of $\prepG$ anyway.} we provide an algorithm for obtaining the dependent variables $\prepG_B$ as holomorphic functions of the independent variables $Z^A$, as well as the prepotential $\prepG(Z)$.

In \cite{hitchinfunc} it is shown that a decomposable complex three-form is completely determined by its real part through the Hitchin function.
If we take $\Re \Omega = x^A \alpha_A - v_A \beta^A$, then the Hitchin function is given by\footnote{To be completely precise, the Hitchin function and $\Im \Omega$ are only determined in terms of $\Re \Omega$ up to a sign choice. The other choice of sign corresponds to plugging $H \rightarrow -H$, $\Im Z^A \rightarrow - \Im Z^A$, $\widetilde{\prepG}(\Re Z^A, \Im Z^A)\rightarrow -\widetilde{\prepG}(\Re Z^A, -\Im Z^A)$ and $\prepG(Z) \rightarrow \overline{\prepG(Z)}|_{\bar{Z}\rightarrow Z}$ in the following equations.}
\eq{
H(x^A,v_A) = \sqrt{-\frac{1}{6} \Tr \, K^2} \, , \quad \text{with} \;\;
K^m{}_n = \frac{1}{2}(\Re \Omega)_{n p_1 p_2} \wedge \frac{1}{3!} (\Re \Omega)_{p_3 p_4 p_5} \epsilon^{m p_1 p_2 p_3 p_4 p_5} \, ,
}
where $\epsilon$ is associated to the canonical volume-form $-\tilde{\omega}^0$.
For $K$ defined as above, one shows that $K^m{}_n K^n{}_p = c \, \delta^m{}_p$.
Then $\Omega$ is decomposable if and only if
$c<0$, which allows for making $K$ into an almost complex structure by normalizing. The variation of the Hitchin
function is given by \cite{hitchinfunc}
\eq{
\delta H(\Re \Omega) \, \tilde{\omega}^0 = \delta \Re \Omega \wedge \Im \Omega(\Re \Omega) \, .
}
One finds then that $\Im \Omega = y^A \alpha_A - w_A \beta^A$ is given in terms
of the independent variables $(x^A,v_A)$ in $\Re \Omega$ as
\eq{
 w_A = - \frac{\partial H(x^A,v_A)}{\partial x^A} \, , \qquad y^A = \frac{\partial H(x^A,v_A)}{\partial v_A} \label{phider}\, .
}

We would however like to find a different description, where $\prepG_A = v_A + i w_A$ are the dependent variables given in
terms of the independent $Z^A = x^A + i y^A$. Let us therefore define the following Legendre transform
\eq{\label{Gtdef}
\widetilde{\prepG}(\Re Z^A, \Im Z^A) = - H(\Re Z^A, \Re \prepG_A(\Re Z^A, \Im Z^A)) + \Im Z^A \Re \prepG_A(\Re Z^A, \Im Z^A) \, ,
}
where, as usual, $\Re \prepG_A(\Re Z^A, \Im Z^A)$ is found by extremizing with respect to $\Re \prepG_A$, or equivalently
by solving the second equation of \eqref{phider} for $v_A$. Let us remark that this is only possible if
\eq{\label{seconddernondegen}
\det \left( \frac{\partial^2 H}{\partial v_A \partial v_B}\right) \neq 0 \, .
}
Now $\widetilde{\prepG}$ determines $\prepG_A$ from $Z^A$ as follows:
\eq{\label{Gtprop}
 \Im \prepG_A = \frac{\partial \widetilde{\prepG}}{\partial(\Re Z^A)} \, , \qquad
 \Re \prepG_A = \frac{\partial \widetilde{\prepG}}{\partial(\Im Z^A)} \, .
}
$\widetilde{\prepG}$ is still a real function, while what we are looking for is a holomorphic function $\prepG(Z)$
such that
\eq{
\label{Gprepot}
\prepG_A = \frac{\partial \prepG\,}{\partial Z^A} \, .
}
However, some calculation shows that, under the condition that \eqref{seconddernondegen} is fulfilled,
$\widetilde{\prepG}$ is a suitable candidate to be the imaginary part of $\prepG$, i.e.\ it satisfies Laplace's equation,
which follows from the integrability of the Cauchy--Riemann equations, and using these Cauchy--Riemann equations
\eqref{Gprepot} implies \eqref{Gtprop}. We can then calculate the full holomorphic $\prepG$
from $\Im \prepG(\Re Z^A, \Im Z^A)$ using a formula from \cite{holofromreal},
\eq{
\prepG(z) - \overline{\prepG(\bar{a})} \,=\, 2 i \, \Im \prepG\Big(\frac{Z^A+a}{2},\frac{Z^A-a}{2i}\Big) \, ,
}
which we can evaluate for $a \rightarrow 0$. Ultimately, \eqref{seconddernondegen} becomes the condition for
the existence of a holomorphic prepotential in the chosen symplectic basis.

%%%%%%%%%%%%%%%%%%%%%%%%%%%%%%%%%%%%%%%%%%%%%%%%%%%%%%%%%%%%%%%%%%%%%%%%%%
\section{Details on the dimensional reduction}\label{sec:DetailsDimRed}

In this appendix we provide the details of the dimensional reduction of the 11D supergravity action \eqref{11daction} on our SU(3)-structure 7D manifolds. In section \ref{subsec:CalcRicciScalar} we reduce the 11D Einstein--Hilbert term, while in section \ref{DetailsRedForm} we turn to the form sector.

\subsection{The Einstein--Hilbert term}\label{subsec:CalcRicciScalar}

\subsubsection*{The higher-dimensional Ricci scalar}

We start by deriving a formula for the higher-dimensional Ricci scalar. The reduction
ansatz is more general than in the rest of the paper, so the expression provided here can also be useful in different setups.

Suppose we have a compactification from $d+D$ to $d$ dimensions with a certain
canonical internal metric $g_{0}$, which has $L$ Killing vectors $k_\Lambda$ ($\Lambda=1,\ldots, L$). Let us associate
a vielbein $e^\alpha$ to the canonical metric and suppose we can choose it such that the
Killing vectors are constant in flat coordinates $\partial_m k^\alpha{}_\Lambda=0$. For a coset
manifold left-invariant vectors will indeed give rise to such Killing vectors \cite{casromanssym},
but the result is not restricted to these manifolds. It was used for instance for generic tri-Sasakian
manifolds in \cite{caskoerbertrisak}. We take the following reduction ansatz for the metric:
\eq{\label{generalmetricansatz}
\d s^2 = e^{2 \varphi} \d s_{d}^2 + g_{\alpha\beta}(x) \left(e^\alpha + k^\alpha_\Lambda A^\Lambda(x)\right) \big(e^\beta+k^\beta_\Sigma A^\Sigma(x)\big) \, ,
}
where $\d s_{d}^2$ indicates an arbitrary metric in the $d$ dimensions, and $g_{\alpha\beta}(x), A^\Lambda(x)$ are respectively
scalar and vector fields in $d$ dimensions. Furthermore, we choose $\varphi$ in order to obtain the Einstein frame in $d$
dimensions,
\eq{
\varphi = - \frac{1}{2(d-2)} \log \frac{\det g}{\det g_{0}} \, .
}

A long calculation leads to the following total Ricci scalar:
\begin{multline}\label{Ricciscalar}
R \,=\,  e^{-2 \varphi} R_{d} + R_{D} - \frac{1}{4} e^{-4 \varphi} (k_\Lambda \cdot k_\Sigma) F^\Lambda_{\mu\nu} F^{\Sigma\,\mu\nu} -\frac{1}{4} e^{-2 \varphi} D_{\mu} g_{\alpha\beta} D^{\mu} g_{\gamma\delta} g^{\alpha\gamma} g^{\beta\delta} \\
 -2 e^{-2\varphi} \nabla^{\mu} \partial_\mu \varphi - (d-2) e^{-2\varphi} \partial_{\mu} \varphi \partial^{\mu} \varphi \, ,\quad
\end{multline}
where $R_{d}$ and $R_{D}$ are the external and internal Ricci scalar (with the latter being calculated as if $g_{\alpha\beta}$ were constant and $A^\Lambda=0$), and the $d$-dimensional indices $\mu,\nu$ indices are raised using the metric in $ds^2_d$.

Here, the covariant derivative and the field strengths are given by:
\eq{\spl{\label{MetricCovDer}
D_{\mu} g_{\alpha\beta} \,=\, \partial_{\mu} g_{\alpha\beta} + 2 \, A^\Lambda_{\mu} k_\Lambda^\gamma\, g_{\gamma\delta} \, \omega^\delta{}_{(\alpha|m} e^{m}{}_{\beta)} \,=\, \partial_{\mu} g_{\alpha\beta} - (\mathcal{L}_{k_\Lambda} g)_{\alpha\beta} A^\Lambda_{\mu}  \, , \\[2pt]
F^\Lambda \,=\, \d A^\Lambda - \frac{1}{2} f^\Lambda{}_{\Sigma\Omega} A^\Sigma \wedge A^\Omega \, , \qquad \text{where} \;\; [k_\Lambda,k_\Sigma] = f^\Omega{}_{\Lambda\Sigma} k_\Omega \, .
}}
Here $\omega^\alpha{}_\beta$ are the spin connection one-forms associated to the coframe $e^\alpha$. Note that the covariant derivative reduces to the ordinary derivative if $k_\Lambda$ is a Killing vector
of the full metric $g_{\alpha\beta}$ (as opposed to only of the canonical metric).

In this paper we have $d=4,D=7$, and there is only one Killing vector, $k$. This is the dual of the one-form $\theta$, and is related to the defining vector of the SU(3)-structure $X$ as $k = e^{V} X$.
Our canonical metric has volume form $-\tilde \omega^0 \wedge\theta$, so for the Weyl factor $\varphi$ we find
\eq{
e^{2 \varphi} = \frac{\sqrt{\det g_{0(D)}}}{\sqrt{\det g_{(D)}}} \,=\, \frac{-\tilde \omega^0 \wedge\theta}{\frac{1}{6}J \wedge J\wedge J \wedge \eta} \,=\, e^{2V} \mathcal{K}^{-1} \, .
}
Then the metric ansatz \eqref{generalmetricansatz} reduces to the one given in the main text in \eqref{KKmetric}.
We will look at calculating the different parts of expression \eqref{Ricciscalar} next.

\subsubsection*{The internal Ricci scalar}

One of the ingredients in the expression \eqref{Ricciscalar} is the internal Ricci scalar $R_{7}$, which we would like to
express in terms of SU(3) torsion classes, and ultimately in terms of 4D scalar fields and geometric fluxes. Taking our cue from \cite{held}
we will make use of the fact that the 7D Ricci scalar has already been calculated in terms of G$_2$ torsion classes \cite{bryantG2} and there exists a one-parameter family of $\text{G}_2$-structures containing our SU(3)-structure (and thus leading to the same metric), given by \eqref{G2fromSU3}.
In terms of the G$_2$ torsion classes, the Ricci scalar is given by \cite{bryantG2}:
\eq{
\label{R7G2}
R_7 = -12 *_7 \d *_7 \tau_1 + \frac{21}{8} \tau_0^2 + 30 (\tau_1)^2 - \frac{1}{2}(\tau_2)^2 - \frac{1}{2}(\tau_3)^2 \, .
}
Plugging the relation \eqref{G2tclSU3} between G$_2$- and SU(3)-structure classes into the above expression, we find
\eq{\label{R7SU3sim}\spl{
R_7 \;=\;& -4\, \eta \lrcorner \d(\Re E ) +\frac{15}{2} |W_1|^2 - \frac{3}{2} R^2 - \frac{14}{3}(\Re E)^2 + 6 R (\Im E) \\
& - \frac{1}{2} (T_2)^2 -\frac{1}{2} |W_2|^2 - \frac{1}{2}(T_1)^2 - \frac{1}{2} (W_3)^2 - \frac{1}{4}|S|^2 \, .
}}
In order to simplify the expression, we used the following quadratic constraints on the torsion classes:
\eq{\label{tclconstr1}\spl{
& 6\, W_1 R - T_1 \cdot W_2 = 0 \, , \\
& 6 \,\eta \lrcorner \d W_1 - 2 W_1 E + 4 W_1 \bar{E} + W_2 \cdot T_2 - i S \cdot W_3 = 0 \, ,
}}
which can be obtained by acting with an exterior derivative on \eqref{SU3tcl} and imposing consistency. Note furthermore
that the above expression for the Ricci scalar is manifestly invariant under the U(1) rotation induced by shifting $\alpha$ in \eqref{G2fromSU3} (as it should be since the metric is also invariant).

To reexpress \eqref{R7SU3sim} in terms of the geometric fluxes, it turns out to be convenient to first
rewrite it as
\eq{\spl{\label{R7onlysingletclasses}
R_7 &\;=\; *_7 \left[\frac{1}{2} (\eta \lrcorner \d J)\wedge (\eta \lrcorner \d J) \wedge J
\,+\, \frac{i}{4} (\eta \lrcorner \d \Omega) \wedge (\eta \lrcorner \d \bar{\Omega})
\,+\, \frac{1}{2} \, \d \eta \wedge \d \eta \wedge J\right] \\
& -\frac{1}{2} \big| \d \Omega|_{\eta=0}\big|^2 -\frac{1}{2} \big| \d J|_{\eta=0}\big|^2  - 8 (\Re E)^2 - 2 (\Im E)^2 + 6 R\, \Im E - \frac{9}{2} R^2 + 18 |W_1|^2 .
}}
Plugging in the expressions for the scalar torsion classes \eqref{TorsionParam} and further working out the derivatives on $J$ and $\Omega$
in terms of geometric fluxes by using the expansions in section \ref{subsec:expansionforms}, we obtain in the end the expression given in~\eqref{Vgeo} below.

\subsubsection*{The kinetic terms}

The kinetic terms in the effective action come from the third to last term in eq.~\eqref{Ricciscalar} (with the proper
prefactor $\frac{1}{2}\, e^{2 \varphi}$). Recalling that in the metric ansatz \eqref{KKmetric} we have just one vector, $k$, and that the 4D gauge vector is called $A^0$, from the third term we find immediately the gauge kinetic term
\eq{ \label{graphotkinterm}
- \frac{1}{8} e^{-2 \varphi} (k_\Lambda \cdot k_\Sigma) F^\Lambda_{\mu\nu} F^{\Sigma\,\mu\nu} = - \frac{1}{4} \mathcal{K} (dA^0)_{\mu\nu}(dA^0)^{\mu\nu} .
}

Next we will study the 4D kinetic terms for the scalars from the internal metric by working out the fourth term in \eqref{Ricciscalar},
\eq{\label{kinterms}
-\frac{1}{8} D_{\mu} g_{\alpha\beta} D^{\mu} g_{\gamma\delta} \,g^{\alpha\gamma} g^{\beta\delta}\,,
}
in terms of $J$, $\Omega$ and $\eta$. Let us first construct the action of the covariant derivative on these forms, starting from
its action \eqref{MetricCovDer} on the metric. Again setting the vector SU(3) torsion classes to zero, we obtain from \eqref{SU3tcl}
\eq{\label{covderforms}\spl{
D_{\mu} \eta &\,=\, \partial_\mu \eta - (\mathcal{L}_k \eta) A^0_\mu \,\,=\, \partial_{\mu} \eta  \, , \\
D_{\mu} J &\, =\, \partial_\mu J - (\mathcal{L}_k J) A^0_\mu \,= \,\partial_{\mu} J -  e^V \left(\tfrac{2}{3} \Re E \, J + T_2 \right) A^0_{\mu} \, , \\
D_{\mu} \Omega &\,=\, \partial_\mu \Omega - (\mathcal{L}_k \Omega) A^0_\mu \,=\, \partial_{\mu} \Omega - e^V \left(E \,\Omega + S  \right) A^0_{\mu} \, .
}}
Plugging in eqs.~\eqref{LISU3str} and \eqref{GenExtAlg}, the Lie derivatives of the forms can be computed in terms of the metric moduli and of the geometric fluxes, and we find the expressions for $Dv^i$, $DZ^A$, $DG_A$ given in the main text in eq.~\eqref{covderv}. Note that the transformation of $Z$, which $A^0_{\mu}$ gauges,
\eq{\label{deltaOmega}
\delta \Omega = -e^V \left(E \,\Omega + S  \right) \, ,
}
is a proper complex deformation since $S$ is a (2,1)-form. This implies that the functional form of the $\prepG_A(Z)$ does not change.

The treatment of \eqref{kinterms} is standard, see e.g.~\cite{candelas}. Although \cite{candelas} considered Calabi--Yau manifolds, their analysis of the kinetic terms \eqref{kinterms} should also apply to our case, since it does not depend on internal exterior derivatives. However, our formulae are going to work even without integrating on the compact manifold; this guarantees that the reduction goes through not only at the level of the action, but also at the level of the equations of motion.
Working out the kinetic terms we find
\begin{multline}
-\frac{1}{8}  D_{\mu} g_{\alpha\beta} D^{\mu} g_{\gamma\delta} g^{\alpha\gamma} g^{\beta\delta} =  \\
-  \frac{1}{2}\partial_{\mu} V \partial^{\mu} V - \frac{1}{4}  \Tr (D_{\mu} g)_{(2,0)} g^{-1} (D^{\mu} g)_{(0,2)} g^{-1}
- \frac{1}{8} \Tr (D_{\mu} g)_{(1,1)} g^{-1} (D^{\mu} g)_{(1,1)} g^{-1} \, .
\end{multline}
Here $(D^{\mu} g)_{(2,0)+(2,0)}$ is associated to the deformations of $\Omega$, while $(D^{\mu} g)_{(1,1)}$ is
associated to the deformation of $J$.

Let us start with the deformations of $\Omega$. Following \cite{candelas} we define the (2,1)-forms $\chi_a$ as
\eq{
\frac{\partial \Omega}{\partial z^a} = \kappa_a \Omega + \chi_a \, .
}
One can then show, analogously to \cite{candelas}, that
\eq{
(D_\mu g)_{(0,2)\alpha\beta} = \frac{1}{8} \chi_{a\, \alpha\gamma\delta} \overline{\Omega}_\beta{}^{\gamma\delta} D_{\mu} z^a \, ,
}
and finally
\eq{
- \frac{1}{4}  \Tr (D_{\mu} g)_{(2,0)} g^{-1} (D^{\mu} g)_{(0,2)} g^{-1} \,=\,
- \frac{1}{8}  \chi_{a} \cdot \bar{\chi}_{\bar{b}} \, D_{\mu} z^a D^{\mu} \bar{z}^{\bar{b}} \,=\,
- g_{a \bar b} D_\mu z^a D^\mu {\bar z}^{\bar b} \, ,
}
where $g_{a \bar b}$ is the special K\"ahler metric associated to the K\"ahler potential $K_{\Omega}$ defined in
\eqref{csKP}.

Turning to the (1,1)-part we find:
\bea
- \frac{1}{8}  \Tr (D_{\mu} g)_{(1,1)} g^{-1} (D^{\mu} g)_{(1,1)} g^{-1} =
- \frac{1}{4}  D_{\mu} J \cdot D^{\mu} J = - \frac{1}{4}  *_7 (D_{\mu} J \wedge *_7 D^{\mu} J)\nonumber \\ [2mm]
 = -\frac{3}{4} \partial_\mu V \partial^\mu V + \frac{1}{2\,\mathcal K}\partial_\mu V D^\mu\mathcal K - g_{ij} D_\mu v^i D^\mu v^j,\quad
\eea
where to write the last line we used the expansion of $J$ introduced in \eqref{LISU3str} and the definition~\eqref{defgij} of the metric $g_{ij}$.

Eventually for the scalar kinetic terms we arrive at
\be
\frac{1}{8}g^{\alpha\gamma} g^{\beta\delta} D_\mu g_{\alpha\beta}D^\mu g_{\gamma\delta} + \partial_\mu\varphi \partial^\mu\varphi\,=\, \partial_\mu \phi \, \partial^\mu \phi + g_{ij}D_\mu v^i D^\mu v^{j} + g_{a \bar b} D_\mu z^a D^\mu {\bar z}^{\bar b}\,,
\ee
where we introduced the 4D dilaton
\be
e^{2\phi} := e^{3V} {\cal K}^{-1} = e^{2\varphi + V}.
\ee

\subsubsection*{The 4D action}

Using the above results, the 11D Einstein--Hilbert term reduces to the following 4D action:

\eq{\spl{\label{Sgeo}
S_{\rm EH} \,=\, \frac{1}{\kappa_4^2} \int \Big( &  \frac{1}{2}R_4 *\!1 - \frac{1}{4} \mathcal K \, dA^0 \wedge * dA^0 + g_{ij}Dv^i\wedge * D v^{j} + g_{a \bar b} Dz^a \wedge *D{\bar z}^{\bar b} \\
& + d \phi \wedge *d \phi - V_{\rm geo}*\! 1
\Big),
}}
where $\kappa^{-2}_4 = \kappa_{11}^{-2} \int (-\tilde\omega^0 \wedge \theta) $, and the Hodge star is with respect to the 4D metric $ds^2_4$. The contribution from the 11D Einstein--Hilbert term to the 4D scalar potential, $V_{\rm geo}$, follows from the internal Ricci scalar $R_7$ as
\eq{
V_{\rm geo} = -\frac{1}{2}e^{2V}\mathcal K^{-1} R_7\,,
}
and is found to be
\eq{\label{Vgeo}\spl{
V_{\rm geo}
 = &  \;\, 8 \,e^K g_{ij} v^k q_k{}^i v^l q_l{}^j + \, \frac{1}{8} e^{-K + 4\phi} g_{ij} m^i m^j
- 2\, e^{K+ 2\phi} v^i v^j Q_i{}^T \sym \mathbb{M} Q_j  \\
&
 + 4\, e^{K + \KOm + 2\phi} (g^{ij}-4v^iv^j) \left(Q_i{}^T \sym Z\right) \left(Q_j{}^T \sym \bar{Z}\right) \\
& + 8i \, e^{K + \KOm} \, Z^T \mathbb{U}^T \sym \mathbb{U} \bar{Z}
+ 8\,e^{K + 2 \KOm}  (Z^T \sym \mathbb{U}\bar{Z})^2 + 4\, e^{K + \KOm + 2\phi} (m^i \mathcal{K}_i)(Z^T \sym \mathbb{U} \bar{Z}).
}}

\subsection{The form sector}\label{DetailsRedForm}

In this section we describe the reduction of the kinetic and Chern--Simons terms of the 11D supergravity three-form. This is done by plugging the expansions \eqref{threeformexp}--\eqref{G4flux} in the 11D action \eqref{11daction}. We also checked the result by reducing the 11D equations of motion and reconstructing the corresponding 4D action.

By evaluating the exterior derivative of the 11D three-form $A_3$, expanded as in \eqref{threeformexp}, we find that the field strength \eqref{ansatzG4}, \eqref{G4flux} is
{\setlength\arraycolsep{1pt}
\eq{\spl{ \label{G4KK}
G_4 = dA_3 + G_4^{\rm flux} \,&=\, H_4 + dB \wedge (\theta + A^0) + H_2^i \wedge \omega_i + Db^i \wedge \omega_ i \wedge (\theta + A^0) \\
&\,  + D\xi ^A \wedge \alpha_A -  D \tilde{\xi}_A \wedge \beta^A +  \chi_i \tilde{\omega}^i \\
&\, + \left[ (b^I Q_I + \mathbb{U}\xi)^A \alpha_A -  (b^I Q_I + \mathbb{U}\xi)_A \beta^A \right] \wedge (\theta + A^0) \,,
}}
}where we introduced the following combinations of 4D fields and charges:
\eq{\spl{
H_4 &= dC_3 + B \wedge dA^0 \,, \\[3pt]
H_2^i &=  - dA^i + A^0 \wedge A^j \, q_j{}^i+ m^i B + b^i dA^0 \,,\\[3pt]
\chi_i  \,& =\,  e_i +  {\cal K}_{ijk} m^j b^k + Q_i^T \sym \xi \,,
\label{4dBianchisint}
}}
together with $b^I \equiv \Re X^I = (1,b^i)$, and with the scalar derivatives given in \eqref{covDers}.
As already mentioned in the main text, we dualize the three-form $C_3$ to a constant $e_0$, following e.g.~\cite{louismicu1}; this leads to
\eq{\label{dualH4}
H_4 \,=\, \mathcal K^{-1} e^{4\phi} \left( b^I \mathcal E_I +\tfrac{1}{2} {\cal K}_{ijk} m^i b^j b^k  \right)*\!1 \,,
}
where we introduced the combination
\eq{
{\cal E}_I \,=\, e_I  + Q_I^T \sym \xi - \tfrac 12\delta_I^0\,  \xi^T \sym \mathbb U \xi\,.
}

In the end we obtain the 4D action
\be\label{SG4}
S_{G_4} = \frac{1}{\kappa_4^2} \int \left[\, {\cal L}_{\textrm{kin,$\,G_4$}} + {\cal L}_{\textrm{top}} - V_{G_4}*\!1\,\right],
\ee
where the kinetic plus instanton contributions are
\eq{\spl{ \label{lagkinG4}
{\cal L}_{\textrm{kin,$G_4$}} & =
-\tfrac{1}{4} \KK_{ijk} b^i H_2^j \wedge H_2^k
+\tfrac{1}{4}  \KK_{ijk} b^i b^j H_2^k \wedge dA^0
-\tfrac{1}{12}  \KK_{ijk} b^i b^j b^k dA^0 \wedge dA^0 \\[2pt]
+&\tfrac{1}{4} e^{-4\phi} dB \wedge *dB
- \KK g_{ij} H_2^i \wedge *H_2^j
+ g_{ij} Db^i \wedge *Db^j  -\tfrac{1}{4} e^{2\phi} (D\xi)^T \!\wedge *  (\sym \mathbb{M} D\xi),
}}
while the other topological terms read
\eq{\spl{
{\cal L}_{\textrm{top}}\; = &\,
-\tfrac{1}{4} dB \wedge \left[ \left(2e_I + Q_I^T \sym \xi \right) A^I - {\xi}^T \sym D\xi \right]\\[2pt]
&\,
+\tfrac{1}{4}\, q_{i}{}^l \KK_{jkl} A^i \wedge A^j \wedge (dA^k - m^k B )  -\tfrac{1}{4} e_i m^i B \wedge B \,,
}}
and the scalar potential is
\eq{\spl{ \label{formSPot}
V_{G_4} \,= &\;
 -\tfrac{1}{4}\,e^{2\phi} {\cal K}^{-1} (b^I Q_I^T + \xi^T\mathbb{U}^T) \sym \mathbb{M} (b^J Q_J + \mathbb{U}\xi) \\[2pt]
&\; +\tfrac{1}{4}\, e^{4\phi} \mathcal K^{-1} \left( b^I{\cal E}_I + \tfrac{1}{2} {\cal K}_{ijk} m^i b^j b^k \right)^2 + \tfrac{1}{16}\, e^{4\phi} {\cal K}^{-1} g^{ij} \chi_i \chi_j \,.
}}

We can now recast some of the above expressions in a form that is suitable for matching the $N=2$ supergravity formalism. In terms of the field strengths defined in \eqref{N=2fieldstrengths}, we have
\eq{
H_2^i \,=\,  -F^i +  b^i F^0 \,.
}
Using this, one can rewrite the gauge kinetic and topological terms in \eqref{lagkinG4} as
\be
-\tfrac{1}{4} \KK_{ijk} b^i H_2^j \wedge H_2^k
+\tfrac{1}{4}  \KK_{ijk} b^i b^j H_2^k \wedge dA^0
-\tfrac{1}{12}  \KK_{ijk} b^i b^j b^k dA^0 \wedge dA^0
= \tfrac{1}{4} {\rm Re}\,\mathcal N_{IJ} F^I \wedge F^J
\ee
and
\be
- \KK g_{ij} H_2^i \wedge *H_2^j \,=\, \tfrac{1}{4} {\rm Im}\,\mathcal N_{IJ} F^I \wedge *F^J + \tfrac{1}{4}\mathcal K F^0 \wedge *F^0\,,
\ee
where the gauge kinetic matrix $\mathcal N_{IJ}$ is precisely the special K\"ahler geometry period matrix \eqref{eq:ReNImN} obtained from the prepotential \eqref{cubicprepot}. Using the same matrix, and recalling \eqref{KahlerPotJ} as well as the last in \eqref{4dBianchisint}, we can also rewrite $V_{G_4}$ in \eqref{formSPot} as \eq{\spl{ \label{formSPotRewr}
V_{G_4} \,= &\;\; -2\,e^{K + 2\phi} (b^I Q_I^T + \xi^T\mathbb{U}^T) \sym \mathbb{M} (b^J Q_J + \mathbb{U}\xi) \\[2pt]
&\;\;  -\tfrac{1}{4} \, e^{4\phi}  \big({\cal E}_I -\Re{\cal N}_{IK} m^K \big)  (\Imag {\cal N})^{-1 IJ} \big({\cal E}_J - \Re{\cal N}_{JL} m^L \big) \,.
}}
Note that this is positive-definite, because both $\Im \mathcal{N}$ and $\sym \mathbb{M}$ are negative-definite.

After these reformulations are implemented, and the contribution \eqref{Sgeo} from the 11D Einstein--Hilbert term derived above is added, the 4D action takes exactly the form given in section \ref{sec:4Daction}.

%%%%%%%%%%%%%%%%%%%%%%%%%%%%%%%%%%%%%%%%%%%%%%%%%%%%%%%%%%%%%%%%%%%%%%%%%%
\section{Matching the $N=2$ scalar potential}\label{CheckScalarPot}

In this section, we show that the scalar potential determined by the $N=2$ gauging described in section \ref{sec:N=2sugra} matches the
expression \eqref{fullVfromDimRed} obtained from the dimensional reduction.

The general form of the scalar potential in 4D $N=2$ gauged supergravity for an electric-magnetic gauging of the type \eqref{generalscalarcovder} is \cite{reviewN2sugra,Michelson,N=2tensor1,N=2tensor2,samtlebenmag}:
\eq{\spl{\label{generalpotential}
V \,=\,
\,\,2\,\Big\{&e^K g_{k l}\, k_I^k\,  \bar k_J^{l}\bar X{}^I  X^J +
4\, e^K h_{uv} k^u \bar k^v \\[3pt]
\,& - \left[ \tfrac{1}{2} ({\rm Im}\,\mathcal N)^{-1 \,IJ} + 4 e^K X^I \bar{X}{}^J \right] ( \mathcal P_I^x - \mathcal N_{IK}\tilde{\mathcal P}^{xK}) ( \mathcal P_J^x - \overline{\mathcal N}_{JL}\tilde{\mathcal P}^{xL}) \Big\}\,, \quad
}}
where we used the symplectic invariant combination
$\,k^u = X^I k_I^u - \prepF_I\tilde{k}^{Iu}$, and the overall factor of 2 is due to the fact that our Killing vectors and Killing prepotentials are rescaled by a factor of $\sqrt 2$ with respect to the ones in e.g.\ \cite{reviewN2sugra}, see footnote \ref{rescalingN=2vecs}.

We now plug in the $N=2$ data found in section~\ref{sec:N=2sugra}. With no extra effort, we can perform the computation by allowing for an arbitrary flux charge $m^0$, so that $m^I = (m^0,m^i)$. This vanishes in 11D supergravity, while it corresponds to the Romans mass in type IIA.

We find that the contribution from the vector multiplet gauging is
\be
2\,e^K g_{k \bar l}\, k_I^k\,  k_J^{\bar l}\bar X{}^I  X^J\, =\, 8\,e^K g_{ij} v^k q_k{}^i v^l q_l{}^j,
\ee
while for the hypermultiplet gauging we get from the first line of \eqref{generalpotential}
\bea
8\, e^K h_{uv} k^u \bar k^v &=&  8\, e^K g_{a\bar b}k^a \bar k^b -\, 2\, e^{K+2\phi} \left( X^I Q_I^T + \xi^T\mathbb{U}^T \right) \sym \mathbb{M} \left( Q_J \bar X^J + \mathbb{U}\xi\right) \nonumber \\
&& +\, 2 \, e^{K+ 4\phi}\left| X^I e_I + X^I Q_I^T\sym \xi - \prepF_I m^I - \tfrac 12 \xi^T \sym \mathbb{U} \xi \right|^2,\eea
where $g_{a\bar b}k^a \bar k^b$ still needs to be evaluated. By expressing $g_{a\bar b}$ in terms of the derivatives of the K\"ahler potential
$\KOm = -\log [i( \bar Z^A \prepG_A - Z^A \bar \prepG_A)]$,
and identifying that $z^a = Z^a/Z^0$, one derives the following relation:
\eq{
g_{ab}\delta z^a \delta \bar z^{b} \,=\, e^{\KOm} \left( \Imag \prepG_{AB} + 2 e^{\KOm} \Imag \prepG_{AC} \bar Z^C \Imag \prepG_{BD} Z^D\right) \delta Z^A \delta \bar Z^B\,.
}
For a variation as in \eqref{Uisometry}, recalling the consistency condition \eqref{consistencyU}, we obtain then
\be
g_{a\bar b}k^a \bar k^b \,=\, i\, e^{\KOm} Z^T \mathbb{U}^T \sym \mathbb{U} \bar Z + \,e^{2\KOm} \left( \bar Z^T \sym \mathbb{U} Z \right)^2.
\ee
Finally, (2 times) the second line of \eqref{generalpotential} gives the expression
\eq{\spl{\label{IntermediateForV}
& 4\, e^{K+\KOm + 2\phi} ( g^{ij} - 4v^iv^j  ) (Z^T \sym Q_i) ( \bar Z^T \sym Q_j) \\[2pt]
&-\tfrac 14 e^{4\phi} ({\rm Im}\,\mathcal N)^{-1 \,IJ} \left(\mathcal E_I - \mathcal N_{IK}m^K\right) \left(\mathcal E_J - \overline{\mathcal N}_{JL}m^L\right)  \\[2pt]
&-2 \, e^{K+ 4\phi} \left| X^I e_I + X^I Q_I^T\sym \xi - \prepF_I m^I - \tfrac 12 \,\xi^T \sym \mathbb{U} \xi  \right|^2  \\[2pt]
& +\,4 \, e^{K + \KOm + 2\phi} (m^i\mathcal K_i -  m^0 b^i \mathcal K_i)  (Z^T \sym \mathbb{U} \bar Z)\,,
}}
with $\mathcal E_I$ as in \eqref{defE_I}, and
where in evaluating the $x =1,2$ contribution (yielding the first line in \eqref{IntermediateForV}) it was useful to notice that for cubic prepotentials one has
\be
\tfrac{1}{2} ({\rm Im}\,\mathcal N)^{-1 \,IJ} + 2\, e^K (X^I \bar{X}{}^J + \bar{X}^I X^J) \,=\,  -\delta^I_k\delta^J_l e^K (g^{kl} - 4 v^kv^l) .
\ee
Adding everything up, and setting $m^0=0$, we obtain precisely the expression for $V$ in \eqref{fullVfromDimRed}.

%%%%%%%%%%%%%%%%%%%%%%%%%%%%%%%%%%%%%%%%%%%%%%%%%%%%%%%%%%%%%%%%%%%%%%%%%%
\section{The cosets} \label{sec:cosets}

Here we would like to collect several details about the construction of the left-invariant forms of the compact homogeneous spaces that we consider in this paper. For each coset, we also construct the corresponding LI SU(3)-structure and its associated metric. Although some of this data have already appeared in the literature, we present here a self-contained summary.

Recall that,  a global, left (say)  action of the group $G$ is defined on a homogeneous manifold $G/H = \{gH: g\in G \}$. This generates motions on the manifold along which a frame is rotated.  Tensors that are invariant under this action are called {\it left-invariant} (LI) tensors. These are globally-defined, and have the property that their value at any point is determined by the value they take at the origin of the group.
Given a $D$-dimensional homogeneous space, one can then study which $G$-structures can be constructed using LI forms. This is determined by the embedding of $H$ in SO$(D)$. In this section, we are going to study 7D homogeneous spaces which admit left-invariant SU(3)-structures. Within our assumptions, these are the spaces collected in table \ref{cosetlist}.

We start by briefly reviewing the notion of left-invariance on coset manifolds $G/H$ (for more details see e.g.\ \cite{cosetreview1}). On a patch with coordinates $y$, take a coset representative $L(y)$. Let $\{ \mathcal{H}_\sigma\}$ be a basis of generators of the algebra of $H$, and $\{ \mathcal{K}_{\alpha}\}$ a basis of the complement of that algebra in the algebra of $G$.
The decomposition of the Lie-algebra valued one-form
\al{ \label{eq:MC}
L^{-1}\d L \,=\, e^\alpha\mathcal{K}_\alpha + \omega^\sigma\mathcal{H}_\sigma
\,,}
defines a local coframe $e^\alpha(y)$ on $G/H$. Furthermore, in terms of the structure constants $f$ of $G$ one has
\eq{\label{dcommut}
\d e^\alpha \,=\, -\frac{1}{2}f^\alpha{}_{\beta\gamma}e^\beta\wedge e^\gamma-f^\alpha{}_{\sigma\beta}\omega^\sigma\wedge e^\beta \, .
}
One can show that a $p$-form $\phi$ is left-invariant if and only if it can be written as
\eq{
\phi \,=\, \frac{1}{p!}\phi_{\alpha_1\dots \alpha_p}e^{\alpha_1}\wedge\dots \wedge e^{\alpha_p} \, ,
}
with the components $\phi_{\alpha_1\dots \alpha_p}$ being independent of $y$, and satisfying
\eq{\label{leftinv}
f^\beta{}_{\sigma[\alpha_1}\phi_{\alpha_2\dots \alpha_p]\beta}=0~.
}
The left-invariance property is preserved by the exterior derivative. One can also show that harmonic forms must be left-invariant and thus the cohomology of the coset manifold
is isomorphic to the cohomology of left-invariant forms.
The left-invariance condition for a metric on the coset is similar, namely it must be of the form
\be
\d s^2_{G/H} \,=\, g_{\alpha\beta}\, e^\alpha e^\beta\,,
\ee
with $g_{\alpha\beta}$ being independent of $y$, and satisfying
\eq{ \label{invmet}
f^\gamma{}_{\sigma(\alpha} \,g_{\beta)\gamma} = 0 \, .
}

We will now use this technology to determine the LI forms on each coset of table \ref{cosetlist}.

\subsection{$S^7 =$ SU(4)/SU(3)} \label{SU4modSU3}

We start by introducing a basis for the fifteen generators of SU(4) in the fundamental representation:
\begin{equation} \label{GenSU4}
T_{i}= -\frac{i}{2} \mu_i , \;  i=1,\ldots, 7, \quad
T_{7+a}= -\frac{i}{2} \left(\begin{array}{cc}
\lambda_a & 0   \\
0 & 0    \\
\end{array} \right) ,\quad a=1,\ldots, 8,
\end{equation}
where $\lambda_a$, $a=1,\ldots, 8$, are the Gell-Mann matrices, and
%
%{\setlength\arraycolsep{0pt}
\begin{eqnarray} \label{GMmat2}
&
\mu_1= \left(\begin{array}{cccc}
0 & 0 & 0 & 1 \\
0 & 0 & 0 & 0 \\
0 & 0 & 0 & 0 \\
1 & 0 & 0 & 0 \\
\end{array} \right), \;
\mu_2= \left(\begin{array}{cccc}
0 & 0 & 0 & -i \\
0 & 0 & 0 & 0 \\
0 & 0 & 0 & 0 \\
i & 0 & 0 & 0 \\
\end{array} \right), \;
\mu_3= \left(\begin{array}{cccc}
0 & 0 & 0 & 0 \\
0 & 0 & 0 & 1 \\
0 & 0 & 0 & 0 \\
0 & 1 & 0 & 0 \\
\end{array} \right), \;
\nonumber \\
&
\mu_4= \left(\begin{array}{cccc}
0 & 0 & 0 & 0 \\
0 & 0 & 0 & -i \\
0 & 0 & 0 & 0 \\
0 & i & 0 & 0 \\
\end{array} \right), \;
\mu_5= \left(\begin{array}{cccc}
0 & 0 & 0 & 0 \\
0 & 0 & 0 & 0 \\
0 & 0 & 0 & 1 \\
0 & 0 & 1 & 0 \\
\end{array} \right), \;
\mu_6= \left(\begin{array}{cccc}
0 & 0 & 0 & 0 \\
0 & 0 & 0 & 0 \\
0 & 0 & 0 & -i \\
0 & 0 & i & 0 \\
\end{array} \right), \;
\nonumber \\
&
\mu_7= \frac{1}{\sqrt{6}} \left(\begin{array}{cccc}
1 & 0 & 0 & 0 \\
0 & 1 & 0 & 0 \\
0 & 0 & 1 & 0 \\
0 & 0 & 0 & -3 \\
\end{array} \right). \;
\end{eqnarray}
The generators that exponentiate into the coset and the generators of
$H=SU(3)$ are taken to be respectively
\eq{
\label{CosetGenSU4}
\{{\cal K}_\alpha\} = \{T_1, \ldots, T_7 \} \; , \quad  \alpha =1, \ldots , 7 \, , \quad
\{ {\cal H}_\sigma \}= \{T_8, \ldots, T_{15} \} \; , \quad  \sigma =1, \ldots , 8.
}

We can now use this representation to find the structure constants of SU(4), and scan for all possible LI forms using eq.~(\ref{leftinv}).
We then use them to construct the following set of expansion forms, which introducing the vector $k$ dual to $\theta$ can be seen to satisfy
the algebraic restrictions \eqref{comp1}--\eqref{normforms}
\eq{\spl{ \label{invformsSU4}
& \theta \equiv -\sqrt{\tfrac{3}{8}} e^7 \\
&\omega_1 \equiv \tfrac{1}{4} \left( e^{12} + e^{34} + e^{56} \right) \; , \quad
\tilde{\omega}^1 \equiv \tfrac{1}{48} \left( e^{3456} + e^{1256} + e^{1234} \right)
\;, \quad
\tilde{\omega}^0 \equiv -\tfrac{1}{64} e^{123456} \;, \quad \\
& \alpha_0 \equiv \tfrac{1}{16} \left( e^{135} -  e^{146} - e^{236} -  e^{245} \right) , \quad
 \beta^0 \equiv - \tfrac{1}{16} \left( e^{246} -  e^{136} - e^{145} -  e^{235} \right)  \!\!\! \!\!\!
}}
Furthermore they satisfy eq.~(\ref{omomK}) with
\eq{
{\cal K}_{111} =6 \, .
}
Further, using (\ref{dcommut}), these forms can be further checked to close into the following exterior algebra:
\begin{eqnarray} \label{extalgSU4}
d \theta =  2 \omega , \quad
d\omega_1 =
d\tilde{\omega}^1 =
d\tilde{\omega}^0 =  0 , \quad
d\alpha_0 = -4 \theta \wedge \beta^0 , \quad
d\beta^0 = 4 \theta \wedge \alpha_0 \, ,
\end{eqnarray}
from which we immediately read off the geometric fluxes of tables \ref{GeomFluxTable} and \ref{GeomFluxCntdTable}.

We can now use these forms to expand the SU(3)-structure on SU(4)/SU(3) as
\begin{eqnarray} \label{strSU4}
\eta =  e^V  \theta \; , \qquad J = e^{2U_1}  \omega \; , \qquad \Omega = 2 \, e^{3U_1} (\alpha_0 + i \beta^0) \; ,
\end{eqnarray}
where $U_1$, $V$ are two arbitrary constants, that become scalars in the effective theory. In order to make contact with the main text  notation, these may be alternatively reparametrized in terms of two other scalars $u_1$, $\phi$ as
\begin{eqnarray} \label{ChangeSusySU4}
u_1 = U_1 + \tfrac{1}{2} V \; , \quad \phi = -3U_1 \; .
\end{eqnarray}
The scalar $\phi$ corresponds to the 4D dilaton, eq.~\eqref{4Ddilaton}, while, in the notation of eq.~(\ref{LISU3str}), $v^1 \equiv  e^{2u_1}$  lies in a vector multiplet.  Note that, in this and all models with $n_H=1$, only the overall scale of $\Omega$ can change, which leads to a trivial moduli space of almost complex structures.  Moreover, we fix the scale of $\Omega$ through the normalization
below \eqref{SU3stralg}.
Finally, the torsion classes of the SU(3)-structure (\ref{strSU4}) can be calculated. In the notation of (\ref{SU3tcl}), the only active classes in this model are the scalars $R$ and $E$. We omit the details here: see eq.~(\ref{torQpqr}) for $Q^{111}$ from where these can be easily deduced.

Finally, the LI metric on SU(4)/SU(3) associated to the SU(3)-structure \eqref{strSU4} reads
\begin{equation} \label{metSU4}
\d s_7^2 =\tfrac{1}{4} e^{2U_1} \left[ (e^1)^2+(e^2)^2+(e^3)^2+(e^4)^2+(e^5)^2+(e^6)^2 \right] +  e^{2V} \theta^2 .
\end{equation}
For $U_1 = V = 0$ this metric is SE$_7$, canonically normalized such that $R_{mn} = 6 g_{mn}$. We omit the details here, and refer to section \ref{Q111} for a similar derivation.

Finally, as already noted in \cite{caskoerbertrisak}, the truncation associated to SU(4)/SU(3) is identical to the universal SE$_7$ truncation of \cite{vargaun2}. The scalars $U_1$, $V$ are related to the breathing and squashing modes of \cite{maldtach,vargaun2}. Here they are clearly seen to independently rescale the fiber and the base $B_6 =\mathbb{CP}^3$ of $S^7 =\;$SU(4)/SU(3). The truncation does not depend on the details of $\mathbb{CP}^3$, which can be thus be replaced by any K\"ahler--Einstein six-fold, leading to the universal SE$_7$ truncation.

\subsection{$M^{110}$} \label{M110}

$M^{pq0}$ is the coset $G/H$ with $G= {\rm SU}(3)\times {\rm SU}(2)$ and $H= {\rm SU}(2) \times {\rm U}(1)$, with SU(2) embedded into SU(3) as the isospin subgroup, and $p, q$ determining the embedding of the Cartan subgroup U(1)$^2$ of $H$ into $G$. We can make this more precise by introducing the eleven generators of $G$ in terms of Gell-Mann and Pauli matrices in the following block diagonal form, where the first block in the diagonal is 3-by-3 and the second 2-by-2:
\begin{equation} \label{GenSU3SU2}
T_a= -\frac{i}{2} \left(\begin{array}{cc}
\lambda_a & 0   \\
0 & 0    \\
\end{array} \right) , \quad
T_{8+i}= -\frac{i}{2} \left(\begin{array}{cc}
0 & 0   \\
0 & \sigma_i  \\
\end{array} \right) . \quad
\end{equation}
We take the seven generators that exponentiate into the coset $M^{pq0}$ as
\begin{eqnarray} \label{CosetGenMpq0}
\{{\cal K}_\alpha\} = \{ T_4, \; T_5, \; T_6, \; T_7, \; T_9, \; T_{10}, \; Z_1 \equiv \tfrac{1}{\sqrt{3p^2+q^2}} \left( p\sqrt{3} T_8 + q T_{11} \right) , \; \alpha =1, \ldots , 7,
\end{eqnarray}
while those defining $H=SU(2) \times U(1)$ as
\begin{eqnarray}
\{{\cal H}_\sigma\}= \{ T_1, \; T_2, \; T_3,  \; Z_2 \equiv \tfrac{1}{\sqrt{3p^2+q^2}} \left( -q T_8 + p\sqrt{3} T_{11} \right) \}\, , \quad  \sigma =1, \ldots , 4.
\end{eqnarray}

We can now determine the LI forms supported by this coset, proceeding as in the SU(4)/SU(3) case explained previously. For generic $p,q$, these are
\al{ \label{invformsM110a}
& \theta \equiv -\tfrac{3p+q}{4\sqrt{3p^2+q^2}} e^7 \; , \qquad
\omega_1 \equiv \tfrac{3p(3p+q)}{16(3p^2+q^2)} \left( e^{12} + e^{34} \right) , \qquad  \omega_2 \equiv \tfrac{q(3p+q)}{8(3p^2+q^2)}  e^{56} \nonumber \\[8pt]
& \tilde{\omega}^1 \equiv \tfrac{3pq(3p+q)^2}{2^8(3p^2+q^2)^2} \left( e^{1256} + e^{3456} \right) \, , \tilde{\omega}^2 \equiv \tfrac{9p^2(3p+q)^2}{2^8(3p^2+q^2)^2}  e^{1234} \, ,  \tilde{\omega}^0 \equiv -\tfrac{9p^2q(3p+q)^3}{2^{11} (3p^2+q^2)^3}  e^{123456} ,\quad
}
where the numerical coefficients have been chosen for convenience. Further, for $p=q$, there are
two LI three-forms $\alpha_0$, $\beta^0$:
\begin{eqnarray} \label{invformsM110b}
\alpha_0 \equiv \tfrac{3\sqrt{2}}{128} \left( e^{135} -  e^{146} - e^{236} -  e^{245} \right) \; , \quad
\beta^0 \equiv - \tfrac{3\sqrt{2}}{128} \left( e^{246} -  e^{136} - e^{145} -  e^{235} \right) .
\end{eqnarray}
We see therefore that for $p \neq q$ the space $M^{pq0}$ does not support LI three-forms and thus no LI SU(3)-structure.  We thus set $p=q$ and, within this class, focus without loss of generality on $M^{110}$, which does admit a LI SU(3)-structure. We find then the relation (\ref{omomK}) with the only non-vanishing coefficient (up to symmetric permutations of the indices)
\eq{
{\cal K}_{112} = 2 \, ,
}
and the following exterior algebra:
\eq{\spl{ \label{extalgMpqr}
& d \theta =  2 \left( \omega_1 + \omega_2 \right) , \quad
d\omega_1 = d\omega_2 = 0 , \quad
d\tilde{\omega}^1 = d\tilde{\omega}^2 = 0 , \quad
d\tilde{\omega}^0 = 0  ,  \\
& d\alpha_0 = -4 \theta \wedge \beta^0 , \quad
d\beta^0 = 4 \theta \wedge \alpha_0 ,
}}
leading to the corresponding entries in tables \ref{GeomFluxTable} and \ref{GeomFluxCntdTable}.

For the LI SU(3)-structure, we take the following expansion
\begin{eqnarray} \label{strMpp0}
\eta =  e^V  \theta \; , \qquad J = e^{2U_1}  \omega_1 +e^{2U_2}  \omega_2  \; , \qquad \Omega = 2 \, e^{2U_1+U_2} (\alpha_0 + i \beta^0) \; ,
\end{eqnarray}
where $U_1$, $U_2$, $V$ become scalar fields in the effective theory. Another useful parametrization is in terms of the scalars that enter definite supermultiplets, $u_1$, $u_2$, $\phi$,
\begin{eqnarray} \label{ChangeSusyM110}
u_1 = U_1 + \tfrac{1}{2} V \; , \quad
u_2 = U_2 + \tfrac{1}{2} V \; , \quad
\phi = -2U_1 - U_2 \; ,
\end{eqnarray}
so that $v^i \equiv e^{2u_i}$, $i=1,2=n_V$, and $\phi$ is the 4D dilaton. At a generic point $(U_1, U_2, V)$ in LI moduli space, the only active classes of the SU(3)-structure (\ref{strMpp0}) are $R$, $E$ and $T_1$. We omit the details here, and refer again to $Q^{111}$, which we discuss in the next section. This SU(3)-structure leads to the metric

\begin{equation} \label{metMpp0}
ds_7^2 =\tfrac{3}{16}  e^{2U_1} \left[ (e^1)^2+(e^2)^2+(e^3)^2+(e^4)^2 \right] + \tfrac{1}{8}  e^{2U_2} \left[(e^5)^2+(e^6)^2 \right] +  e^{2V} \theta^2.
\end{equation}
It can be verified that $U_1=U_2=V=0$ is the canonically normalized SE$_7$ point: see next section for details on the similar $Q^{111}$ case.

Finally, observe that the effect of the three scalars $U_1, U_2, V$ in the metric is to independently rescale the fiber and the factors $\mathbb{CP}^2  \times \mathbb{CP}^1$ in the base. By similar arguments as in the previous subsection, $\mathbb{CP}^2$ can be replaced by any compact, 4D K\"ahler--Einstein manifold, leading to exactly the same truncation as $M^{110}$. The only other regular, compact K\"ahler--Einstein four-folds are del Pezzo surfaces.

\subsection{$Q^{111}$}  \label{Q111}

$Q^{pqr}$ is the coset $G/H$ with $G= {\rm SU}(2)^3$ and $H={\rm U}(1)^2$, and $p,q,r$ describing the embedding of $H$ into $G$. Using Pauli matrices $\sigma^i$, $i=1,2,3$, we write the nine generators of $G= {\rm SU}(2)^3$ in the fundamental in the following block diagonal form:
\begin{equation} \label{SU2CubedGen}
T_i= -\frac{i}{2} \left(\begin{array}{ccc}
\sigma^i & 0 & 0  \\
0 & 0 & 0   \\
0 & 0 & 0   \\
\end{array} \right) , \quad
T_{3+i}= -\frac{i}{2} \left(\begin{array}{ccc}
0 & 0 & 0  \\
0 & \sigma^i  & 0   \\
0 & 0 & 0   \\
\end{array} \right) , \quad
T_{6+i}= -\frac{i}{2} \left(\begin{array}{ccc}
0 & 0 & 0  \\
0 & 0 & 0   \\
0 & 0 & \sigma^i    \\
\end{array} \right) . \quad
\end{equation}
The seven generators that exponentiate into the coset are
\eq{ \label{GenQpqr}
\{{\cal K}_\alpha\} = \{ T_1, \; T_2, \; T_4, \; T_5, \; T_7, \; T_8, \; Z_1 \equiv \tfrac{1}{\sqrt{p^2 + q^2 + r^2}} \left( p T_3 + q T_6 +r T_9 \right)  \}  , \quad \alpha =1, \ldots , 7,
}%
while the generators of $H={\rm U}(1)^2$ are taken to be
\eq{\spl{ \label{GenHQpqr}
 \{{\cal H}_\sigma\} = \{ &
Z_2 \equiv \tfrac{1}{\sqrt{p^2 + q^2}} \left( q T_3 -p T_6 \right) , \\
&  Z_3 \equiv \tfrac{1}{\sqrt{p^2 + q^2 + r^2} \sqrt{p^2 + q^2}} \left( pr T_3 + qr T_6 -(p^2 + q^2) T_9 \right) \} \; , \qquad   \sigma =1, 2.
}}

For generic $p,q,r$, $Q^{pqr}$ has the following LI forms:
\eq{\spl{ \label{invformsQpqra}
& \theta \equiv -\tfrac{\sqrt{3}}{4} e^7 \; , \qquad
\tilde{\omega}^0   \equiv - \tfrac{3\sqrt{3}}{2^9} \tfrac{pqr}{(p^2+q^2+r^2)^{3/2}} e^{123456}   \\[8pt]
&
\omega_1 \equiv  \tfrac{\sqrt{3}}{8} \tfrac{p}{\sqrt{p^2+q^2+r^2}} e^{12} \; , \quad
\omega_2 \equiv  \tfrac{\sqrt{3}}{8} \tfrac{q}{\sqrt{p^2+q^2+r^2}} e^{34} \; , \quad
\omega_3 \equiv  \tfrac{\sqrt{3}}{8} \tfrac{r}{\sqrt{p^2+q^2+r^2}} e^{56}   \\[8pt]
&
\tilde{\omega}^1 \equiv  \tfrac{3}{64} \tfrac{qr}{p^2+q^2+r^2} e^{3456} \; , \quad
\tilde{\omega}^2 \equiv  \tfrac{3}{64} \tfrac{pr}{p^2+q^2+r^2} e^{1256} \; , \quad
\tilde{\omega}^3 \equiv  \tfrac{3}{64} \tfrac{pq}{p^2+q^2+r^2} e^{1234}   \\
}}
where the numerical coefficients have been chosen for convenience. Further, for $p=q=r$, there are
two LI three-forms $\alpha_0$, $\beta^0$:
\begin{eqnarray} \label{invformsQpqrb}
\alpha_0 \equiv \tfrac{\sqrt{2}}{64} \left( e^{135} -  e^{146} - e^{236} -  e^{245} \right) , \quad
\beta^0 \equiv - \tfrac{\sqrt{2}}{64} \left( e^{246} -  e^{136} - e^{145} -  e^{235} \right).
\end{eqnarray}
Just like $M^{pq0}$, the space $Q^{pqr}$ can only have a LI SU(3)-structure if $p=q=r$, which we thus fix henceforth. The invariant forms (\ref{invformsQpqra}), (\ref{invformsQpqrb})
 can be seen to satisfy the algebraic constraints  (\ref{comp1})--(\ref{normforms}), and the relation (\ref{omomK}) with
\eq{
{\cal K}_{123} =1
}
as the only non-vanishing coefficient (up to symmetric permutations of the indices). Furthermore we find the following exterior algebra:
\eq{\spl{ \label{extalgQpqr}
& d\theta =  2 \left( \omega_1 + \omega_2 + \omega_3 \right) , \quad
d\omega_1 = d\omega_2 = d \omega_3= 0 , \quad
d\tilde{\omega}^1 = d\tilde{\omega}^2 = d\tilde{\omega}^3 = 0 ,  \quad
  \\
& d\tilde{\omega}^0 = 0 , \quad
 d\alpha_0 = -4 \theta \wedge \beta^0 , \quad
d\beta^0 = 4 \theta \wedge \alpha_0 ,
}}
from which the geometric fluxes are read off.

The coset $Q^{111}$ supports the following LI SU(3)-structure:
\begin{eqnarray} \label{strQpqr}
\eta =  e^V  \theta \; , \; J = e^{2U_1}  \omega_1 + e^{2U_2}  \omega_2 + e^{2U_3}  \omega_3  \; , \; \Omega = 2e^{U_1+U_2+U_3} (\alpha_0 + i \beta^0) \; ,
\end{eqnarray}
where the $U_1$, $U_2$, $U_3$, $V$ become scalar fields in the 4D theory. The alternative parametrization
\begin{eqnarray} \label{ChangeSusyQ111}
u_1 = U_1 + \tfrac{1}{2} V \; , \quad
u_2 = U_2 + \tfrac{1}{2} V \; , \quad
u_3 = U_3 + \tfrac{1}{2} V \; , \quad
\phi = -U_1 - U_2- U_3 \; .\end{eqnarray}
is useful to make contact with supermultiplets: here we have  $v^i \equiv e^{2u_i}$, $i=1,2,3=n_V$, and $\phi$ the effective 4D dilaton. For this coset we will provide some explicit details on the intrinsic torsion. The only non-vanishing torsion classes of the SU(3)-structure (\ref{strQpqr}) are
\eq{\spl{ \label{torQpqr}
& R =  \frac{2}{3} \, e^{V} \big( e^{-2U_1} + e^{-2U_2} + e^{-2U_3} \big) \; , \qquad  E = 4ie^{-V}  \; ,   \\[8pt]
&T_1 =
\frac{2}{3} \, e^{V}  \Big[ \left( 2- e^{2U_1-2U_2} - e^{2U_1-2U_3} \right) \omega_1
%  \\
%&+&
%
+\left( 2 - e^{-2U_1+2U_2}  - e^{2U_2-2U_3} \right) \omega_2   \\
& \qquad  \qquad \; +  \left(2 - e^{-2U_1 +2U_3} - e^{-2U_2+2U_3}  \right) \omega_3 \Big] \; ,
}}
in agreement with (\ref{TorsionParam}).

The LI metric on $Q^{111}$ corresponding to the above LI SU(3)-structure is
\be \label{metQpqr}
ds_7^2 =
\tfrac{1}{8}  e^{2U_1} \left[ (e^1)^2+(e^2)^2 \right]
+ \tfrac{1}{8}  e^{2U_2} \left[ (e^3)^2+(e^4)^2 \right]
+ \tfrac{1}{8}  e^{2U_3} \left[ (e^5)^2+(e^6)^2 \right] +  e^{2V}\theta^2 \, .
\ee
The effect of the scalars in the metric  (\ref{metQpqr}) is to independently rescale the fiber and the three $\mathbb{CP}^1$ factors in the base. Also, we will present the Ricci tensor corresponding to the metric (\ref{metQpqr}). Although we have not specified the general expression of the Ricci tensor in terms of the torsion classes in (\ref{SU3tcl}), we can nevertheless compute it by means of  coset technology. We find the following non-vanishing components for the Ricci tensor,
\eq{\spl{ \label{RicQpqr}
& R_{11} = R_{22} = 1-\tfrac{1}{4} e^{-2U_1} , \quad
R_{33} = R_{44} = 1-\tfrac{1}{4} e^{-2U_2} , \quad
 R_{55} = R_{66} = 1-\tfrac{1}{4} e^{-2U_3} \; , \\
& R_{77} = \tfrac{3}{8} e^{4V}\left( e^{-2U_1} + e^{-2U_2}+ e^{-2U_3} \right) \,.
}}
Note that, at the point $U_1=U_2=U_3=V=0$, the metric becomes Einstein with canonical normalization, $R_{mn} = 6 g_{mn}$, and the torsion classes (\ref{torQpqr}) become SE$_7$, as given in (\ref{SE7stru}).

Finally, note that the metric, the Ricci tensor and the torsion classes for $M^{110}$ and SU(4)/SU(3) can be recovered from the $Q^{111}$ ones by setting $U_3=U_1$, and $U_3=U_1=U_2$, respectively.

\subsection{$V_{5,2}$} \label{V52}

The Stiefel manifold $V_{5,2}$ is the coset $G/H$ with $G={\rm SO}(5)$ and $H={\rm SO}(3)$ canonically embedded into ${\rm SO}(5)$, i.e.\ the embedding is such that $\mathbf{5} \rightarrow \mathbf{4} + \mathbf{1} \rightarrow \mathbf{3} + \mathbf{1}+ \mathbf{1}$ under ${\rm SO}(5) \supset {\rm SO}(4) \supset {\rm SO}(3)$. The generators $M_{ij}=-M_{ji}$, $i,j=1,\ldots,5$, in the fundamental of SO(5) can be
taken to be $(M_{ij})^m{}_n \equiv 2 \delta_{[i}^{m} \delta_{j]n}$, with
\begin{eqnarray} \label{GenV52}
\{ {\cal K}_\alpha \} = \{ M_{14}, \; M_{15}, \; M_{24}, \; M_{25}, \; M_{34}, \; M_{35}, \; M_{45}  \} \; , \quad  \alpha =1, \ldots , 7,
\end{eqnarray}
exponentiating into the coset, while
\begin{eqnarray} \label{SO5HGen}
\{ {\cal H}_\sigma \}= \{ M_{12}, \; M_{13}, \; M_{23} \} \; , \quad  \sigma =1, 2,3,
\end{eqnarray}
define $H=\;$SO(3).

We find the following LI forms for $V_{5,2}$:
\eq{\spl{ \label{invformsV52}
& \theta \equiv \tfrac{3}{4} e^7 \; , \; \\
& \omega_1 \equiv  \tfrac{3}{8} \left( e^{12} + e^{34} +  e^{56} \right)  , \quad
\tilde{\omega}^1 \equiv  \tfrac{3}{64} \left( e^{1234} + e^{1256} +  e^{3456} \right), \quad
\tilde{\omega}^0 \equiv  -\tfrac{27}{2^9} e^{123456}   \\[8pt]
& \alpha_0 \equiv \tfrac{3\sqrt{3}}{16\sqrt{2}} e^{135} , \qquad
\beta^0 \equiv -\tfrac{3\sqrt{3}}{16\sqrt{2}}   e^{246} , \\[8pt]
& \alpha_1 \equiv \tfrac{3}{16\sqrt{2}} \left( e^{136} + e^{145} +  e^{235} \right)  , \qquad
\beta^1 \equiv \tfrac{3}{16\sqrt{2}} \left(  e^{146} + e^{236} +  e^{245} \right).
}}
They satisfy the algebraic constraints  (\ref{comp1})--(\ref{normforms}), and the relation (\ref{omomK}) with
\eq{
{\cal K}_{111} =6.
}
Furthermore, they close into the following exterior algebra
\eq{\spl{ \label{extalgV52}
& \d \theta =  2 \omega , \qquad
\d\omega_1  = 0 , \qquad
\d\tilde{\omega}^1 = 0 , \qquad
\d\tilde{\omega}^0 = 0 ,  \\
& d\alpha_0 = -\tfrac{4}{\sqrt{3}} \theta \wedge \alpha_1 , \qquad
d\beta^0 = -\tfrac{4}{\sqrt{3}} \theta \wedge \beta^1 , \\
& d\alpha_1 = \tfrac{4}{\sqrt{3}} \theta \wedge \alpha_0 - \tfrac{8}{3} \theta \wedge \beta^1 , \qquad
d\beta^1 = \tfrac{8}{3} \theta \wedge \alpha_1 + \tfrac{4}{\sqrt{3}} \theta \wedge \beta^0  \, ,
}}
from which we can read off the geometric fluxes presented in the tables \ref{GeomFluxTable} and \ref{GeomFluxCntdTable}.

For this model $n_H=2$ so that the moduli space of almost complex deformations is non-trivial. Using the algorithm
of appendix \ref{howtoprepot} one straightforwardly finds the prepotential associated to the basis of $(\alpha,\beta)$-forms
of~\eqref{extalgV52}:
\eq{ \label{CSprepotV52}
\mathcal{G} = -\frac{1}{3\sqrt{3}} \frac{(Z^1)^3}{Z^0} \, .
}
It is well-known that the c-map for this prepotential leads to the quaternionic--K\"ahler manifold $\frac{G_{2(2)}}{SO(4)}$.

We can now parameterize the LI SU(3)-structure as follows:
\eq{ \label{strV52}
\eta =  e^V  \theta \; , \quad J =  e^{2U_1} \omega_1 , \quad
\Omega = e^{3U_1+3\varphi} \left(\alpha_0 + \sqrt{3} z \alpha_1 + \sqrt{3} z^2 \beta^1 - z^3 \beta^0 \right) \; ,
}
where we introduced the 4D scalar fields $V, U_1$ and
\eq{ \label{eq:zV52}
z = \frac{Z^1}{\sqrt{3} Z^0} =  \chi + i e^{-2\varphi} \, .
}
As an alternative to the first two fields it is sometimes useful to introduce
\eq{ \label{ChangeSusyV52}
u_1 = U_1 + \tfrac{1}{2} V \; , \quad
\phi = -3U_1 \; ,
}
where $v^1 \equiv e^{2u_1}$ and $\phi$ is the 4D dilaton. The LI metric associated to this SU(3)-structure
takes the following block-form in the $e^\alpha$-basis:
\eq{
g = \frac{3}{8} \, e^{2 U_1} \left( \begin{array}{cccc} L^T L & 0 & 0 & 0\\ 0 & L^T L & 0 & 0\\ 0 & 0 & L^T L & 0 \\ 0&0&0&\tfrac{3}{2}e^V
\end{array} \right) , \qquad \text{with} \quad L = \left(\begin{array}{cc} e^\varphi & 0\\\chi & e^{-\varphi} \end{array}\right) \, .
}

The non-vanishing torsion classes of the SU(3)-structure (\ref{strV52}), in terms of the parameters $(U_1, z =\chi + i e^{-2\varphi} , V)$ and the LI forms (\ref{invformsV52}) are: 
\eq{\spl{\label{torV52}
R  = & \,2 \, e^{-2U_1+V} \; ,  \qquad E =   \,-4 e^{-V} \frac{(1+|z|^2)}{(z-\bar{z})} \; , \\
S = & \, e^{3U_1+3\varphi-V} \big[ \left(4z -e^V E \right) \alpha_0
+\tfrac{1}{\sqrt{3}} \left(4(2z^2-1) -3ze^V E \right) \alpha_1  \\
& \qquad + z^2 \left(4 + ze^V E \right) \beta^0
+\tfrac{1}{\sqrt{3}} z \left( 4\, z^2-8 -3 z e^V E \right) \beta^1 \big] \, .
}}
For $U_1=\varphi=\chi=V=0$, these become SE$_7$ torsion classes, as given in (\ref{SE7stru}), and the metric accordingly becomes Einstein, with canonical normalization $R_{mn}=6 \, g_{mn}$.

Finally, we remark that we find the Sasaki--Einstein truncation of section \ref{SU4modSU3} as a subtruncation by keeping only
the following linear combinations of the three-forms:
\eq{
(\alpha'_0, \beta^{'0}) = (\alpha_0 -\sqrt{3} \beta^1, \beta^0 + \sqrt{3} \alpha_1) \, ,
}
which amounts to putting $z=i$ or $\chi=\varphi=0$.

\subsection{$S^7 =\;$Sp(2)/Sp(1)} \label{Sp2modSp1}

The seven-sphere considered as the coset space Sp(2)/Sp(1) actually supports a tri-Sasakian structure. The reduced theory is equivalent to the 4D, $N=4$ supergravity obtained from the universal truncation on any tri-Sasaki space \cite{caskoerbertrisak, triS}. In line with our present $N=2$ interest, here we will determine the largest $N=2$ truncation of the universal $N=4$ theory. It will turn out that
we can find a set of expansion forms obeying the constraints of section \ref{sec:SU3str} by imposing invariance under a certain
discrete symmetry group on the expansion forms of the $N=4$ reduction. Just like the tri-Sasakian reduction this $N=2$ reduction will
be valid for a generic tri-Sasakian manifold.

As in section \ref{V52}, we take the generators $M_{ij}=-M_{ji}$, $i,j=1,\ldots,5$ of $\mathfrak{sp}(2) \approx \mathfrak{so}(5)$ to be $(M_{ij})^m{}_n \equiv 2 \delta_{[i}^{m} \delta_{j]n}$. Introducing the following $\mathfrak{sp}(1) + \mathfrak{sp}(1)$ generators $J_1,J_2,J_3$, $L_1,L_2,L_3$,
\begin{eqnarray}
J_1 = \tfrac{1}{2} ( M_{23} +M_{14} ) , \quad
J_2 = \tfrac{1}{2} ( -M_{13}+M_{24} ), \quad
J_3 =  \tfrac{1}{2} ( M_{12} +M_{34} ), \nonumber \\
L_1 = \tfrac{1}{2} ( M_{23} -M_{14} ), \quad
L_2 = \tfrac{1}{2} ( -M_{13} -M_{24} ), \quad
L_3 = \tfrac{1}{2} ( M_{12} -M_{34} ) ,
\end{eqnarray}
the generators that define the coset are then
\begin{eqnarray} \label{GenSp2}
\{{\cal K}_\alpha \}= \{ M_{15}, \; M_{25}, \; M_{35}, \; M_{45}, \; J_1, \; J_2, \;   J_3 \} \; , \quad  \alpha =1, \ldots , 7,
\end{eqnarray}
while
\begin{eqnarray} \label{Sp1Gen}
\{ {\cal H}_\sigma \}= \{  L_1, \; L_2, \;  L_3 \} \; , \quad  \sigma =1, 2, 3.
\end{eqnarray}
define $H=\;$Sp(1).

The LI forms are
\eq{\spl{
& \{ \eta^I \} = \{ \tfrac{1}{\sqrt{2}} e^6, \;\tfrac{1}{\sqrt{2}} e^5,\; \tfrac{1}{\sqrt{2}} e^7 \} \, , \\
& \{ J^I \} = \{ \tfrac{1}{4} (-e^{13}+e^{24}),\; \tfrac{1}{4} (e^{14}+e^{23}),\; \tfrac{1}{4} (e^{12}+e^{34}) \}
\, , \qquad I=1,2,3\, ,
}}
and wedge product thereof. These form a LI tri-Sasakian structure, and an expansion in these forms leads
to the $N=4$ theory of \cite{caskoerbertrisak}. We will now construct the largest subset of forms, which obeys
the constraints of section \ref{sec:SU3str}, and thus leads to an $N=2$ theory. We start by choosing a direction
and singling out $\theta = \eta^3$ (choosing any other direction leads to an equivalent theory). It turns
out that we can find the rest of the expansion forms by imposing a discrete $\mathbb{Z}_2$-symmetry, which
reads in the $e^{\alpha}$-basis
\eq{
e^1 \rightarrow e^1, \quad e^2 \rightarrow e^2, \quad e^3 \rightarrow -e^3, \quad  e^4 \rightarrow -e^4,
\quad e^5 \rightarrow -e^5, \quad e^6 \rightarrow -e^6, \quad e^7 \rightarrow e^7.
}
It follows that the subtruncation based on the surviving expansion forms is consistent.
In fact, this discrete symmetry can be expressed entirely in terms of the tri-Sasakian structure as
follows
\eq{
\eta^1 \rightarrow -\eta^1, \quad \eta^2 \rightarrow -\eta^2, \quad \eta^3 \rightarrow \eta^3, \quad
J^1 \rightarrow -J^1,\quad  J^2 \rightarrow -J^2, \quad J^3 \rightarrow J^3,
}
which implies it can be imposed for a generic tri-Sasakian manifold.

In the end we take the following set of expansion forms
\eq{\spl{ \label{invformsSp2}
& \theta \equiv \tfrac{1}{\sqrt{2}} e^7 \; , \\
& \omega_1 \equiv  \tfrac{1}{4} \left( e^{12} + e^{34} \right) \; , \;
\omega_2 \equiv  \tfrac{1}{2}  e^{56}  \; , \;
\tilde{\omega}^1 \equiv  \tfrac{1}{16} \left( e^{1256} + e^{3456} \right) \; , \;
\tilde{\omega}^2 \equiv  \tfrac{1}{16}  e^{1234}   \; , \;
\tilde{\omega}^0 = -\tfrac{1}{32} e^{123456} \\[7pt]
& \alpha_0 \equiv \tfrac{1}{8\sqrt{2}} \left(e^{135} - e^{245} \right) - \tfrac{1}{8\sqrt{2}} \left(e^{146} + e^{236} \right) , \;
\beta^0 \equiv   \tfrac{1}{8\sqrt{2}} \left(e^{136} - e^{246} \right) + \tfrac{1}{8\sqrt{2}} \left(e^{145} + e^{235} \right),   \\[7pt]
& \alpha_1  \equiv  \tfrac{1}{8\sqrt{2}} \left(e^{135} - e^{245} \right) + \tfrac{1}{8\sqrt{2}} \left(e^{146} + e^{236} \right), \;
\beta^1  \equiv \tfrac{1}{8\sqrt{2}} \left(e^{136} - e^{246} \right) - \tfrac{1}{8\sqrt{2}} \left(e^{145} + e^{235} \right),
}}
which obey the algebraic constraints  (\ref{comp1})--(\ref{normforms}) (with $n_H=n_V=2$) and (\ref{omomK}) with the only non-vanishing coefficient (up
to permutations of the indices)
\eq{
{\cal K}_{112} =2 \, .
}
Furthermore, they close into the differential algebra
\eq{\spl{ \label{extalgSp2}
& d\theta =  2 ( \omega_1 + \omega_2 ) , \qquad
\d\omega_1  = -4 \alpha_1  , \quad
\d\omega_2  = 4 \alpha_1, \quad
\d\tilde{\omega}^1 =  d\tilde{\omega}^2 = 0 \; , \qquad d\tilde{\omega}^0 = 0 , \\
& \d\alpha_0 = -4\theta \wedge \beta^0 \ , \qquad
 \d\beta^0 =  4 \theta \wedge \alpha_0  ,  \qquad
 \d\alpha_1 = 0  , \qquad
\d\beta^1 = 4 (\tilde{\omega}^1 - \tilde{\omega}^2) \, ,
}}
from which as usual we read off the geometric fluxes.

Since $n_H=2$ the moduli space of almost complex structure is non-trivial. Using the algorithm of appendix \ref{howtoprepot}
we find the prepotential
\begin{eqnarray} \label{eq:prepot}
\mathcal{G} =-\frac{i}{2} \left( (Z^0)^2 - (Z^1)^2 \right) \, .
\end{eqnarray}
According to the results of \cite{FS}, the resulting special quaternionic K\"ahler manifold is SO(4,2)/(SO(4)$\,\times\,$SO(2)).

We can now parameterize the LI SU(3)-structure as follows:
\eq{\spl{ \label{strSp2}
& \eta =  e^{V}  \theta \; , \qquad
J =  e^{2U_1} \omega_1 +\epsilon e^{V_1+V_2} \omega_2  \; , \\
& \Omega = e^{2U_1+V_1} \left[ (1-i z) \alpha_0 + (1+i z) \alpha_1 +( i +z) \beta^0 + (-i+z) \beta^1 \right] \; ,
}}
where we introduced a sign $\epsilon= \pm 1$, the 4D scalar fields $V, U_1, V_1+ V_2$ and the almost complex structure modulus
\eq{ \label{eq:zSp2Sp1}
z = \chi + i \epsilon e^{\varphi} \, ,
}
with $\varphi \equiv V_2-V_1$. The choice of $\epsilon$ corresponds to two different branches of the $N=2$ model,
as we explain in more detail in section \ref{sec:features}.
As an alternative to these fields it is sometimes useful to introduce
\eq{\spl{ \label{ChangeSusySp2}
& u_1 = U_1 + \tfrac{1}{2} V \; , \quad
u_2 = \tfrac{1}{2}(V_1 +V_2 +V ) \; , \\
& \phi = -2U_1 - \tfrac{1}{2} (V_1 +V_2)   \; ,
\quad \varphi= V_2- V_1 \, .
}}
where $v^i \equiv e^{2u_i}$, $i=1,2$ and $\phi$ is the 4D dilaton.

The LI metric associated to this SU(3)-structure
is given by:
\eq{\label{frameSp2}
\d s_7^2 = \tfrac{1}{4} \left( (e^1)^2+(e^2)^2+(e^3)^2+(e^4)^2\right)
+ \tfrac{1}{2} e^{2 V_1}  \left( (e^5)^2 + 2\,\chi e^5e^6 + (\chi^2 + e^{2\varphi})(e^6)^2\right)
+ e^{2V} \theta^2 \, .
}

We conclude our description of this coset by explicitly giving some relevant torsion classes of the SU(3)-structure (\ref{strSp2}):
\eq{\spl{ \label{torSp2}
R & = \tfrac{2}{3} \left(2 e^{V-2U_1} + \epsilon e^{V-V_1-V_2}  \right) \; ,  \\
E & = i e^{-V-V_1-V_2} \left(e^{2 V_2}+2 e^{V_1+V_2}+e^{2 V_1} \left(1+\chi ^2\right)\right) \; , \\
W_1 &=  \tfrac{2}{3} (z-i)\left(-e^{V_1-2U_1} + \epsilon e^{-V_2}  \right) \; ,  \\
T_1 & = \tfrac{2}{3} e^{V} \big[ \left(1 -\epsilon e^{2U_1-V_1-V_2} \right) \omega_1 + 2 \left(1 -\epsilon e^{-2U_1 +V_1+V_2} \right) \omega_2 \big] \; ,  \\
W_2 & = \tfrac{2}{3} (z-i) e^{V_1} \big[ -\left(2 +\epsilon e^{2U_1-V_1-V_2} \right) \omega_1 + 2 \left(1 +2\epsilon e^{-2U_1 +V_1+V_2} \right) \omega_2 \big] \; ,
}}
and we omitted the complicated expressions for $W_3$ and $S$.

We find two important points in LI moduli space.
The first one is  $U_1=U_2=V_1=V_2=V=\chi=0$ (the round point), where the metric defined by (\ref{frameSp2}) becomes
Einstein. For $\epsilon=+1$ the torsion classes obey the conditions for a Sasaki-Einstein
structure as given in (\ref{SE7stru}), while for $\epsilon=-1$ they obey the conditions of \eqref{halfflat7d}.
This means that the supersymmetry of the round solution, which is $N=3$ in the $N=4$ theory, is broken to respectively
$N=2$ and $N=1$ for the branches $\epsilon=\pm 1$.

The other relevant point is $e^{2V_1} =e^{2V_2} =e^{2V} =\tfrac{1}{5} e^{2U_1}$ (the squashed point),
which is also an Einstein point for the metric defined by (\ref{frameSp2}).
At this point, and for $\epsilon =-1$ only, the classes (\ref{torSp2}) become as given in (\ref{halfflat7d}),
thus giving $N=1$ at this vacuum. In the branch with $\epsilon=1$ the supersymmetry of this vacuum becomes invisible.

\subsection{$N(k,l)$} \label{Nklsec}

We finally  consider the Aloff--Wallach spaces $N(k,l)={\rm SU}(3)/{\rm U}(1)$, 
introduced in the supergravity literature in \cite{casnpqr} (where they were called $N^{pqr}$).
The integers $k,l$ specify the embedding of U(1) into SU(3) according to
\be\label{embeddingU1inSU3}
e^{i\theta} \quad\rightarrow\quad{\rm diag}(e^{ik\theta},\,e^{il\theta}\,,e^{-i(k+l)\theta})\,.
\ee
Accordingly, we chose the coset generators to be
\begin{eqnarray} \label{GenNkl}
\{ {\cal K}_\alpha \}= \{ \frac{1}{2i}\left( \lambda_4 , \- \lambda_5 , \ \lambda_6 , \ \lambda_7 , \ \lambda_1 , \ \lambda_2 , \   \tfrac{1}{2\sqrt{k^2+kl+l^2}}\left(\sqrt{3}(l+k)\lambda_3 + (l-k)\lambda_8\right) \right) \} \; , %\ \alpha =1, \ldots , 7,
\end{eqnarray}
in terms of the Gell-Mann matrices $\lambda_a$, $a=1,\ldots, 8$, and the generator of $H= {\rm U}(1)$  to be
\begin{eqnarray}  \label{GenHNkl}
{\cal H}_1= \tfrac{1}{4i\sqrt{k^2+kl+l^2}} \left( (l-k)\lambda_3 - \sqrt{3}(l+k) \lambda_8  \right).
\end{eqnarray}
The integers $k,l$ are taken relatively prime, which guarantees the space to be simply connected. Moreover, from \eqref{embeddingU1inSU3} we see that there is a permutation symmetry $(k,l)\to(l,k)$ and $(k,l)\to(k,-k-l)$.

For generic $k$, $l$, we chose the LI forms as follows
\eq{\spl{ \label{invformsNkl}
& \theta \equiv \frac{\sqrt{k^2+kl+l^2}}{\sqrt 3} e^7 \; , \\[8pt]
& \omega_1 \equiv  e^{12} \; , \;
\omega_2 \equiv  e^{34} \; , \;
\omega_3 \equiv  e^{56}   \, , \quad
\tilde{\omega}^1 \equiv   e^{3456} \; , \;
\tilde{\omega}^2 \equiv   e^{1256} \; , \;
\tilde{\omega}^3 \equiv   e^{1234}  \; , \;
\tilde{\omega}^0 \equiv e^{123456} \; ,\\[8pt]
& \alpha_0 \equiv \tfrac{1}{2} \left( e^{135} -  e^{146} - e^{236} -  e^{245} \right) , \quad
\beta^0 \equiv  \tfrac{1}{2} \left( e^{246} -  e^{136} - e^{145} -  e^{235} \right) \, ,
}}
leading to a truncation with $n_V=3$, $n_H=1$. We find furthermore for the tensor ${\cal K}_{ijk}$ in (\ref{omomK})
\eq{
\mathcal{K}_{123} = 1 \, .
}
These LI forms were also found in \cite{Micu:2006ey}, but that reference seems to have missed the fact that,
just like in the case of $M^{pqr}$ and $Q^{pqr}$, for certain values of $k$, $l$ an enhancement in the number of LI forms occurs.
We find two special cases. For $k=l=1$ we find $n_V=3$, $n_H=2$, while for  $k=-l=1$ we find $n_V=5$ and $n_H=2$.
We will omit the details for $N(1,1)$, which leads to an effective theory consisting of the mere addition of
a Betti vector multiplet to the Sp(2)/Sp(1) model discussed in section \ref{Sp2modSp1},
and instead give some details for $N(1,-1)$. We find then two extra two-forms (and corresponding dual four-forms)
\eq{\spl{ \label{invformsN1m1}
& \omega_4 \equiv  \left( e^{13} + e^{24} \right)  \; , \qquad
\omega_5 \equiv  \left( e^{14} - e^{23} \right)  \; , \\[8pt]
&
\tilde{\omega}^4 \equiv  -\tfrac{1}{2} \left( e^{13} + e^{24} \right) \wedge e^{56} \; , \qquad \tilde{\omega}^5 \equiv  -\tfrac{1}{2} \left( e^{14} - e^{23} \right)  \wedge e^{56} \,.
}}
The LI forms on $N(k,l)$ for generic $k$, $l$ obey the algebra
\eq{\spl{ \label{extalgN1m1}
& d\theta  =  \tfrac{1}{2} \left[ l\, \omega_1 + k\, \omega_2 - (k+l)\, \omega_3 \right] , \quad 
 d\omega_1 = d\omega_2 = d\omega_3 = \alpha_0 \,, \qquad   \\%[8pt]
& d\alpha_0 = 0\, , \quad
d\beta^0 = - \tilde{\omega}^1 - \tilde{\omega}^2
- \tilde{\omega}^3 , \quad %\\%[8pt]
 d \tilde{\omega}^1 = d \tilde{\omega}^2 =  d\tilde{\omega}^3 = 0 \, , \quad   d\tilde{\omega}^0 = 0\,,\quad
}}%
giving as usual directly the values for the geometric fluxes.

For $N(1,-1)$ the extra forms give rise to the additional exterior algebra
\eq{\spl{ \label{extalg1m1}
& d\omega_4 = - 3\, \theta\wedge \omega_5  \; , \qquad
d\omega_5 =   3\, \theta\wedge \omega_4 \; ,  \\
& d\tilde{\omega}_4 = - 3\, \theta\wedge \tilde{\omega}_5  \; , \qquad
d\tilde{\omega}_5 =   3\, \theta\wedge \tilde{\omega}_4 \, ,
}}
leading to $q$-charges and thus to vector multiplet gauging, according to the general discussion.
In our conventions, the only non-vanishing components for the tensor ${\cal K}_{ijk}$ in (\ref{omomK}) for $N(1,-1)$ are (up to symmetric
permutations)
\begin{eqnarray} \label{KijkN1m1}
{\cal K}_{123} = 1 , \quad {\cal K}_{344} = -2 , \quad {\cal K}_{355} = -2 .
\end{eqnarray}

The space $N(1,-1)$ supports an SU(3)-structure specified by
\eq{ \label{strN1m1}
 \eta =  e^V  \theta \; , \quad
 J =  e^{-V} v^i \omega_i   \, , \quad
\Omega = 2 \, e^{U_1+U_2+U_3} \left(\alpha_0 + i \beta^0 \right) \; ,
}
with
\eq{\spl{ \label{N1m1moduli}
& v^1 = e^{2U_1 +V} \cosh t , \quad v^2 = e^{2U_2 +V} \cosh t , \quad  v^3 = e^{2U_3 +V} , \;  \\
& v^4 = -e^{U_1+U_2+V} \sinh t \sin \sigma , \quad  v^5 = e^{U_1+U_2+V} \sinh t \cos \sigma , \; \\
& \phi = -U_1 -U_2 - U_3 .
}}
The LI metric is then given by
\begin{multline}
\d s_7^2 =  e^{2 U_1} \cosh t \left((e^1)^2+(e^2)^2\right) + e^{2 U_2} \cosh t \left((e^3)^2+(e^4)^2\right) \\
+ 2\,e^{U_1 + U_2} \sinh t \left(\cos \sigma (e^1 e^3+e^2 e^4) + \sin \sigma (e^1 e^4 - e^2 e^3) \right)\\
+ e^{2 U_3} \left((e^5)^2 + (e^6)^2\right) + e^V \theta^2 \,.
\end{multline}

The two (supersymmetric) Einstein metrics supported by any member in the $N(k,l)$ family are also found on $N(1,-1)$, but in this case they become isometric \cite{squashedpagepope}, since they are related by an exchange of the values of $U_1$ and $U_2$, which is a symmetry of the coset. The supersymmetric Einstein point is located at
\begin{eqnarray} \label{EinsN1m1}
& e^{2U_1}=\tfrac{1}{30} \left( 5 \pm \sqrt{5} \right), \quad
 e^{2U_2}= \tfrac{1}{30} \left( 5 \mp \sqrt{5} \right),  \quad
 e^{2U_3}= \tfrac{1}{5}, \quad
 e^{2V} = \tfrac{16}{45} , \nonumber \\[5pt]
& t=0 , \quad \theta \ \textrm{arbitrary}
\end{eqnarray}
normalized as $R_{mn} = 6 g_{mn}$. In addition, we numerically find an Einstein locus with $t\neq 0$.
Because of this non-diagonal entry, this solution was missed in the old supergravity literature,
where only diagonal LI metric deformations associated with rescalings of the canonical vielbeins were considered.
This Einstein metric is the same as the one found in the mathematical literature in \cite{Einstein7},
and is analogous to the (non-supersymmetric) Einstein  metric on the 5D $T^{1,1}$ coset space found in
\cite{Alekseevsky, T11red}.
In \cite{Einstein7}, a search for Einstein metrics taking into account the off-diagonal left-invariant modes
on $N(1,1)$ and $N(1,-1)$ was performed, with the conclusion that both these manifolds admit exactly two inequivalent homogeneous Einstein metrics. Hence on $N(k,l)$ there are exactly two inequivalent Einstein metrics for any choice of $k$ and $l$.

\section{Supersymmetric spectra} \label{sec:susyspectra}

Here we give the spectrum for each model at all supersymmetric, $N=2$ or $N=1$, points, and arrange it into representations of OSp$(4|N)$. The full, untruncated, Kaluza-Klein spectrum of 11D supergravity on $M^{110}$ and $Q^{111}$ has been given in \cite{Fabbri:1999mk} and \cite{Merlatti:2000ed}, respectively. We find perfect agreement with those references. Besides the long vector multiplet of the universal truncation, the extra modes in these models arrange themselves into one and two massless (Betti) vector multiplets, respectively. Similarly, the full $V_{5,2}$ spectrum has been given in \cite{Ceresole:1999zg}. As noted there, in this case, multiplet shortening can occur for multiplets such that $\Delta = |R|$, where $\Delta$ is the conformal dimension and $R$ the R-charge of the lowest component. We do find this shortening occurring, producing two chiral multiplets in the spectrum besides the universal long vector multiplet. We find a small disagreement with \cite{Ceresole:1999zg} in that our chiral multiplets are, by construction, singlets under SO(5), rather than lying on a non-trivial representation. We also present the spectra for the $N=2$ and $N=1$ points of the squashed $S^7$ and $N(1,1)$ models with, respectively, $\epsilon =+1$ and $\epsilon =-1$ in the notation of appendix \ref{Sp2modSp1}. Further, we present the (numerical, for some states) spectrum for a representative ($k=2$, $l=1$), of the generic $N(k,l)$ model, for both $N=1$ points I and II. We conclude with the (again numerical for some states) $N(1,-1)$ spectrum at the only $N=1$ point.  We believe that these $N(k,l)$ and $N(1,-1)$ spectra are new. The spectrum for SU(4)/SU(3) coincides with the one given in \cite{vargaun2}.

We have also included the vector spectrum, which we have computed in the symplectic frame of the action of section \ref{sec:4Daction}. In this frame, the two-form $B$ is massive and, in fact, it acquires the same mass across all models, for both the $N=2$ and $N=1$ points. The corresponding entry in the tables has been labeled as spin 1, since this would correspond to a massive vector in a purely electric frame.  Horizontal lines in the tables separate different OSp multiplets, a summary of which has been brought to  tables \ref{table:SummaryN=2Points}, \ref{table:SummaryN=1Points} in the main text. Note that some of the subtruncation patterns discussed in section \ref{sec:TheModels} can be appreciated from the tables: these translate into setting to zero all eigenstates in certain OSp multiplets. 

Our mass matrix conventions are those of \cite{vargaun2}. Scalar eigenstates follow the notation of  appendix \ref{sec:cosets}, except for the $N(1,-1)$ scalars $(t, \sigma)$, for which the mass matrix degenerates and a field redefinition to new (real) scalars $(x,y)$ needs to be performed as, for example, 
\begin{equation}
x+ i y = \frac{1}{2} e^{i \sigma} \tanh \frac{t}{2} ,
\end{equation}
in order to obtain the right masses in this sector. We have used  the geometric fluxes quoted in tables \ref{GeomFluxTable} and \ref{GeomFluxCntdTable}, except for the $N(2,1)$ and $N(1,-1)$ models, where computationally more conventient choices were made. Note that only the eigenstates, not the masses, may depend on the choice of geometric fluxes. We have always chosen the largest root, $\Delta_+$, of $\Delta (\Delta -3) = m^2 L^2$, where $L$ is the radius of AdS. When the smallest root $\Delta_-$ is also needed to fill out a multiplet, we have denoted the corresponding entry as $(\Delta_+ , \Delta_-)$. Finally, besides the supersymmetric spectra, we also include the non-supersymmetric spectra at the corresponding skew-whiffed points.  These are attained for $e_0 >0$ and $e_0 <0$, respectively.

%%%%%%%%%%%%%%%%%%%%%%%%%%%%%%%%%%%%%%%%%%%%%%%%%%%%%%%%%%%%%
%%%%%%%%%%%%%%%%%%%%%%%%%%%%%%%%%%%%%%%%%%%%%%%%%%%%%%%%%%%%%
%%%%%%%%%%%%%%%%%%%%%%%%%%%%%%%%%%%%%%%%%%%%%%%%%%%%%%%%%%%%%
%%%%%%%%%%%%%%%%%%%%%%%%%%%%%%%%%%%%%%%%%%%%%%%%%%%%%%%%%%%%%

\begin{sidewaystable}%[!h]
\begin{center}
{\scriptsize

\scalebox{0.910}{
\begin{tabular}{cccccccccccc}
\hline
spin	
& 	
$S^7=\frac{SU(4)}{SU(3)}$
&   	
$M^{110}$
&	
$Q^{111}$
&	
$V_{5,2}$
&	
$S^7=\frac{Sp(2)}{Sp(1)}$
&	
$N(1,1)$
&
$U(1)$
&			
\multicolumn{2}{c}{susy}
&
\multicolumn{2}{c}{skew} \\
\\[-14pt]

& 	
mass eigenstate
&   	
mass eigenstate
&	
mass eigenstate
&	
mass eigenstate
&	
mass eigenstate
&	
mass eigenstate
&
R-charge
&
$m^2 L^2$
&
$\Delta$
&
$m^2 L^2$
&
$\Delta$
\\
\hline
\\[-10pt]
2
& 	
$g_{\mu \nu}$
&   	
$g_{\mu \nu}$
&	
$g_{\mu \nu}$
&	
$g_{\mu \nu}$
&	
$g_{\mu \nu}$
&	
$g_{\mu \nu}$
&
0
&
0
&
3
&
0
&
3
\\
1
& 	
$A^0$
&   	
$A^0$
&	
$A^0$
&	
$A^0$
&	
$A^0$
&	
$A^0$
&
0
&
0
&
$2$
&
0
&
$2$
\\
\hline
\\[-10pt]
1
& 	
$B$
&   	
$B$
&	
$B$
&	
$B$
&	
$B$
&	
$B$
&
0
&
$12$
&
$5$
&
$12$
&
$5$
\\
0
& 	
$6 U_1 +V$
&   	
$4 U_1 +2U_2 +V$
&	
$2 U_1 +2U_2 +2U_3 + V$
& 	
$6 U_1 +V$
& 	
$4 U_1 + V_1 + V_2 + V_3$
&   	
$2 U_1 + 2 U_2 + V_1 + V_2 + V_3$
&
0
&
18
&
$6$
&
18
&
$6$
\\
0
& 	
$3b^1$
&   	
$2b^1+b^2$
&	
$b^1+b^2+b^3$
& 	
$3b^1$
& 	
$2b^1+b^2$
&   	
$b^1+b^2+b^3$
&
0
&
10
&
$5$
&
$-2$
&
$2$
\\
0
& 	
$\xi^0$
&   	
$\xi^0$
&	
$\xi^0$
& 	
$\xi^{\prime 0} \equiv \tfrac{1}{2}\xi^0 +\tfrac{\sqrt{3}}{2} \tilde{\xi}_1$
& 	
$\xi^0$
&   	
$\xi^0$
&
2
&
10
&
$5$
&
$-2$
&
$2$
\\
0
& 	
$\tilde\xi_0$
&   	
$\tilde\xi_0$
&	
$\tilde\xi_0$
& 	
$\tilde{\xi}^{\prime}_{0} \equiv  -\tfrac{\sqrt{3}}{2} \xi^1 + \tfrac{1}{2} \tilde{\xi}_0$
& 	
$\tilde\xi_0$
&   	
$\tilde\xi_0$
&
$-2$
&
10
&
$5$
&
$-2$
&
$2$
\\
0
& 	
$3U_1-3V$
&   	
$2U_1+U_2 -3V$
&	
$U_1+U_2 +U_3 -3V$
& 	
$3U_1-3V$
& 	
$2U_1-2V_3$
&   	
$U_1 + U_2 -2V_3$
&
0
&
4
&
$4$
&
4
&
$4$
\\
\hline
\\[-10pt]
1
&
--
&
--
&
--
&
--
& 	
$2A^1 - 2A^2$
&   	
$A^1 + A^2 -2A^3 $
&
0
&
$6$
&
$4$
&
$6$
&
$4$
\\
0
&
--
&
--
&
--
&
--
& 	
$2b^1 - \tilde{\xi}_1$
&   	
$b^1 + b^2 - \tilde{\xi}_1$
&
0
&
$10$
&
$5$
&
$-2$
&
$2$
\\
0
&
--
&
--
&
--
&
--
& 	
$2U_1 - V_1 - V_2$
&   	
$U_1 +U_2 - V_1 - V_2$
&
0
&
4
&
$4$
&
$4$
&
$4$
\\
0
&
--
&
--
&
--
&
--
& 	
$\varphi \equiv V_2 - V_1$
&   	
$\varphi \equiv  V_2 - V_1$
&	
2
&
4
&
$4$
&
4
&
$4$
\\
0
&
--
&
--
&
--
&
--
& 	
$\chi$
&   	
$\chi$
&	
$-2$
&
4
&
$4$
&
$4$
&
$4$
\\
0
&
--
&
--
&
--
&
--
& 	
$2b^1 -2b^2 +2  \tilde{\xi}_1$
&   	
$b^1 +b^2 -2b^3 +2  \tilde{\xi}_1$
&	
0
&
0
&
$3$
&
18
&
$6$
\\
\hline
\\[-10pt]
1
& 	
--
&   	
$A^1- A^2$
&	
$A^1- A^2$
&
--
& 	
--
&	
$A^1- A^2$
&
0
&
0
&
$2$
&
0
&
$2$
\\
0
& 	
--
&   	
$U_1-U_2$
&	
$U_1-U_2$
&
--
& 	
--
&	
$U_1-U_2$
&
0
&
$-2$
&
$(2,1)$
&
$-2$
&
$2$
\\
0
& 	
--
&   	
$b^1-b^2$
&	
$b^1-b^2$
&
--
&
--
&	
$b^1-b^2$
&
0
&
$-2$
&
$(2,1)$
&
$4$
&
$4$
\\
\hline
\\[-10pt]
1
& 	
--
&   	
--
&	
$A^1- A^3$
&
--
&
--
&
--
&
0
&
$0$
&
$2$
&
$0$
&
$2$
\\
0
& 	
--
&   	
--
&	
$U_1-U_3$
&
--
&
--
&
--
&
0
&
$-2$
&
$(2,1)$
&
$-2$
&
$2$
\\
0
& 	
--
&   	
--
&	
$b^1-b^3$
&
--
&
--
&
--
&
0
&
$-2$
&
$(2,1)$
&
$4$
&
$4$
\\
\hline
\\[-8pt]
0
&
--
&
--
&
--
& 	
$\varphi$
&
--
&
--
&
$-\tfrac{8}{3}$
&
$-\tfrac{20}{9}$
&
$\tfrac{5}{3}$
&
$-\tfrac{20}{9}$
&
$\tfrac{5}{3}$
\\[4pt]
0
&
--
&
--
&
--
& 	
$ \xi^{\prime 1} \equiv -\tfrac{\sqrt{3}}{2}\xi^0 +\tfrac{1}{2} \tilde{\xi}_1$
&
--
&
--
&
$-\tfrac{2}{3}$
&
$-\tfrac{14}{9}$
&
$\tfrac{2}{3}$
&
$\tfrac{22}{9}$
&
$\tfrac{11}{3}$
\\
\hline
\\[-8pt]
0
&
--
&
--
&
--
& 	
$\chi$
&
--
&
--
&
$-\tfrac{4}{3}$
&
$-\tfrac{20}{9}$
&
$\tfrac{5}{3}$
&
$-\tfrac{20}{9}$
&
$\tfrac{5}{3}$
\\[4pt]
0
&
--
&
--
&
--
& 	
$ \tilde{\xi}^{\prime}_{1} \equiv -\tfrac{1}{2}\xi^1 -\tfrac{\sqrt{3}}{2} \tilde{\xi}_0$
&
--
&
--
&
$\tfrac{2}{3}$
&
$-\tfrac{14}{9}$
&
$\tfrac{2}{3}$
&
$\tfrac{22}{9}$
&
$\tfrac{11}{3}$
\\
\hline
\\[-10pt]
\end{tabular}
}
\caption[SEspectra]{The bosonic LI spectrum on the $N=2$ and skew-whiffed points across all models. A hyphen indicates an absent state. Horizontal lines separate the different OSp$(4|2)$ multiplets at the susy point. From top to bottom, these are: gravity multiplet, two long vector multiplets, two massless vector multiplets and two chiral multiplets.}\label{table:SEspectra}
}\normalsize
\end{center}
\end{sidewaystable}

%%%%%%%%%%%%%%%%%%%%%%%%%%%%%%%%%%%%%%%%%%%%%%%%%%%%%%%%%%%%%
%%%%%%%%%%%%%%%%%%%%%%%%%%%%%%%%%%%%%%%%%%%%%%%%%%%%%%%%%%%%%
%%%%%%%%%%%%%%%%%%%%%%%%%%%%%%%%%%%%%%%%%%%%%%%%%%%%%%%%%%%%%
%%%%%%%%%%%%%%%%%%%%%%%%%%%%%%%%%%%%%%%%%%%%%%%%%%%%%%%%%%%%%

%%%%%%%%%%%%%%%%%%%%%%%%%%%%%%%%%%%%%%%%%%%%%%%%%%%%%%%%%%%%%
%%%%%%%%%%%%%%%%%%%%%%%%%%%%%%%%%%%%%%%%%%%%%%%%%%%%%%%%%%%%%
%%%%%%%%%%%%%%%%%%%%%%%%%%%%%%%%%%%%%%%%%%%%%%%%%%%%%%%%%%%%%
%%%%%%%%%%%%%%%%%%%%%%%%%%%%%%%%%%%%%%%%%%%%%%%%%%%%%%%%%%%%%
\begin{table}%[!h]
\begin{center}
{\scriptsize
\begin{tabular}{ccccccc}
\hline
spin	
& 	
$S^7=\frac{Sp(2)}{Sp(1)}$
&   	
$N(1,1)$
&			
\multicolumn{2}{c}{susy}
&
\multicolumn{2}{c}{skew} \\
\\[-14pt]

& 	
mass eigenstate
&   	
mass eigenstate
&
$m^2 L^2$
&
$\Delta$
&
$m^2 L^2$
&
$\Delta$
\\
\hline
\\[-10pt]
2
& 	
$g_{\mu \nu}$
&   	
$g_{\mu \nu}$
&
0
&
3
&
0
&
3
\\
\hline
\\[-10pt]
1
& 	
$B$
&   	
$B$
&
$12$
&
$5$
&
$12$
&
$5$
\\[4pt]
1
& 	
$2A^1 - 2A^2$
&   	
$A^1 + A^2 - 2A^3$
&
$6$
&
$4$
&
$6$
&
$4$
\\
\hline
\\[-10pt]
0
& 	
$4 U_1 + V_1 + V_2 + V_3$
&   	
$2 U_1 + 2 U_2 + V_1 + V_2 + V_3$
&
18
&
$6$
&
18
&
$6$
\\[4pt]
0
& 	
$2b^1 -5b^2 +2\tilde{\xi}_1$
&   	
$b^1 +b^2 -5b^3 +2\tilde{\xi}_1$
&
10
&
$5$
&
$-2$
&
$2$
\\
\hline
\\[-10pt]
0
& 	
$V_1 + V_2-2V_3$
&   	
$V_1 + V_2-2V_3$
&
$\tfrac{52}{9}$
&
$\tfrac{13}{3}$
&
$\tfrac{52}{9}$
&
$\tfrac{13}{3}$
\\[4pt]
0
& 	
$2b^1 -\tilde{\xi}_1$
&   	
$b^1 + b^2 -\tilde{\xi}_1$
&
$\tfrac{10}{9}$
&
$\tfrac{10}{3}$
&
$\tfrac{190}{9}$
&
$\tfrac{19}{3}$
\\[4pt]
\hline
\\[-8pt]
0
& 	
$\varphi \equiv V_2-V_1$
&   	
$\varphi \equiv V_2-V_1$
&
$\tfrac{52}{9}$
&
$\tfrac{13}{3}$
&
$\tfrac{52}{9}$
&
$\tfrac{13}{3}$
\\[4pt]
0
& 	
$\xi^0$
&   	
$\xi^0$
&
$\tfrac{10}{9}$
&
$\tfrac{10}{3}$
&
$\tfrac{190}{9}$
&
$\tfrac{19}{3}$
\\[4pt]
\hline
\\[-8pt]
0
& 	
$\chi$
&   	
$\chi$
&
$\tfrac{52}{9}$
&
$\tfrac{13}{3}$
&
$\tfrac{52}{9}$
&
$\tfrac{13}{3}$
\\[4pt]
0
& 	
$\tilde{\xi}_0$
&   	
$\tilde{\xi}_0$
&
$\tfrac{10}{9}$
&
$\tfrac{10}{3}$
&
$\tfrac{190}{9}$
&
$\tfrac{19}{3}$
\\[4pt]
\hline
\\[-8pt]
0
& 	
$ 2b^1 +30 b^2 +2 \tilde{\xi}_1$
&   	
$ b^1 + b^2 + 30 b^3 +2 \tilde{\xi}_1$
&
$-\tfrac{8}{9}$
&
$\tfrac{8}{3}$
&
$\tfrac{10}{9}$
&
$\tfrac{10}{3}$
\\[4pt]
0
& 	
$6U_1-2V_1-2V_2-2V_3$
&   	
$3U_1 + 3U_2 -2V_1 -2V_2  -2V_3$
&
$-\tfrac{20}{9}$
&
$\tfrac{5}{3}$
&
$-\tfrac{20}{9}$
&
$\tfrac{5}{3}$
\\
\hline
\\[-8pt]
1
& 	
$A^0$
&   	
$A^0$
&
0
&
$2$
&
0
&
$2$
\\
\hline
\\[-8pt]
1
& 	
--
&	
$A^1- A^2$
&
$0$
&
$2$
&
$0$
&
$2$
\\
\hline
\\[-8pt]
0
& 	
--
&   	
$ b^1 - b^2 $
&
$\tfrac{70}{9}$
&
$\tfrac{14}{3}$
&
$-\tfrac{20}{9}$
&
$\tfrac{5}{3}$
\\[4pt]
0
& 	
--
&   	
$U_1 - U_2$
&
$\tfrac{22}{9}$
&
$\tfrac{11}{3}$
&
$\tfrac{22}{9}$
&
$\tfrac{11}{3}$
\\[4pt]
\hline
\end{tabular}
\caption[Sp2Sp1Sqspectra]{The bosonic LI spectrum on the squashed ($N=1$) and skew-whiffed points in the $S^7=Sp(2)/Sp(1)$ and $N(1,1)$ models.}\label{table:Sp1Sp1Sqspectra}
}\normalsize
\end{center}
\end{table}
%%%%%%%%%%%%%%%%%%%%%%%%%%%%%%%%%%%%%%%%%%%%%%%%%%%%%%%%%%%%%
%%%%%%%%%%%%%%%%%%%%%%%%%%%%%%%%%%%%%%%%%%%%%%%%%%%%%%%%%%%%%
%%%%%%%%%%%%%%%%%%%%%%%%%%%%%%%%%%%%%%%%%%%%%%%%%%%%%%%%%%%%%
%%%%%%%%%%%%%%%%%%%%%%%%%%%%%%%%%%%%%%%%%%%%%%%%%%%%%%%%%%%%%

%%%%%%%%%%%%%%%%%%%%%%%%%%%%%%%%%%%%%%%%%%%%%%%%%%%%%%%%%%%%%
%%%%%%%%%%%%%%%%%%%%%%%%%%%%%%%%%%%%%%%%%%%%%%%%%%%%%%%%%%%%%
%%%%%%%%%%%%%%%%%%%%%%%%%%%%%%%%%%%%%%%%%%%%%%%%%%%%%%%%%%%%%
%%%%%%%%%%%%%%%%%%%%%%%%%%%%%%%%%%%%%%%%%%%%%%%%%%%%%%%%%%%%%
\begin{table}%[!h]
\begin{center}
{\scriptsize
\begin{tabular}{cccccc}
\hline
spin	
& 	
$N(2,1)_I$
&   	
\multicolumn{2}{c}{susy}
&
\multicolumn{2}{c}{skew} \\
\\[-14pt]

& 	
mass eigenstate
&
$m^2 L^2$
&
$\Delta$
&
$m^2 L^2$
&
$\Delta$
\\
\hline
\\[-10pt]
2
& 	
$g_{\mu \nu}$
&
0
&
3
&
0
&
3
\\
\hline
\\[-10pt]
1
& 	
$B$
&   	
$12$
&
$5$
&
$12$
&
$5$
\\
1
& 	
$c_i A^i$
&   	
$6$
&
$4$
&
$6$
&
$4$
\\
\hline
\\[-10pt]
0
& 	
$2U_1 + 2U_2 + 2U_3 + V$
&
18
&
$6$
&
18
&
$6$
\\
0
& 	
$b^1 +b^2  +b^3 -2\tilde{\xi}_0$
&
10
&
$5$
&
$-2$
&
$2$
\\
\hline
\\[-10pt]
0
& 	
$x_i^\prime b^i +  \tilde{\xi}_0$
&
$m^2L^2$
&
$\Delta_1 + 1$
&
$m^2L^2$
&
$\Delta_1 - 2$
\\
0
& 	
$x^i U_i + V$
&
$m^2L^2$
&
$\Delta_1$
&
$m^2L^2$
&
$\Delta_1$
\\
\hline
\\[-10pt]
0
& 	
$y^i U_i + V$
&
$m^2L^2$
&
$\Delta_2 +1$
&
$m^2L^2$
&
$\Delta_2 +1$
\\
0
& 	
$y_i^\prime b^i +  \tilde{\xi}_0$
&
$m^2L^2$
&
$\Delta_2$
&
$m^2L^2$
&
$\Delta_2 + 3$
\\
\hline
\\[-10pt]
0
& 	
$z_i^\prime b^i +  \tilde{\xi}_0$
&
$m^2L^2 $
&
$(\Delta_3 + 1, 2-\Delta_3)$
&
$m^2L^2 $
&
$5-\Delta_3$
\\
0
& 	
$z^i U_i + V$
&
$m^2L^2$
&
$(3- \Delta_3, \Delta_3)$
&
$ m^2L^2$
&
$3- \Delta_3$
\\
\hline
\\[-10pt]
1
& 	
$A^0$
&
0
&
$2$
&
0
&
$2$
\\
\hline
\\[-10pt]
1
&   	
$c^\prime_i A^i$
&	
0
&
$2$
&
0
&
$2$
\\
\hline
\end{tabular}
\caption[N21Ispectrum]{{\scriptsize The bosonic LI spectrum on the $N=1$ and skew-whiffed critical point I of the $N(2,1)$ model. We have defined $(\Delta_1 , \Delta_2 , \Delta_3) = (3.88744, 3.14403, 1.25659)$, $(x_1,x_2,x_3) = (3.17844, 1.62696, -5.8054)$, $(y_1,y_2,y_3) = (-0.176003, -0.837448, 0.013451)$, $(z_1,z_2,z_3) = (-3.43257, 3.09891, -0.666343)$ and, for $i=1,2,3$, $(x_i^\prime , y_i^\prime , z_i^\prime) = (x^i + 1 , y^i + 1 , z^i + 1)$. Scalar masses not explicitly indicated are related to their corresponding $\Delta$, to their right, via $m^2L^2 = \Delta (\Delta-3)$. We omit the specification of the vector mixings $c_i$, $c^\prime_i$, $i=1,2,3$.} }\label{table:N21Ispectrum}
}\normalsize
\end{center}
\end{table}
%%%%%%%%%%%%%%%%%%%%%%%%%%%%%%%%%%%%%%%%%%%%%%%%%%%%%%%%%%%%%
%%%%%%%%%%%%%%%%%%%%%%%%%%%%%%%%%%%%%%%%%%%%%%%%%%%%%%%%%%%%%
%%%%%%%%%%%%%%%%%%%%%%%%%%%%%%%%%%%%%%%%%%%%%%%%%%%%%%%%%%%%%
%%%%%%%%%%%%%%%%%%%%%%%%%%%%%%%%%%%%%%%%%%%%%%%%%%%%%%%%%%%%%

%%%%%%%%%%%%%%%%%%%%%%%%%%%%%%%%%%%%%%%%%%%%%%%%%%%%%%%%%%%%%
%%%%%%%%%%%%%%%%%%%%%%%%%%%%%%%%%%%%%%%%%%%%%%%%%%%%%%%%%%%%%
%%%%%%%%%%%%%%%%%%%%%%%%%%%%%%%%%%%%%%%%%%%%%%%%%%%%%%%%%%%%%
%%%%%%%%%%%%%%%%%%%%%%%%%%%%%%%%%%%%%%%%%%%%%%%%%%%%%%%%%%%%%
\begin{table}%[!h]
\begin{center}
{\scriptsize
\begin{tabular}{cccccc}
\hline
spin	
& 	
$N(2,1)_{II}$
&   	
\multicolumn{2}{c}{susy}
&
\multicolumn{2}{c}{skew} \\
\\[-14pt]

& 	
mass eigenstate
&
$m^2 L^2$
&
$\Delta$
&
$m^2 L^2$
&
$\Delta$
\\
\hline
\\[-10pt]
2
& 	
$g_{\mu \nu}$
&
0
&
3
&
0
&
3
\\
\hline
\\[-10pt]
1
& 	
$B$
&   	
$12$
&
$5$
&
$12$
&
$5$
\\
1
& 	
$c_i A^i$
&   	
$6$
&
$4$
&
$6$
&
$4$
\\
\hline
\\[-10pt]
0
& 	
$2U_1 + 2U_2 + 2U_3 + V$
&
18
&
$6$
&
18
&
$6$
\\
0
& 	
$b^1 +b^2  +b^3 +2\tilde{\xi}_0$
&
10
&
$5$
&
$-2$
&
$2$
\\
\hline
\\[-10pt]
0
& 	
$x_i^\prime b^i -  \tilde{\xi}_0$
&
$m^2 L^2$
&
$\Delta_1 + 1$
&
$m^2 L^2$
&
$\Delta_1 -2$
\\
0
& 	
$x^i U_i +  V$
&
$m^2 L^2$
&
$\Delta_1$
&
$m^2 L^2$
&
$\Delta_1$
\\
\hline
\\[-10pt]
0
& 	
$y^i U_i +  V$
&
$m^2 L^2$
&
$\Delta_2 + 1$
&
$m^2 L^2$
&
$\Delta_2 + 1$
\\
0
& 	
$y_i^\prime b^i -  \tilde{\xi}_0$
&
$m^2 L^2$
&
$\Delta_2$
&
$m^2 L^2$
&
$\Delta_2 + 3$
\\
\hline
\\[-10pt]
0
& 	
$z_i^\prime b^i -  \tilde{\xi}_0$
&
$m^2 L^2$
&
$\Delta_3 + 1$
&
$m^2 L^2$
&
$\Delta_3 + 2$
\\
0
& 	
$z^i U_i +  V$
&
$m^2 L^2$
&
$\Delta_3$
&
$m^2 L^2$
&
$\Delta_3$
\\
\hline
\\[-10pt]
1
& 	
$A^0$
&
0
&
$2$
&
0
&
$2$
\\
\hline
\\[-10pt]
1
&   	
$c^\prime_i A^i$
&	
0
&
$2$
&
0
&
$2$
\\
\hline
\end{tabular}
\caption[N21IIspectrum]{{\scriptsize The bosonic LI spectrum on the $N=1$ and skew-whiffed critical point II of the $N(2,1)$ model. We have defined $(\Delta_1 , \Delta_2 , \Delta_3) = (3.67378, 3.32855, 1.65477)$, $(x_1,x_2,x_3) = (23.5626, -26.065, 1.50232)$, $(y_1,y_2,y_3) = (-0.0118381, 0.00857999, -0.996742)$, $(z_1,z_2,z_3) = (-1.68486, -1.33023, 2.0151)$ and, for $i=1,2,3$, $(x_i^\prime , y_i^\prime , z_i^\prime) = (x^i +1 , y^i + 1 , z^i + 1)$. Scalar masses not explicitly indicated are related to their corresponding $\Delta$, to their right, via $m^2L^2 = \Delta (\Delta-3)$.  We omit the specification of the vector mixings $c_i$, $c^\prime_i$, $i=1,2,3$.}}\label{table:N21IIspectrum}
}\normalsize
\end{center}
\end{table}
%%%%%%%%%%%%%%%%%%%%%%%%%%%%%%%%%%%%%%%%%%%%%%%%%%%%%%%%%%%%%
%%%%%%%%%%%%%%%%%%%%%%%%%%%%%%%%%%%%%%%%%%%%%%%%%%%%%%%%%%%%%
%%%%%%%%%%%%%%%%%%%%%%%%%%%%%%%%%%%%%%%%%%%%%%%%%%%%%%%%%%%%%
%%%%%%%%%%%%%%%%%%%%%%%%%%%%%%%%%%%%%%%%%%%%%%%%%%%%%%%%%%%%%

%%%%%%%%%%%%%%%%%%%%%%%%%%%%%%%%%%%%%%%%%%%%%%%%%%%%%%%%%%%%%
%%%%%%%%%%%%%%%%%%%%%%%%%%%%%%%%%%%%%%%%%%%%%%%%%%%%%%%%%%%%%
%%%%%%%%%%%%%%%%%%%%%%%%%%%%%%%%%%%%%%%%%%%%%%%%%%%%%%%%%%%%%
%%%%%%%%%%%%%%%%%%%%%%%%%%%%%%%%%%%%%%%%%%%%%%%%%%%%%%%%%%%%%
\begin{table}%[!h]
\begin{center}
{\scriptsize
\scalebox{0.92}{
\begin{tabular}{cccccc}
\hline
spin	
& 	
$N(1,-1)$
&   	
\multicolumn{2}{c}{susy}
&
\multicolumn{2}{c}{skew} \\
\\[-14pt]

& 	
mass eigenstate
&
$m^2 L^2$
&
$\Delta$
&
$m^2 L^2$
&
$\Delta$
\\
\hline
\\[-10pt]
2
& 	
$g_{\mu \nu}$
&
0
&
3
&
0
&
3
\\
\hline
\\[-10pt]
1
& 	
$B$
&   	
$12$
&
$5$
&
$12$
&
$5$
\\
1
& 	
$c_i A^i$
&   	
$6$
&
$4$
&
$6$
&
$4$
\\
\hline
\\[-10pt]
0
& 	
$2U_1 + 2U_2 + 2U_3 + V$
&
18
&
$6$
&
18
&
$6$
\\
0
& 	
$b^1 +b^2  +b^3 -2\tilde{\xi}_0$
&
10
&
$5$
&
$-2$
&
$2$
\\
\hline
\\[-10pt]
0
& 	
$x_i^\prime b^i +  \tilde{\xi}_0$
&
$m^2 L^2$
&
$\Delta_1 + 1$
&
$m^2 L^2$
&
$\Delta_1 -2$
\\
0
& 	
$x^i U_i +V$
&
$m^2 L^2$
&
$\Delta_1$
&
$m^2 L^2$
&
$\Delta_1$
\\
\hline
\\[-10pt]
0
& 	
$y^i U_i +V$
&
$m^2 L^2$
&
$\Delta_2 + 1$
&
$m^2 L^2$
&
$\Delta_2 + 1$
\\
0
& 	
$y_i^\prime b^i +  \tilde{\xi}_0$
&
$m^2 L^2$
&
$\Delta_2$
&
$m^2 L^2$
&
$\Delta_2 + 3$
\\
\hline
\\[-10pt]
0
& 	
$z_i^\prime b^i +  \tilde{\xi}_0$
&
$m^2 L^2$
&
$\Delta_3 + 1$
&
$m^2 L^2$
&
$5-\Delta_3$
\\
0
& 	
$z^i U_i +V$
&
$m^2 L^2$
&
$\Delta_3$
&
$m^2 L^2$
&
$\Delta_3$
\\
\hline
\\[-10pt]
1
& 	
$A^0$
&
0
&
$2$
&
0
&
$2$
\\
\hline
\\[-10pt]
1
&   	
$c^\prime_i A^i$
&	
0
&
$2$
&
0
&
$2$
\\
\hline
\\[-8pt]
1
& 	
$ A^4 $
&
$\tfrac{405}{64}$
&
$\tfrac{3}{2} + \tfrac{\sqrt{421}}{8}$
&
$\tfrac{405}{64}$
&
$\tfrac{3}{2} + \tfrac{\sqrt{421}}{8}$
\\[4pt]
0
& 	
$x$
&
$\tfrac{277}{64}$
&
$\tfrac{3}{2} + \tfrac{\sqrt{421}}{8}$
&
$\tfrac{277}{64}$
&
$\tfrac{3}{2} + \tfrac{\sqrt{421}}{8}$
\\[4pt]
\hline
\\[-8pt]
1
& 	
$ A^5 $
&
$\tfrac{405}{64}$
&
$\tfrac{3}{2} + \tfrac{\sqrt{421}}{8}$
&
$\tfrac{405}{64}$
&
$\tfrac{3}{2} + \tfrac{\sqrt{421}}{8}$
\\[4pt]
0
& 	
$y$
&
$\tfrac{277}{64}$
&
$\tfrac{3}{2} + \tfrac{\sqrt{421}}{8}$
&
$\tfrac{277}{64}$
&
$\tfrac{3}{2} + \tfrac{\sqrt{421}}{8}$
\\[4pt]
\hline
\end{tabular}
}
\caption[N1m1spectrum]{{\scriptsize The bosonic LI spectrum on the $N=1$ and skew-whiffed critical point of the $N(1,-1)$ model. We have defined $(\Delta_1 , \Delta_2 , \Delta_3) = (3.72876, 3.28909, 1.56033)$, $(x_1,x_2,x_3) = (-9.69772, 1.52283, 7.17488)$, $(y_1,y_2,y_3) = (0.0177555, -0.968581, -0.0491748)$, $(z_1,z_2,z_3) = (-1.02824, 2.15395, -2.12571)$ and, for $i=1,2,3$, $(x_i^\prime , y_i^\prime , z_i^\prime) = (x^i + 1 , y^i + 1 , z^i + 1)$. Scalar masses not explicitly indicated are related to their corresponding $\Delta$, to their right, via $m^2L^2 = \Delta (\Delta-3)$. For the susy point, these masses are $m^2 =\mu$, with $\mu$  the roots of the cubics $\mu^3-24\mu^2 -56 \mu +2160$ and $\mu^3-32\mu^2 -40 \mu +560$.  We omit the specification of the vector mixings $c_i$, $c^\prime_i$, $i=1,2,3$.}}\label{table:N1m1spectrum}
}\normalsize
\end{center}
\end{table}
%%%%%%%%%%%%%%%%%%%%%%%%%%%%%%%%%%%%%%%%%%%%%%%%%%%%%%%%%%%%%
%%%%%%%%%%%%%%%%%%%%%%%%%%%%%%%%%%%%%%%%%%%%%%%%%%%%%%%%%%%%%
%%%%%%%%%%%%%%%%%%%%%%%%%%%%%%%%%%%%%%%%%%%%%%%%%%%%%%%%%%%%%
%%%%%%%%%%%%%%%%%%%%%%%%%%%%%%%%%%%%%%%%%%%%%%%%%%%%%%%%%%%%%

\newpage

\bibliography{7dcosets}
\end{document}